\def\arxivvariant{1}

\ifdefined \arxivvariant
\newcommand{\arxivonly}[1]{{#1}}
\else
\newcommand{\arxivonly}[1]{}
\fi

\documentclass{article}
\usepackage[utf8]{inputenc}
\usepackage{amsfonts}
\usepackage{amsmath}
\usepackage{colonequals}
\usepackage{graphicx}
\usepackage{epigraph}
\usepackage{jheppub}
\usepackage{array,longtable}
\usepackage{listings}
\newcolumntype{L}{>{$}l<{$}}
\newcolumntype{R}{>{$}r<{$}}
\newcolumntype{G}{>{$\scriptscriptstyle}c<{$}}

\usepackage{color}
\usepackage{hyperref}

\hypersetup{colorlinks = true, urlcolor = blue, linkcolor = black}

\newcommand{\xcite}[2]{\href{#2}{{\NoHyper}\cite{#1}{\endNoHyper}}}

\allowdisplaybreaks

\lstdefinestyle{pystyle}{
  basicstyle=\footnotesize
  }

\begin{document}

\begin{center}

\mathversion{bold}
{\bf\Large $SO(8)$ Supergravity and the Magic of Machine Learning}
\mathversion{normal}

\bigskip\bigskip\medskip

{\bf Iulia M. Comsa, Moritz Firsching, Thomas Fischbacher}
\vspace{.1cm}

{\em Google Research\\
  Brandschenkestrasse 110, 8002 Z\"urich, Switzerland\\
}
{\small \texttt{\{iuliacomsa,firsching,tfish\}@google.com}}

\end{center}

\setlength{\epigraphwidth}{0.8\textwidth}

\bigskip
\bigskip
\medskip

\begin{abstract}
\noindent
  Using de Wit-Nicolai $D=4\;\mathcal{N}=8\;SO(8)$ supergravity as an
  example, we show how modern Machine Learning software libraries
  such as Google's TensorFlow can be employed to greatly simplify
  the analysis of high-dimensional scalar sectors of some M-Theory
  compactifications.
  We provide detailed information on the location, symmetries, and
  particle spectra and charges of~192 critical points on the scalar
  manifold of SO(8)~supergravity, including one newly discovered
  $\mathcal{N}=1$ vacuum with~$SO(3)$ residual symmetry, one new
  potentially stabilizable non-supersymmetric solution, and examples for
  ``Galois conjugate pairs'' of solutions, i.e.\ solution-pairs that
  share the same gauge group embedding into~$SO(8)$ and minimal
  polynomials for the cosmological constant. Where feasible,
  we give analytic expressions for solution coordinates and
  cosmological constants.

  As the authors' aspiration is to present the discussion in a form
  that is accessible to both the Machine Learning and String Theory
  communities and allows adopting our methods towards the study of
  other models, we provide an introductory overview over the relevant
  Physics as well as Machine Learning concepts. This includes short
  pedagogical code examples. In particular, we show how to formulate a
  requirement for residual Supersymmetry as a Machine Learning loss
  function and effectively guide the numerical search towards
  supersymmetric critical points. Numerical investigations suggest
  that there are no further supersymmetric vacua beyond this newly
  discovered fifth solution.
\end{abstract}

\renewcommand{\thefootnote}{\arabic{footnote}}
\setcounter{footnote}{0}

\newpage

% Alphabets, as in "Fourteen new stationary points".

%%% % SO(8) vector
\def\va{{a}}
\def\vb{{b}}
\def\vc{{c}}
\def\vd{{d}}
\def\ve{{e}}
\def\vf{{f}}
\def\vg{{g}}
\def\vh{{h}}
%%% % SO(8) spinor
\def\sa{{\alpha}}
\def\sb{{\beta}}
\def\sc{{\gamma}}
\def\sd{{\delta}}
% SO(8) co-spinor
\def\ca{{\dot\alpha}}
\def\cb{{\dot\beta}}
\def\cc{{\dot\gamma}}
\def\cd{{\dot\delta}}
%%% % SO(8) symmetric traceless spinor
\def\sstab{{\underline{(\alpha\beta)}}}
\def\sstcd{{\underline{(\gamma\delta)}}}
%%% % SO(8) symmetric traceless co-spinor
\def\sctab{{\underline{(\dot\alpha\dot\beta)}}}
\def\sctcd{{\underline{(\dot\gamma\dot\delta)}}}
%%% % SU(8) vector (mnemonic: unitary)
\def\ua{{i}}
\def\ub{{j}}
\def\uc{{k}}
\def\ud{{l}}
\def\ue{{m}}
\def\uf{{n}}
\def\ug{{p}}
\def\uh{{q}}
\def\uA{{I}}
\def\uB{{J}}
\def\uC{{K}}
\def\uD{{L}}
%%% %\def\uap{{a'}}
%%% %\def\ubp{{b'}}
%%% %\def\ucp{{c'}}
%%% %\def\udp{{d'}}
%%% %\def\uep{{e'}}
%%% %\def\ufp{{f'}}
%%% %\def\ugp{{g'}}
%%% %\def\uhp{{h'}}
%%%
%%% % SU(8) adjoint
\def\Ua{{A}}
\def\Ub{{B}}
\def\Uc{{C}}
\def\Ud{{D}}
%%% % E7 fundamental
% \def\efa{{\tilde{\hbox{$\mathcal A$}}}}
% \def\efb{{\tilde{\hbox{$\mathcal B$}}}}
% \def\efc{{\tilde{\hbox{$\mathcal C$}}}}
% \def\efd{{\tilde{\hbox{$\mathcal D$}}}}
\def\efa{A}
\def\efb{B}
\def\efc{C}
\def\efd{D}
%%% % E7 adjoint
\def\eaa{{\mathcal A}}
\def\eab{{\mathcal B}}
\def\eac{{\mathcal C}}
\def\ead{{\mathcal D}}
%%% % "Machine" auxiliary indices
\def\ma{{\tt a}}
\def\mb{{\tt b}}
\def\mc{{\tt c}}
\def\md{{\tt d}}
\def\mm{{\tt m}}
\def\mn{{\tt n}}
\def\mp{{\tt p}}
\def\mq{{\tt q}}
\def\mi{{\tt i}}
\def\mj{{\tt j}}
\def\mk{{\tt k}}
\def\ml{{\tt l}}

\epigraph{
   At the moment, the $\mathcal{N}=8$ Supergravity Theory is the only
   candidate in sight. There are likely to be a number of crucial
   calculations within the next few years that have the possibility of
   showing that the theory is no good. If the theory survives these
   tests, it will probably be some years more before we develop
   computational methods that will enable us to make predictions and
   before we can account for the initial conditions of the universe as
   well as the local physical laws. These will be the outstanding
   problems for theoretical physics in the next twenty years or so.

   But to end on a slightly alarmist note, they may not have much more
   time than that. At present, computers are a useful aid in research, but
   they have to be directed by human minds. If one extrapolates their
   recent rapid rate of development, however, it would seem quite
   possible that they will take over altogether in theoretical
   physics. So, maybe the end is in sight for theoretical physicists,
   if not for theoretical physics.}{\textit{S. Hawking, Conclusion of
   his 1981 Inaugural lecture\xcite{Hawking:1981sq}{https://iopscience.iop.org/article/10.1088/0031-9112/32/1/024/pdf}\\
   ``Is the End in Sight for Theoretical Physics?''}}

\section{Introduction\protect\footnote{%
An expanded introduction that provides more context on M-Theory and
Machine Learning to interested readers without a deep background in
one of these subjects is available in version~3 of the arXiv preprint
of this work at: \href{https://arxiv.org/abs/1906.00207v3}{\texttt{https://arxiv.org/abs/1906.00207v3}}.
}}

Google's primary open source library for Machine Learning,
TensorFlow~\xcite{Abadi2016}{https://www.tensorflow.org/},
has many potential uses beyond Machine Learning. In
this article, we want to show how it also is an excellent tool to
address one specific technically challenging M-Theory research
problem: Finding static field configurations of dimensionally reduced
models with known but structurally complicated potentials, such
as~$SO(8)$
supergravity~\xcite{de1982nwithlocal}{https://doi.org/10.1016/0370-2693(82)91194-7}~\xcite{de1981local}{https://cds.cern.ch/record/985488/files/CM-P00062104.pdf},
which we study here, as well as determining their stability properties.
The underlying computational methods can be readily generalized to other models,
including for example maximal five-dimensional
supergravity~\xcite{gunaydin1985gauged}{https://doi.org/10.1016/0370-2693(85)90361-2}.

For the impatient reader, there is an open sourced Google Colab
at~\xcite{superstringscolab}{https://colab.sandbox.google.com/github/google-research/google-research/blob/master/m_theory/dim4/so8_supergravity_extrema/colab/so8_supergravity.ipynb}
that runs an efficient search for vacuum candidates of
$SO(8)$~supergravity and can be used interactively from within a web
browser, alongside additional Python code to analyze numerical
solutions at~\xcite{googlemtheory}{https://github.com/google-research/google-research/tree/master/m_theory}.

\subsection{On M-Theory}

``M-Theory'', or ``The Theory Formerly Known as
Strings''~\xcite{duff1998theory}{https://www.nikhef.nl/pub/services/biblio/bib_KR/sciam14395569.pdf}
is a so far only partially explored and understood unifying framework
for studying (some) field theoretic models of quantum gravity. The
five known (very likely) consistent ten-dimensional Superstring
theories (including compactifications to lower dimensions), as well as
11-dimensional supergravity are understood to be different limits
of M-Theory
dynamics~\xcite{witten1995string}{https://doi.org/10.1016/0550-3213(95)00158-O}.
If supersymmetry~\xcite{wess1974supergauge}{https://doi.org/10.1016/0550-3213(74)90355-1}
is part of the answer why the observed fundamental
laws of physics are the way they are (and it seems to have some good
answers to problems that arise in Planck-scale physics), then, due to
the existence of gravity, there is no way to escape the conclusion
that a viable theory must contain
supergravity~\xcite{Freedman:1976xh}{https://doi.org/10.1103/PhysRevD.13.3214},~\xcite{Deser:1976eh}{https://doi.org/10.1016/0370-2693(76)90089-7}
and in particular a supersymmetric partner to the graviton with
spin-3/2, the gravitino.
While the question is still not settled whether one can construct
a theory of supergravity that not only works as an effective low
energy theory but is well-behaved at every length scale
\xcite{bern2009ultraviolet}{https://doi.org/10.1103/PhysRevLett.103.081301},
\xcite{zvi2015surprises}{http://bapts.lbl.gov/Bern.pdf}, it is
generally thought that problems that arise in simple models of
supergravity~\xcite{PhysRevLett.38.527}{https://doi.org/10.1103/PhysRevLett.38.527}
are ultimately resolved by the notion of ``point particles'' in
quantum field theory breaking down
at very high energies~\xcite{witten2018every}{https://doi.org/10.1007/978-3-319-64813-2_8},
i.e.\ superstring theory. Now, if one accepts superstring theory, there
is no way to avoid the conclusion that there also needs to be a way to
describe its non-perturbative strong coupling limit, which then
inevitably leads us to M-Theory~\xcite{Weinberg:2000cr}{https://www.cambridge.org/de/academic/subjects/physics/theoretical-physics-and-mathematical-physics/quantum-theory-fields-volume-3}.

Despite the remarkable success of the Standard Model (SM) of
Elementary Particle
Physics~\xcite{tanabashi2018review}{https://doi.org/10.1103/PhysRevD.98.030001},
which quantitatively describes the properties and interactions of
matter and force particles so well that the LHC at the time of this
writing did not to come up with clear evidence of ``new''
(beyond-the-SM) physics, there are a number of unsolved problems, for
example non-observation of the particles that constitute Dark
Matter~\xcite{bertone2005particle}{https://doi.org/10.1016/j.physrep.2004.08.031},
or explaining why the neutron's electric dipole moment is too small to
be measurable~\xcite{peccei2008strong}{https://doi.org/10.1007/978-3-540-73518-2_1}.
The most puzzling such problem of theoretical physics currently is
perhaps explaining the observed positive -- but from a quantum field
theory perspective extremely small -- vacuum energy
density~\xcite{carroll2001cosmological}{https://doi.org/10.12942/lrr-2001-1}
of the universe. M-Theory currently struggles to give an answer to
how this could arise naturally. Even if M-Theory ultimately turned out
to not be the correct answer to the question how to quantize gravity,
it already by now has made major contributions to uncovering
interesting hidden connections in pure mathematics, of which we here
only want to mention the geometric Langlands correspondence
as one example,~\xcite{Witten:2009at}{https://arxiv.org/abs/0905.2720}.

\subsubsection{Unification}

The unification of Quantum Electrodynamics with Quantum Chromodynamics
(QCD) and the Weak Force into the SM is highly
successful from a theoretical perspective, with both QCD and
Electroweak theory individually being afflicted by problems that
cancel in the SM~\xcite{adler1969axial}{https://doi.org/10.1103/PhysRev.177.2426}~\xcite{bilal2008lectures}{https://arxiv.org/pdf/0802.0634}.
A key property is that in the SM, all forces are described in an uniform
way by vector gauge bosons. Since spacetime symmetries (rotations and
boosts) affect all vector gauge bosons in the same way, one can
consider superposition quantum states between them. Indeed, the SM Photon
emerges as a specific quantum superposition of a Weak Force's particle
and another ``hidden'' force's particle termed~$U(1)_Y$ in a way that is
governed by properties of the Brout-Englert-Higgs Boson. This also
sets -- for example -- the relative strengths of the Electromagnetic
and Weak forces. It is quite plausible that at even (much!) higher
energies (but somewhat below the quantum gravity energy scale), the
same mechanism also is at play in the form of ``Grand
Unification''~\xcite{Georgi:1974sy}{https://journals.aps.org/prl/abstract/10.1103/PhysRevLett.32.438}~\xcite{Pati:1974yy}{https://journals.aps.org/prd/abstract/10.1103/PhysRevD.10.275},
merging the Strong Force with the Electroweak Force.

The main new discoveries described in the current article are also
about this ``Higgs
effect''~\xcite{englert2014nobel}{https://doi.org/10.1103/RevModPhys.86.843}~\xcite{higgs2014nobel}{https://doi.org/10.1103/RevModPhys.86.851},
leading to some particular sets of
particles and interactions in ``low energy'' physics, so in terms of
basic mechanisms this is rather similar to what is now well established
SM physics. Our setting, however, is that of a model that
might actually describe Planck-scale physics well, so if this
construction ever turned out to somehow actually be related to the
SM (see
e.g.~\xcite{nicolai19853}{https://doi.org/10.1016/0550-3213(85)90643-1}~\xcite{meissner2015standard}{https://doi.org/10.1103/PhysRevD.91.065029}~\xcite{kleinschmidt2015standard}{https://doi.org/10.1016/j.physletb.2015.06.005}
for some speculation in this direction), interesting Physics that is not
well understood yet would need to happen in the gap of~$\sim 17$
orders of magnitude between the Higgs boson energy scale and the
quantum gravity energy scale. In particular, there is a
discrepancy on gauge groups, cosmological constant, and particle
chirality properties.

More likely, this work describes a collection of (stable as well as
unstable background-)solutions to the M-Theory field equations in a
very different corner than the one that describes the Standard
Model. Given our still limited understanding of M-Theory, it
nevertheless seems useful to get a better idea about relevant
properties, mechanisms, and phenomena by investigating solutions that
are accessible with current technology. In the past, studying
solutions to the 4-dimensional field theory equations and especially
their embedding into M-Theory has taught us some very useful lessons
about compactification mechanisms and about 11-dimensional dynamics. One in
particular notes that~$SO(8)$ supergravity also describes the physics
of a stack of M2-branes~\xcite{Bobev:2009ms}{https://doi.org/10.1088/1126-6708/2009/09/043}.

\subsubsection{Kaluza-Klein Supergravity}

While the observation, originally by Kaluza~\xcite{kaluza1921unitatsproblem}{https://doi.org/10.1142/S0218271818700017}
and Klein~\xcite{klein1987quantum}{https://doi.org/10.1007/BF01397481},
that four-dimensional physics can be understood in the
context of dynamics in higher dimension was certainly interesting, the
question remained how to find guidance on what higher dimensional
dynamics to start from. A major step forward happened in~1979, when
Cremmer, Julia, and Scherk set out to construct the four-dimensional field theory
with the maximum possible amount of boson-fermion symmetry~\xcite{Cremmer:1978km}{https://doi.org/10.1016/0370-2693(78)90894-8}~\xcite{Cremmer:1978ds}{https://doi.org/10.1016/0370-2693(78)90303-9}~\xcite{cremmer1979so}{https://doi.org/10.1016/0550-3213(79)90331-6}.
Due to the
complicated structure of the (Higgs-like) scalar boson interactions,
this was achieved by first realizing that such a theory should be
obtainable via compactification of a higher dimensional ancestor theory,
and that there can only be one graviton with helicity~$\pm 2$ and no
interacting massless particles with higher helicity (due to nonexistence
of a suitable source current), and that the highest dimension in which
such a symmetry can exist is eleven (since otherwise dimensional
reduction to four dimensions would give rise to too many
supersymmetry generators that require going to helicities
beyond~$+2$ that cannot be consistently coupled). Constructing
the 11-dimensional model first
in~\xcite{Cremmer:1978km}{https://doi.org/10.1016/0370-2693(78)90894-8},
the maximally symmetric four-dimensional model in which all matter and
force particles are unified succeeded
in~\xcite{Cremmer:1978ds}{https://doi.org/10.1016/0370-2693(78)90303-9}
via Kaluza-Klein reduction on a 7-dimensional torus.

The ``auxiliary'' 11-dimensional field theory, originally introduced as
a mathematical trick, soon was found to be very interesting in itself. For
example, it so turns out that symmetry constraints completely
determine its structure, and there is no way to adjust its parameters
or field content. It almost certainly describes the low energy limit
of an as yet unknown 11-dimensional theory of supersymmetric membranes
(and perhaps other dynamical degrees of freedom) which, upon
dimensional reduction on a circle also wrapped by one direction of the
membrane, produces 10-dimensional Superstring
Theory~\xcite{witten1995string}{https://doi.org/10.1016/0550-3213(95)00158-O}.
This unknown (perhaps) 11-dimensional theory of (likely) supermembranes has been
given the provisional name ``M-Theory''.

As explained, a key ingredient in the effort to unify force and matter
quantum fields is boson-fermion symmetry, or ``Supersymmetry'', as it is
commonly called. This is presently is a somewhat esoteric topic
outside of quantum field theory and some branches of differential
topology, plus perhaps the theory of stochastic dynamical
systems~\xcite{parisi1982supersymmetric}{https://doi.org/10.1016/0550-3213(82)90538-7}.

\subsubsection{Supergravity in eleven and four dimensions}

A supersymmetry transformation changes the
helicity of a particle by 1/2 and so -- in a theory of gravity -- it must
connect the helicity-2 graviton with a helicity-3/2 fermionic
particle, the ``gravitino''. It is possible to construct models with
more than the minimal amount of supersymmetry that then fuse more helicity
states~\xcite{Ferrara:1976iq}{https://doi.org/10.1016/0550-3213(77)90161-4}~\xcite{Fradkin:1979cw}{https://doi.org/10.1016/0370-2693(79)90774-3}~\xcite{Fayet:1975yi}{https://doi.org/10.1016/0550-3213(76)90458-2},
allowing one to
start with a helicity-2 graviton, apply a supersymmetry transformation
to step down to the helicity-3/2 gravitino, and use another
independent supersymmetry transformation to further step down to a helicity-1
photon-like particle that couples with gravitational strength (a
``graviphoton''~\xcite{scherk1979antigravity}{https://doi.org/10.1016/0370-2693(79)90463-5}).
A maximally supersymmetric theory in four dimensions, as obtainable
through dimensional reduction of 11-dimensional supergravity,
has~$\mathcal{N}=8$ independent supersymmetries that connect all
quantum states from the helicity~$+2$ graviton down to the
oppositely-polarized helicity~$-2$ graviton, with (according to
simple combinatorics) $\binom{8}{k}$ particles of
helicity~$2-k/2$, so in total one graviton, eight gravitini, $28$~photon-
or gluon-like force carriers, $56$~spin-$1/2$ fermions,
and $70$~Higgs-Boson-like scalars. It was the discovery of
this~$\mathcal{N}=8$ supergravity, which manages to unify all
particles and interactions starting from only a symmetry
principle as input, that made S. Hawking suggest in his
1981 inaugural lecture~\xcite{Hawking:1981sq}{https://iopscience.iop.org/article/10.1088/0031-9112/32/1/024/pdf}
that within perhaps 20 years, we would know the ``Theory of Everything''.

A peculiar feature of this construction is that it interprets the
70~Higgs fields as parametrizing a very special 70-dimensional coset
manifold. Just as the sphere can be regarded as the manifold of
3-dimensional rotations (read: orthogonal bases) modulo another rotation
(around the outward-pointing direction), i.e.\ $S^2=SO(3)/SO(2)$, and the
hyperbolic plane\footnote{For a game that allows one to develop some
  intuition about living on a hyperbolic plane,
  see~\xcite{kopczynskihyperrogue}{https://roguetemple.com/z/hyper/}.}
can be regarded as the coset space~$SO(2,1)/SO(2)=SL(2)/SO(2)$, the
relevant scalar manifold of $\mathcal{N}=8$~Supergravity is\footnote{Strictly
 speaking, it is actually~$E_{7(7)}/(SU(8)/{\mathbb Z}_2)$, as the~$SU(8)$
 group element that maps an~$\mathbf{8}$-vector
 (or~$\bar{\mathbf{8}}$-vector) to minus itself gets represented as
 the identity operation when acting on the scalar manifold, while this
 does not happen for any other element from the center of
 $SU(8)$, apart from the identity itself.}
$E_{7(7)}/SU(8)$, where $SU(8)$ is the group of complex $8\times 8$
matrices with unit determinant and the maximal compact subgroup
of the $133$-dimensional non-compact exceptional Lie
group~$E_{7(7)}$, which we describe in appendix~\ref{app:e7}.

In Cremmer and Julia's original
construction~\xcite{Cremmer:1978ds}{https://doi.org/10.1016/0370-2693(78)90303-9},
which compactified 11-dimensional Supergravity on a
7-dimensional torus, there are 28~photon-like gauge boson
particles. Soon after, it was realized that one can also obtain a
four-dimensional theory with the maximal amount of supersymmetry by
dimensionally reducing on the ``round'' 7-dimensional
sphere\footnote{The 7-sphere is rather special, as there are many (28
in total) different spaces that all are topologically 7-spheres, but
not diffeomorphic to one
another~\xcite{milnor1956manifolds}{https://doi.org/10.2307/1969983}~\xcite{brieskorn1966examples}{https://doi.org/10.1073/pnas.55.6.1395}.}
instead~\xcite{de1982n}{https://doi.org/10.1016/0550-3213(82)90120-1}.
Here, one ends up with the 28 vector gauge bosons belonging
to the non-abelian gauge symmetry~$SO(8)$, which may superficially be
thought of as some sort of more complicated Quantum Chromodynamics
(but with very differently behaving ``quarks'' and ``gluons'', and perhaps
without confinement due to vanishing $\beta$-function). Clearly,
given that such nonabelian gauge symmetries do play an important role
in the SM, this looks like a major step in the right
direction, but unfortunately, the group~$SO(8)$ is too small to embed
the~$SU(3)_{\rm QCD}\times SU(2)_{\rm Weak}\times U(1)_{\rm Y}$ gauge
group of the SM into it. Also, the experimentally observed
left-right asymmetry of the SM (``chirality'') cannot be obtained by
compactifying non-chiral 11-dimensional supergravity on a
manifold~\xcite{Witten:1983ux}{https://doi.org/10.1142/9789814542340_0091}.

The early literature on this topic contemplated scenarios in which
the observed particles would emerge as composite, being made of
more fundamental ``preons'', somewhat along the lines of how QCD at
lower energies gives rise to baryon and meson bound states.
Given that there are in terms of
energy perhaps $17$~orders of magnitude between the quantum gravity
scale and the Higgs boson energy scale, this may not be entirely
unplausible. Still, considering in particular the problems associated
e.g.\ with chiral fermions, it is nowadays generally regarded as more
promising to investigate scenarios in which the SM's gauge
symmetry directly emerges from some large higher-dimensional symmetry.
Going from 11-dimensional M-Theory to Superstring Theory first,
and then down to four dimensions, there are by now multiple options
to directly get a large ``Grand Unification'' gauge symmetry into
which the SM gauge symmetry can be embedded.

The main focus of the current article, however, is not to provide more
insights into how experimental particle physics might be related to
M-Theory. Rather, we want to allow deeper investigations into the
structure of M-Theory by both expanding our knowledge on what possible
background solutions to its field equations can look like, and also by
providing tools that allow one to come to grips with some of the
technical complications that arise in particular when working with
high-dimensional scalar manifolds. In the past, we have time and again
seen the study of models of quantum gravity produce highly surprising
and useful insights even if they were not at all focused on the
four-dimensional world we inhabit. Most notably, there was the
realization in
1997~\xcite{maldacena1999large}{https://doi.org/10.1023/A:1026654312961}
that the partition function (/generating functional) of a
Conformal Field Theory (CFT) can be the same as the partition function
of a supersymmetric theory of gravity with \emph{negative}
cosmological constant (hence in ``anti
de Sitter (AdS) space'') in a different spacetime dimension (i.e.\ with
an extra spatial direction), the so-called AdS/CFT
correspondence~\xcite{gubser1998gauge}{https://doi.org/10.1016/S0370-2693(98)00377-3}~\xcite{witten1998anti}{https://dx.doi.org/10.4310/ATMP.1998.v2.n2.a2}~\xcite{maldacena1999large}{https://doi.org/10.1023/A:1026654312961}.
One example for a rather surprising further development of this idea
is the insight that Quantum Field Theory may provide a lower bound for
the ratio of shear viscosity to entropy density of \emph{any}
liquid~\xcite{kovtun2005viscosity}{https://doi.org/10.1103/PhysRevLett.94.111601}.
Another modern development that could not
have been anticipated is the application of this idea of gauge/gravity
duality to study superconductivity~\xcite{gubser2008breaking}{https://doi.org/10.1103/PhysRevD.78.065034}~\xcite{hartnoll2008holographic}{https://doi.org/10.1088/1126-6708/2008/12/015}~\xcite{hartnoll2008building}{https://doi.org/10.1103/PhysRevLett.101.031601}.
To give another example, ``holographic duality'' has been employed to map
solution-generating symmetries of the Einstein
equations~\xcite{Ehlers:1957zz}{https://inspirehep.net/record/45502/}~\xcite{Geroch:1970nt}{https://doi.org/10.1063/1.1665681}
to solution-generating symmetries for the Navier-Stokes
equation~\xcite{berkeley2013navier}{https://doi.org/10.1007/JHEP04(2013)092}~\xcite{bhattacharyya2008nonlinear}{https://doi.org/10.1088/1126-6708/2008/02/045}.

Concerning specifically~$\text{AdS}_4/\text{CFT}_3$ duality, the
holographic dual of the~$SO(8)$ supergravity studied here is
(the~$k=1$ case of) three-dimensional
ABJM~theory~\xcite{aharony2008n}{https://doi.org/10.1088/1126-6708/2008/10/091},
which describes the dynamics of M2-branes. This was used e.g.\
in~\xcite{Bobev2018}{https://doi.org/10.1007/JHEP03(2018)050}
to construct new supersymmetric~$AdS_4$ black holes and provide an
explanation for their Bekenstein-Hawking entropy, exploiting
the relation between mass-deformed ABJM theory with~$\mathcal{N}=2$
supersymmetry and the~$AdS_4$ vacuum
with~$\mathcal{N}=2\,SU(3)\times U(1)$ symmetry, which in this work
is called solution S0779422.

\subsection{On Machine Learning}

Artificial Intelligence (AI) is a broad field concerned with crafting
algorithms for solving problems that require some form of
human-like intelligence. To avoid any misconceptions, we clarify that
the main concern of AI is not finding ways to allow algorithms to
perform introspective reasoning on par with or exceeding human ability.
Indeed, as famously noted by Alan Turing, the question of whether machines
can think is
ill-posed~\xcite{machinery1950computing}{http://cogprints.org/499/1/turing.html}.

The earliest forms of AI consisted of explicit, manually-crafted
rules. Machine Learning (ML) introduced a new perspective on creating
artificially intelligent algorithms. This field was pioneered by
Arthur Samuel, who demonstrated that a computer program can learn to
play the game of checkers better than the person who programmed
it~\xcite{Samuel1959some}{https://doi.org/10.1147/rd.33.0210}. Instead
of operating with pre-adjusted rules and fixed numeric values, the
algorithm would instead tune itself in order to solve the
problem. In other words, given a function of an input space that
represents the problem data and an output space that represents the
problem solution, the challenge becomes to learn the parameters of
this function in such a way that its results (the output, or the
solution of the algorithm) is as close as possible to the correct
solution. Usually, the learned parameters are of numeric form. The
field of ML is thus primarily concerned with the pragmatic problem of
finding and efficiently refining functions that usually have a large
number of adjustable (``learnable'') parameters, with the purpose of
solving challenging problems that often involve real-world data. ML
methods are suitable whenever facing a problem that is difficult to
put into words or fixed rules.

In a way, Machine Learning (ML) and physics can be regarded as intellectual antipodes:
Physics tries to understand fundamental processes and important mechanisms
underlying the functioning of a system, while ML tries solve a particular
problem as well as possible, while eschewing the
need to fully understand it. In fact, the implicit knowledge obtained
by an ML algorithm by solving a problem is often difficult to
analyze. Understanding how certain highly-successful ML algorithms
manage to solve highly difficult problems and visualizing various parts
of the learned function in order to produce an intuitive understanding
of the problem and the solution space is an active field of
research~\xcite{Olah2017feature}{https://distill.pub/2017/feature-visualization/}.

Example problems that have, sometimes surprisingly so, turned out to
be amenable to ML approaches include
text~\xcite{Genzel2015paper}{https://ai.googleblog.com/2015/05/paper-to-digital-in-200-languages.html}
or object~\xcite{krizhevsky2012imagenet}{https://papers.nips.cc/paper/4824-imagenet-classification-with-deep-convolutional-neural-networ}~\xcite{szegedy2015going}{https://www.cv-foundation.org/openaccess/content_cvpr_2015/html/Szegedy_Going_Deeper_With_2015_CVPR_paper.html}
recognition in images, mapping pictures to textual descriptions of their
content~\xcite{sharma-etal-2018-conceptual}{https://www.aclweb.org/anthology/papers/P/P18/P18-1238/},
machine translation of natural
language~\xcite{vaswani2017attention}{https://papers.nips.cc/paper/7181-attention-is-all-you-need},
scoring possible moves in the game of
Go~\xcite{silver2016mastering}{https://doi.org/10.1038/nature16961}
and Starcraft~\xcite{alphastarblog}{https://deepmind.com/blog/alphastar-mastering-real-time-strategy-game-starcraft-ii/},
and many more. Increasingly, we also see ML methods
being applied to problems that do not strictly follow this pattern,
such as synthesis of realistic-looking
portraits~\xcite{karras2018style}{https://arxiv.org/abs/1812.04948}.

Concerning direct applications of ML to theoretical physics, it can
and indeed has happened in the past that ML demonstrated an ability to
predict a system's behavior well beyond what our current thinking
would have considered possible, indicating the existence of extra
structure that our current models cannot capture well. For
example,~\xcite{peccei2008strong}{https://doi.org/10.1103/PhysRevLett.120.024102}
demonstrated a clever set-up that allows ML to accurately predict the
behavior of a chaotic system over eight Lyapunov times.

One particularly successful family of ML algorithms is that of
Artificial Neural Networks (ANNs). ANNs are loosely inspired by
biological brains, which are made of billions of interconnected neurons
working together to control optimally the behaviour of intelligent organisms.
A simple model of a neuron, called a Perceptron, was proposed by Frank
Rosenblatt in
1958~\xcite{rosenblatt1968perceptron}{https://psycnet.apa.org/doi/10.1037/h0042519},
but the idea of networks composed of multiple layers of Perceptrons only
started becoming popular in the ML community in the
1980s~\xcite{rumelhart1986general}{https://web.stanford.edu/~jlmcc/papers/PDP/Volume\%201/Chap2_PDP86.pdf}.
ANNs consist of artificial ``neurons'', non-linear circuit elements that are
interconnected through directed artificial
``synapses'' that transmit signals with different efficacies,
which act like ``weights'' in directed graphs. The connectivity architecture of
such a network is usually layered, the intuition being that each layer
builds up more abstract concepts than the previous. A fully connected feedforward network
includes connections between every node of a layer and every node of the
next layer, but other variations also exist, such as
recurrent~\xcite{hochreiter1997long}{https://doi.org/10.1162/neco.1997.9.8.1735}
or convolutional~\xcite{lecun1998gradient}{https://doi.org/10.1109/5.726791}
layers.

Such ``layered'' ANNs are popular as they are known to be universal
approximators~\xcite{Hartman1990}{https://doi.org/10.1162/neco.1990.2.2.210}
and have been found to work well for many problems, but it is by no means true
that ML is tied to this particular class of architectures. As long as there
is a way to model a problem in terms of a function that differentiably depends
on many parameters, and parameter-tuning can substantially improve performance,
ML techniques are applicable.

Deep learning, which has recently achieved resounding success in
solving difficult real-world problems like the ones mentioned earlier,
refers to ANNs with a large number of stacked layers especially
designed to apply specialized operations on the input. It was not
trivial to discover that such deep networks can work at all -- until
Hinton's seminal 2006
article~\xcite{hinton2006fast}{https://doi.org/10.1162/neco.2006.18.7.1527},
which sparked the deep learning revolution, common thinking was that networks
with more than two layers were essentially impossible to train, and other
ML approaches, such as kernelized support vector
machines~\xcite{Cortes1995}{https://doi.org/10.1007/BF00994018},
would generally perform better than ANNs. Later progress uncovered a number of
general misconceptions and useful tricks on how to train ANNs, for example the
superior performance of the ``Rectified Linear Unbounded'' (ReLU)
activation function~\xcite{pmlr-v15-glorot11a}{http://proceedings.mlr.press/v15/glorot11a}
in comparison to the classical sigmoid non-linear activation function
used in earlier research.

What type of problems is ML applicable to? Depending on the amount and type
of available data, there are three main paradigms for training
an ML model: supervised, unsupervised and reinforcement
learning. Supervised learning refers to data where the expected result
is known in advance for the data available; for example, given a large
set of images of people, the name of the person appearing in each
image is also given. This type of learning is often used with
classification (``given $m$~labels, pick the correct one'') and regression
(``predict a value in a continuous domain''), but can often be adapted for
other types of problems, for example in assessing the value of each possible
next action in a
game~\xcite{silver2016mastering}{https://doi.org/10.1038/nature16961}.
Supervised learning with ANNs is currently the most widely-used and successful
approach to ML. In contrast, unsupervised learning occurs when no labels exist
for the given data; in this case, the aim is to group the data in such a way that
items similar to each other belong to the same group, or are close to
each other in the output
domain~\xcite{kohonen1982selforganized}{https://doi.org/10.1007/BF00337288}.
Examples where this approach is useful are fetching web pages, songs or videos
similar to the one that an internet user might be viewing or listening to
currently. Nonetheless, such problems can also be expressed as a supervised
problems~\xcite{covington2016deep}{https://doi.org/10.1145/2959100.2959190}
Finally, reinforcement learning is applied when no exact labels exist, but
there is some knowledge on whether a proposed output is good or
bad. This is applicable in particular to automated game playing, where
the algorithm acts as agent that chooses to perform a sequence of
actions with the aim of winning the game; by playing multiple games, it
slowly learns to pick better actions based on whether the past games
resulted in wins or losses. Reinforcement learning can also be combined with
supervised learning: given a very large number of possible actions,
a supervised deep ANN can approximate the value of each possible
action~\xcite{mnih2015human}{https://doi.org/10.1038/nature14236}.

So how does learning in an ML model actually work? A key idea is that learning
involves the minimization of a loss (or error) function. This function is
designed such that, when applied to any given output, it provides
a numerical measure of how far off this output is from an expected answer.
In supervised learning,
this can be thought of as a distance between the actual output and the target
(desired) output of the algorithm. If the value of the
loss function is smaller, the error is smaller, and the algorithm is closer to
the desired output. Thus, instead of deeming an output as either correct or
incorrect, the loss function provides a graded measure of ``wrongness''. The
output of the network can thus be often interpreted as a probability. Crucially,
if the loss function is differentiable with respect to the algorithm parameters
(for example, in case of an ANN, the network weights),
the gradient of the loss function can be used to point towards the direction
of the minimum of this function. The gradient of the loss function (which
usually is estimated on a random selection of examples, see below) can be used to
iteratively tune parameters in order to improve the performance of predictions.
For many problems, there are natural choices of loss functions. For
classification problems with $n$ possible labels
(e.g.\ ``which digit does the image show'' with~$n=10$), the predicted
probabilities can be regarded as dual to chemical potentials, represented
as (linearly) accumulated evidence~$E_j$ for or against a particular
classification label~$p_j$, i.e.\ $p_j = exp(-E_j) / \sum_{i=1}^{n} exp(-E_i)$~\xcite{dunne1997pairing}{https://citeseerx.ist.psu.edu/viewdoc/summary?doi=10.1.1.49.6403}.
However, any type of loss function can be employed as long as it indicates
the correct solution to the given problem, is differentiable with respect to the
learnable parameters, and has a reasonable shape -- for example, not too many
`bad' local minima. One of the surprising insights of the Deep
Learning revolution was that a simple non-linear activation function
with discontinuous derivative that reduces to the identity in the
activation region and to the null function outside that region,
i.e.~$\mathrm{ReLU}\,(x)\colonequals (x+|x|)/2$, allows training deeply nested
transformations to extract high-level information such as whether there is
a face in an image. In some situations, finding good loss functions to represent
important aspects of a problem is less straightforward, and may need some
experimenting. In this work, for example, we show how the desire to have
unbroken vacuum supersymmetry can be represented concisely through a ML loss
function.

A key idea that made ANN-based learning possible is that, when given a
computer program that computes a~$\mathbb{R}^{n}\to\mathbb{R}$
function~$f$, it is possible to automatically transform this into
another program that computes the $n$-component gradient~$\nabla f$ at
any given point with relatively small computational effort that is
independent of~$n$~\xcite{Speelpenning:1980:CFP:909337}{https://archive.org/details/compilingfastpar1002spee}.
The work on reverse-mode automated differentiation pre-dates and
provides a more general framework than ``error backpropagation'' for
ANNs as it was rediscovered independently in~1985
in~\xcite{rummelhart1987learning}{https://doi.org/10.21236/ADA164453}.
Interestingly, one can also regard ``reverse mode automatic
differentiation'' as a discretized version of the idea underlying
Pontyragin's maximum principle in Optimal Control Theory, i.e.\ the
Hamilton-Jacobi-Bellman (HJB)
equation~\xcite{Bellman231}{https://doi.org/10.1073/pnas.40.4.231}
from the 50's, in the sense that applying reverse-mode AD to the most
basic ODE solver algorithm directly produces the HJB equation.

Given that not all ML applications use a single straightforward
layered ANN architecture, it makes sense for a Machine Learning
library like TensorFlow to provide some form of general-purpose
reverse-mode automatic differentiation capabilities. In principle,
there are three ways to do this, (1) full program analysis, which for
a language as complex as C++ or even only Python is a formidable task
(this has been done for Scheme with
R6RS-AD,~\xcite{siskindAD}{https://github.com/qobi/R6RS-AD}), (2)
implementing some ``domain-specific language (DSL)'' for arithmetic
graphs, and (3) ``Tape-based'' AD, where in the forward pass, the
sequence of arithmetic operations gets recorded on a ``tape'', which
then is replayed in reverse. TensorFlow~1.x uses approach~(2), while
TensorFlow~2.x tries to make the tape-based paradigm the default
choice. In this article, we will exclusively use a graph-based
TensorFlow~1 approach.

\subsection{Tensors in Machine Learning}

To give a rough mental picture of what the training process might look
like at the level of number-crunching, we give an example problem
where the goal is to predict whether a person appears or not in a
given image. We assume that a labeled dataset on the order of a
million images is available for training.  An image could be
represented as a 3-index array~$X[row, col, c]$, with indices
providing row and column pixel coordinates as well as the color
channel~$c$.
The label for each image is represented by the
number~$1$ of there is a person in the image, and the number~$0$ if
not. One would typically start by grouping example images into
sufficiently large (randomized) batches to get reasonable estimates for loss
function gradients with respect to model parameters, perhaps $b=1024$
images per batch. A batch of training images would then be naturally
represented as a $b$-dimensional array of pairs~$(X_b, Y_b)$. It has
become fashionable to call these higher-rank arrays ``tensors'' in ML
terminology, which indeed is a useful notion for expressing the ANN
operations in terms of tensor products and index-splitting
operations. However, symmetry groups to this date play a rather minor
role in ML (with notable exceptions such
as~\xcite{kondor2018covariant}{https://arxiv.org/abs/1801.02144}), and
if they actually do, one often talks about ``equivariant neural
networks'' to discriminate these from networks with less
structure. For a problem such as recognizing whether a picture
contains a person, which evidently benefits from utilizing  symmetry,
the common approach is to factor out translational symmetry by
effectively imposing constraints on network parameters relevant
for detecting the target entity, or elements of it, at different
locations in the image. This is usually done by
``convolutional layers'' that computes convolutions
$C[b, i, j, k] = \sum_{\xi, \eta} X[b, i + \eta, j + \xi, c] S[k, \eta, \xi, c]$
of the example images with a
collection~$S[\cdot, \cdot, \cdot, \cdot]$ of small stencils
represented as an array of trainable parameters. The stencil
parameters then get adjusted in the training phase such that they are
optimally useful for coming up with a good probability prediction for
the image to show a person in any location. Each such stencil will
consist of lines that describe typical features associated with a
person in the image, such as noses, eyes or ears. Intuitively, one
could imagine one of the stencils getting tuned by training to have
large inner product with ``the average shape of all noses'', so
getting specialized to a nose-detector. Each such stencil would, in
turn, be made of lower-level stencils, such as lines with particular
orientations, which, in the right combination, form salient
shapes. Subsequent processing layers in the network would then collect
and combine different such evidence and in the end produce a Bayesian
prediction roughly along the lines of: ``We are highly confident to
have seen a nose in the picture, and we also have moderate confidence
to have seen an eye somewhere, so, with high likelihood, the image
shows a person''. Krizhevsky's seminal
paper~\xcite{krizhevsky2012imagenet}{https://doi.org/10.1145/3065386}
explains in detail one such convolutional ANN architecture for image
processing. Recent work on feature visualization in ANNs has
spectacularly uncovered collections of shapes and patterns that hidden
layers in a network learn to
recognize~\xcite{Olah2017feature}{https://distill.pub/2017/feature-visualization/}.

The above example illustrates why numerical higher-rank arrays are so prevalent
in modern ML. As hinted earlier in this section, one very common primitive
``tensor'' operation in such a setting is batched matrix-multiplication.
For example, linear conversion of a set of
example images from RGB color space to some implicit color space that can be
trained to be optimally useful for solving the problem codified by the
loss function expressed as
\[ X2[b, i, j, d] = \sum_{c} X[b, i, j, c] * M[d, c] \] with trainable
parameters in the matrix~$M$.

\section{$D=4$ $SO(8)$ supergravity and its scalar sector}

Let us briefly review some salient features of supergravity in four
and eleven dimensions before we look into finding equilibrium
solutions to the equations of motion.

Four-dimensional supersymmetry can at most unify all particle states
from the helicity +2~down to the helicity -2~graviton. As there are
eight helicity-1/2 steps between these helicities, we can have at most
eight times the minimal amount of supersymmetry, and as each of these
eight supersymmetry transformations comes as a real (Majorana)
four-dimensional spinor, we are looking at a theory with $8\cdot4=32$
supersymmetry components. The highest spacetime dimension in which we can have a real
32-component spinor is $d=11$ (or perhaps~$d=12$ if we accepted a
second direction of
time~\xcite{Bars:1996dz}{https://doi.org/10.1103/PhysRevD.54.5203}).
A supersymmetric theory of gravity in
$D=11$~dimensions will have~$(D-2)(D-1)/2-1=44$ transversal
graviton polarization states (described by a symmetric
traceless $9\times 9$ matrix), plus~$128$ gravitino degrees of
freedom. The mismatch in the number of degrees of freedom is
compensated by a gauge field with~$84$ degrees of freedom,
describing a higher-dimensional cousin of the photon whose
polarization is not given by a~1-dimensional vector, but by
a 3-dimensional volume(-form) embedded into 9-dimensional transversal
space,~$A_{MNP}$,
with associated (4-form) field strength~$F_{MNPQ}$. With the
``polarization'' being a 3-dimensional object, this (abelian) gauge
field can not be sourced by charged particles (the 1-dimensional
photon polarization couples to the 1-dimensional worldline of
an electron), but by some membrane-like extended object that lives
in eleven dimensions. This is now understood to be
the~M2-brane~\xcite{Bergshoeff:1987cm}{https://doi.org/10.1016/0370-2693(87)91272-X}.
It is amazing to see how starting from one of the three possible gauge
principles, the vector-spinor, in its very own preferred (maximal)
dimension, one automatically obtains a theory that unifies all three
of the possible gauge principles, and furthermore turns out to be
completely fixed, i.e.\ not permit any free parameters.

The Lagrangian of 11-dimensional
Supergravity reads~\xcite{cremmer1989supergravity}{https://doi.org/10.1016/0370-2693(78)90894-8}~\xcite{Duff:1986kaluza}{https://doi.org/10.1016/0370-1573(86)90163-8}
\begin{equation}
  \begin{array}{lcl}
    \mathcal{L}/e &=&
    \frac{1}{4}R_{MN}{}^{AB}e_M{}^Ae_N{}^B-\frac{i}{2}\bar\Psi_M\Gamma^{MNP}D_N\left(\frac{1}{2}(\omega+\tilde\omega)\right)\Psi_P\\
    &&\kern-1.5cm{}-\frac{1}{48}F_{MNPQ}F^{MNPQ}\\
    &&\kern-1.5cm{}+\frac{2}{12^4}\epsilon^{MNPQRSTUVWX}F_{MNPQ}F_{RSTU}A_{VWX}\\
    &&\kern-1.5cm{}+\frac{3}{4\cdot 12^2}\Bigl(\bar\Psi_M\Gamma^{MNWXYZ}\Psi_N+12\bar\Psi^W\Gamma^{XY}\Psi^Z\Bigr)\left(F_{WXYZ}+\tilde F_{WXYZ}\right)\\
         {\rm where} &&\\
         e&\colonequals &{\rm det}\,e_M{}^A\\
         D_M(\omega)&\colonequals &\partial_m-\frac{1}{4}\omega_M{}^{AB}\Gamma_{AB},\\
         \omega_{MAB}&\colonequals &\frac{1}{2}\left(\Omega_{ABM}-\Omega_{MAB}-\Omega_{BMA}\right)+K_{MAB}\\
         K_{MAB}&\colonequals &\frac{i}{4}\left(-\bar\Psi_N\Gamma_{MAB}{}^{NP}\Psi_P\right.\\
         &&\phantom{\frac{i}{4}}\;\,\left.+2\left(\bar\Psi_M\Gamma_B\Psi_A-\bar\Psi_M\Gamma_A\Psi_B+\bar\Psi_B\Gamma_M\Psi_A\right)\right)\\
         \Omega_{MN}{}^{A}&\colonequals &\partial_N e_M{}^A-\partial_M e_N{}^A\\
         \tilde\omega_{MAB}&\colonequals &\omega_{MAB}+\frac{i}{4}\bar\Psi_N\Gamma_{MAB}{}^{NP}\Psi_P\\
         F_{MNPQ}&\colonequals &4\delta_{MNPQ}^{RSTU}\partial_R A_{STU}\\
         \tilde F_{MNPQ}&\colonequals &F_{MNPQ}-3\delta_{MNPQ}^{RSTU}\bar\Psi_R\Gamma_{ST}\Psi_U\;.
  \end{array}
\end{equation}

\subsection{Compactification to four dimensions}

Freund and Rubin noted~\xcite{freund1980dynamics}{https://doi.org/10.1016/0370-2693(80)90590-0}
that this theory preferentially compactifies to four dimensions due to the
presence of the four-form field strength~$F_{ABCD}$. Indeed, a ``flux'' compactification
with~$F_{ABCD}\sim\epsilon_{\mu\nu\rho\sigma}$, i.e.\ flux aligned with
the submanifold of four-dimensional spacetime, will look isotropic
from the four dimensional perspective. Kaluza-Klein compactification
to four spacetime dimensions on a 7-sphere that is the surface of an
8-ball gives the Lagrangian of the de Wit-Nicolai
model~\xcite{de1982n}{https://doi.org/10.1016/0550-3213(82)90120-1}~\xcite{de1982nwithlocal}{https://doi.org/10.1016/0370-2693(82)91194-7}.

For our investigations, we are mostly concerned with the scalar sector
of this~``$SO(8)$ supergravity''. Naturally, polarized fields in
11~dimensions give rise to different types of fields in four
dimensions, depending on how 11-dimensional polarization is oriented
with respect to the split into a seven-dimensional compact manifold
and four-dimensional space-time, just like in original Kaluza-Klein
theory, where the five-dimensional metric gives rise to the
four-dimensional metric (gravitons), vector potential (photons), and
scalar field (Higgs boson). Maximal supersymmetry fixes the particle
content completely, and so Cremmer and Julia's construction of
ungauged four-dimensional maximal supergravity that compactifies on a
7-torus and drops higher Kaluza-Klein modes must give rise to the same
particle content as compactification on the surface of an 8-ball (and
retaining only massless modes). The rather nontrivial input here is
that both constructions actually do lead to maximally supersymmetric
models. In Cremmer and Julia's construction, one gets 35 Higgs fields from
the 11-dimensional ``$A_{MNP}$-photons'' for which the 3-dimensional
polarization is parallel to the direction of the 7-dimensional
compactification manifold. Since reversing the handedness of
three-dimensional space can be expressed as an 11-dimensional rotation
that also reverses the handedness of the 7-dimensional
compactification manifold, which is experienced by a 3-form field as a
sign reversal, these~$7\cdot6\cdot5/3!=35$ scalar fields are
pseudo-scalars, i.e.\ odd under a parity transformation.
Correspondingly, we get~$7\cdot 8/2=28$ scalars from those
polarization states of the 11-dimensional graviton~$g_{MN}$ that are
parallel to the embedding manifold. \emph{However}, we also get seven
four-dimensional-2-form potentials~$A_{\mu\nu P}$ for which only one
of the three 11-dimensional~$A_{MNP}$ polarization directions is parallel
to the compactification manifold. These give
rise to four-dimensional 3-form field strengths~$\sim F_{\mu\nu\rho}$,
which can be dualized to 1-form field
strengths~$G_{\lambda}\sim g_{\lambda\sigma}\epsilon^{\sigma\mu\nu\rho}F_{\mu\nu\rho}$,
which in turn come from scalar potentials~$G=\partial A_G$. So,
dualization~\xcite{Cremmer:1997ct}{https://doi.org/10.1016/S0550-3213(98)00136-9}~\xcite{Cremmer:1998px}{https://doi.org/10.1016/S0550-3213(98)00552-5}
of these 2-forms produces another seven scalar fields which, like the
28 from the graviton, are parity-even, so (proper) scalars. One finds
that these indeed combine into one irreducible representation of
eight-dimensional rotations, so, in this compactification, we have
35~scalars~($35^+$) as well as 35~pseudoscalars~($35^-$) from the
(lowest-energy Kaluza-Klein (``Fourier'') modes of the)
11-dimensional degrees of freedom. One indeed finds that these
$35+35$ scalar fields can be understood at parametrizing the coset
space~$E_{7(7)}/(SU(8)/{\mathbb Z}_2)$.

Subtly, despite ungauged maximal supergravity and $SO(8)$~supergravity
having equivalent particle spectra, and the latter also having a
smooth limit in which the gauge coupling constant is taken to
zero\footnote{This does not hold in general, and in particular not for
maximal supergravity in five or seven dimensions, see
e.g.~\cite{gunaydin1985gauged}.}.,
one can \emph{not} readily identify the 70 scalars of one
construction with the 70 scalars of the
other~\xcite{Duff:1982yw}{https://arxiv.org/abs/1201.0386}. Rather, when
compactifying on a 7-sphere, one has to work out fluctuations around a
compactification background geometry, as explained
e.g.\
in~\xcite{Biran:1982eg}{https://doi.org/10.1016/0370-2693(83)91400-4},
\xcite{Duff:1983gq}{http://inspirehep.net/record/188823?ln=en},
\xcite{deWit:1986oxb}{https://doi.org/10.1016/0550-3213(87)90253-7},
\xcite{nicolai2012consistent}{https://doi.org/10.1007/JHEP03(2012)099}.
In the latter case, there is a straightforward way for the rotational
symmetry of the 8-ball to act on these fluctuations, so all
four-dimensional fields should form~$SO(8)$ irreducible
representations. In the former case, the seven two-forms which one
gets from~$A_{MNP}$ clearly do not form an irreducible representation
of~$SO(8)$, so the symmetry enlargement is an emergent phenomenon.

In general, determining the low-energy field content of Kaluza-Klein
type compactifications of M-Theory will require carefully analyzing
the spectra of generalized Laplace operators which act not on scalar
but tensor-valued fields (see
e.g.~\xcite{Duff:1986kaluza}{https://doi.org/10.1016/0370-1573(86)90163-8},
esp. chapters 4, 5, 9),
whose eigenfunctions can be thought of as generalized spherical
harmonics that live on the compactification manifold rather than the
surface of a 3-dimensional ball (as the spherical harmonics do). On
other compactification manifolds, the low energy particle content of
the theory may be rather different, it may even contain~\emph{more}
Higgs-like fields than this~$SO(8)$ supergravity, as
in the construction discussed
in~\xcite{Duff:1983vj}{https://doi.org/10.1016/0370-2693(83)90724-4},
which in total has $67+67=134$.

The algebra~$\mathfrak{so}(8)$ of eight-dimensional rotations is very special in
that it allows an~$S_3$ group of outer algebra automorphisms which
permute the roles of the three different real eight-dimensional
irreducible representations, the vectors, spinors, and co-spinors. Due
to this phenomenon of ``triality'', we have a choice in how to attach
the ``vector'', ``spinor'' and ``co-spinor'' label to the different
eight-dimensional representations. While it is physically reasonable
and common in the literature to associate the~$35^+$ with the
symmetric traceless matrices over the~$\mathfrak{so}(8)$ vectors~${\bf 35}_v$
(given that they contain the graviton polarization states), we deviate
from this convention in the present work and instead associate the
scalars with the symmetric traceless matrices over the
spinors~$35^+\equiv {\bf 35}_s$, while associating (now in alignment
with the literature) the pseudoscalars with the symmetric traceless
matrices over the co-spinors,~$35^-\equiv {\bf 35}_c$. The advantage
of this approach is that it aligns the defining ${\bf 8}$-representation
of the maximal compact subalgebra~$\mathfrak{su}(8)$ of the~$\mathfrak{e}_{7(7)}$ algebra
with the~$\mathfrak{so}(8)$ vector representation, as well as the~${\bf 35}_{s,c}$
with the self-dual/anti-self-dual four-forms
of~$\mathfrak{su}(8)$, i.e.\ we can use the geometric~$\mathfrak{so}(8)$-invariants
$\gamma^{ijkl}_{\dot\alpha\dot\beta}$, $\gamma^{ijkl}_{\alpha\beta}$
to translate between (anti-)self-dual and symmetric-traceless-matrix
language. We heavily rely on this property to give simple expressions
for the locations of all critical points.

While it would be tempting to give the full general~$\mathcal{N}=8$ Lagrangian
in the general unifying form presented
in~\xcite{de2007maximal}{https://doi.org/10.1088/1126-6708/2007/06/049}
that uses the gauge group embedding tensor framework to also include
some alternative constructions in which the gauge group is a
non-compact group\footnote{%
While one would be inclined
to outright reject the idea of noncompact gauge groups, it turns out
that at least the obvious unitarity problems are avoided in
supergravity as the vector kinetic term has a ``mass matrix'' like
factor involving the scalars that actually fixes the signs for non-compact
directions~\xcite{hull1985structure}{https://doi.org/10.1016/0550-3213(85)90551-6},
and any concerns about renormalizability of
such theories are not much different from standard supergravity.}
such as a different real form of $SO(8)$, i.e.~$SO(p, 8-p)$, or a
contraction
thereof~\xcite{hull1989non}{https://doi.org/10.1016/0370-2693(84)91131-6}~\xcite{hull1985structure}{https://doi.org/10.1016/0550-3213(85)90551-6},
or a ``dyonic''
variant~\xcite{dall2012evidence}{https://doi.org/10.1103/PhysRevLett.109.201301},
it is more straightforward for this work to instead refer to the ``classical''
de~Wit-Nicolai Lagrangian in order to explain the physical role of some
key objects for which this work provides extensive data.

The Lagrangian of~$SO(8)$ supergravity reads~\xcite{de1982n}{https://doi.org/10.1016/0550-3213(82)90120-1}:
\begin{equation}
\label{eq:lagrangianSO8}
\begin{array}{lcl}
   {\mathcal L}/e &=& -\frac{1}{2}\,R(e,\omega)-\frac{1}{2}\epsilon^{\mu\nu\rho\sigma}\,
   \Bigl(\bar\psi_{\mu}^{i}\gamma_{\nu}{D}_{\rho}\psi_{\sigma i}-
    \bar\psi_{\mu}^{i}
    \overleftarrow{D}_{\rho}\gamma_{\nu}\psi_{\sigma i}\Bigr)
    \\
    &&
    -\frac{1}{12}\,\Bigl(\bar\chi^{ijk}\gamma^\mu{D}_\mu \chi_{ijk} - \bar\chi^{ijk}\overleftarrow{D}_\mu \gamma^\mu\chi_{ijk}\Bigr)
    -\frac{1}{96}\,\mathcal{A}_{\mu}^{ijk\ell}\mathcal{A}^{\mu}_{ijk\ell}\\
    &&
    -\frac{1}{8}\,\Bigl(F^+_{\mu\nu\,IJ}\left(2S^{IJ,KL}-\delta^{IJ}_{KL}\right)F^{+\,\mu\nu}{}_{KL}+\mbox{h.c.}\Bigr)\\
    &&
    -\frac{1}{2}\,\Bigl(F^+_{\mu\nu\,IJ}\left(S^{IJ,KL}O^{+\,\mu\nu\,KL}\right)+\mbox{h.c.}\Bigr)\\
    &&
    -\frac{1}{4}\,\Bigl(O^+_{\mu\nu}{}^{IJ}\left(S^{IJ,KL}+u^{ij}{}_{IJ}v_{ijKL}\right)O^{+\,\mu\nu\,KL}+\mbox{h.c.}\Bigr)\\
    &&
    -\frac{1}{24}\,\Bigl(\bar\chi_{ijk}\gamma^\nu\gamma^\mu\psi_{\nu\ell}\left(\hat{\mathcal{A}}_\mu^{ijk\ell}+\mathcal{A}_\mu^{ijk\ell}\right)+\mbox{h.c.}\Bigr)\\
    &&
    -\frac{1}{2}\,\delta^{ij}_{i'j'}\bar\psi_\mu^{i'}\psi_\nu^{j'}\bar\psi^\mu_i\psi^\nu_j\\
    &&
    +\frac{\sqrt2}{4}\,\Bigl(\bar\psi^i_\lambda\sigma^{\mu\nu}\gamma^\lambda\chi_{ijk}\bar\psi_\mu^j\psi_\nu^k+\mbox{h.c.}\Bigr)\\
    &&
    +\Bigl(\frac{1}{144}\,\eta\epsilon_{ijk\ell mnpq}\bar\chi^{ijk}\sigma^{\mu\nu}\chi^{\ell mn}\bar\psi^p_\mu\psi^q_\nu+\Bigr.\\
    &&
    \phantom{-\frac{1}{144}}\,\Bigl.\frac{1}{8}\,\bar\psi^i_\lambda\sigma^{\mu\nu}\gamma^\lambda\chi_{ik\ell}\bar\psi_{\mu j}\gamma_\nu\chi^{jk\ell}+\mbox{h.c.}\Bigr)\\
    &&
    +\frac{\sqrt2\eta}{6\cdot 144}\,\Bigl(\epsilon^{ijk\ell mnpq}\bar\chi_{ijk}\sigma^{\mu\nu}\chi_{\ell mn}\bar\psi^r_\mu\gamma_\nu\chi_{pqr}+\mbox{h.c.}\Bigr)\\
    &&
    +\frac{1}{32}\,\bar\chi^{ik\ell}\gamma^\mu\chi_{jk\ell}\bar\chi^{jmn}\gamma_\mu\chi_{imn}\\
    &&
    -\frac{1}{96}\,\bar\chi^{ijk}\gamma^\mu\chi_{ijk}\bar\chi^{\ell mn}\gamma_\mu\chi_{\ell mn}\\
    &&
    +\sqrt{2}g\,A_1^{ij}\bar\psi_{\mu\,i}\sigma^{\mu\nu}\psi_{\nu\,j}+\frac{1}{6}\,g\,A_2{}^i{}_{jk\ell}\bar\psi_{\mu\,i}\gamma^\mu\chi^{jk\ell}\\
    &&
    +g\,A_3{}^{ijk\,\ell mn}\bar\chi_{ijk}\chi_{\ell mn}+\mbox{h.c.}\\
    &&
    +g^2\,\Bigl(\frac{3}{4}\,A_1^{ij}A_{1\,ij}-\frac{1}{24}\,A_2{}^i{}_{jk\ell}A_{2\,i}{}^{jk\ell}\Bigr).
\end{array}
\end{equation}
In the above Lagrangian,~$\mathcal{O}^\pm_{\mu\nu\,KL}$ is a
bilinear function of the fermionic fields~$\psi$ and $\chi$,
$S^{IJ,KL}$ is a function of the Higgs-fields, $u^{ij}{}_{IJ}$
and~$v_{IJKL}$ are pieces of the~$E_7$ ``vielbein'' in
the~${\bf{56}\times\bf{56}}$ representation that describes a point
on the Higgs scalar manifold, while the~$\mathcal{A}_{\mu}^{ijk\ell}$ are
Higgs-scalar kinetic velocities. For details
cf.~\xcite{de1982n}{https://doi.org/10.1016/0550-3213(82)90120-1},~\xcite{deWit:2002vt}{https://doi.org/10.1016/S0550-3213(03)00059-2},~\xcite{de2007maximal}{https://doi.org/10.1088/1126-6708/2007/06/049}.

This Lagrangian is a \emph{consistent truncation} of 11-dimensional
supergravity~\xcite{nicolai2012consistent}{https://doi.org/10.1007/JHEP03(2012)099},
i.e.\ the Kaluza-Klein modes retained here do not source higher modes, and so any solution
of the four-dimensional field equations can be uplifted to an exact (non-linear) solution of
the equations of motion of 11-dimensional supergravity. This is a
``miraculous'' property of the $S^7$~compactification for which the
$F\wedge F\wedge A$-term in the Lagrangian plays an essential role.
The gauge coupling constant~$g$ here is proportional to the inverse
radius of the compactification manifold~$S^7$ which, in Kaluza-Klein
Supergravity, is not determined.

\subsection{The scalar potential}

In this work, we are mainly concerned with the~$\propto g$
and~$\propto g^2$ terms in the Lagrangian. At order~$g^1$, we see
Yukawa couplings that provide the (naive) gravitino and spin-1/2
fermion mass terms via their coupling to the Higgs-like
scalars,~$\sim gA_1 \bar\psi \sigma\psi$,
~$\sim gA_3 \bar\chi \sigma\chi$, and~$\sim A_2\bar\psi\gamma\chi$.
Here, the ``spin 1/2 fermion mass matrix''~$A_3^{ijk\,\ell mn}$,
is given in terms of the gravitino-fermion Yukawa matrix~$A_2$ as
\begin{equation}
A_3^{ijk\,\ell mn}=\frac{\sqrt{2}}{144}\epsilon^{ijkpqr\ell' m'}A_2{}^{n'}{}_{pqr}\delta_{\ell' m'n'}^{\ell mn}.
\end{equation}

At order~$g^2$, we have the scalar potential
\begin{equation}
V(\phi)\colonequals -g^2 e\,\Big \{\frac{1}{24}  A_{2i}{}^{\!jk\ell}A^{\,\,i}_{2\,jk\ell} -
\frac{3}{4}  A_1^{ij}A_{1\,ij}\Big\}\;.
\end{equation}

Since we are restricting ourselves in this work to the single
case of the \emph{compact} gauge group~$SO(8)$ of the original de
Wit-Nicolai
model~\xcite{de1982nwithlocal}{https://doi.org/10.1016/0370-2693(82)91194-7},
we can ignore a number of subtle aspects of electric/magnetic duality
in four-dimensional supergravity that become relevant when trying to
generalize our investigations to other gaugings in four dimensions,
for details
see~\xcite{deWit:2002vt}{https://doi.org/10.1016/S0550-3213(03)00059-2},~\xcite{deWit:2002vz}{https://arxiv.org/abs/hep-th/0212245},~\xcite{dall2012evidence}{https://doi.org/10.1103/PhysRevLett.109.201301}.
The problem at hand then consists of finding critical points of the
scalar potential~$V(\phi_0,\ldots,\phi_{69})$, parametrized by
70~scalar coefficients of non-compact generators of
the~$\mathfrak{e}_{7(7)}$ algebra. In detail, the computation of the
potential looks as follows, using the notational conventions
of~\xcite{fischbacher2010fourteen}{https://doi.org/10.1007/JHEP09(2010)068},
\emph{apart from index-counting always starting at~$0$ in this work}, in order
to make the correspondence between tensor arithmetic and numerical
code published alongside it even more straightforward.
\begin{equation}
\label{eq:potential}
\begin{array}{lcl}
V/g^2&=&-\frac{3}{4}A_1^{ij}\left(A_1^{ij}\right)^*+\frac{1}{24}A_2{}^{i}{}_{jkl}\left(A_2{}^{i}{}_{jkl}\right)^*\\
\mbox{with:}&&\\
A_1{}^{ij}&=&-\frac{4}{21}T_m{}^{ijm}\\
A_{2\,\ell}{}^{ijk}&=&-\frac{4}{3}T_{\ell}{}^{i'j'k'}\delta_{i'j'k'}^{ijk}\\
T_{\ell}{}^{kij}&=&\left(u^{ij}{}_{IJ}+v^{ijIJ}\right)\left(u_{\ell m}{}^{JK}u^{km}{}_{KI}-v_{\ell mJK}v^{kmKI}\right)\\
&&\\
\mathcal{V}^\efa{}_\efb&=&\exp\left(\sum_\mn\phi_\mn g^{(\mn)}\right)^\efa{}_\efb\\
u_{ij}{}^{IJ}&=&2\,\mathcal{V}^\efa{}_\efb\,\delta_\efa^{\mm}\delta^\efb_{\mn}\delta_{ij}^{\ma\mb}\delta^{IJ}_{\mc\md}\\
&\mbox{for}&\efa< 28,\;\efb< 28,\;(\ma,\mb)=Z(\mm),\;(\mc,\md)=Z(\mn)\\
u^{kl}{}_{KL}&=&2\,\mathcal{V}^\efa{}_\efb\,\delta_\efa^{\mm}\delta^\efb_{\mn}\delta^{kl}_{\ma\mb}\delta_{KL}^{\mc\md}\\
&\mbox{for}&\efa\ge28,\;\efb\ge28,\;(\ma,\mb)=Z(\mm-28),\;(\mc,\md)=Z(\mn-28)\\
v_{ijKL}&=&2\,\mathcal{V}^\efa{}_\efb\,\delta_\efa^{\mm}\delta^\efb_{\mn}\delta_{ij}^{\ma\mb}\delta_{KL}^{\mc\md}\\
&\mbox{for}&\efa<28,\;\efb\ge28,\;(\ma,\mb)=Z(\mm),\;(\mc,\md)=Z(\mn-28)\\

v^{klIJ}&=&2\,\mathcal{V}^\efa{}_\efb\,\delta_\efa^{\mm}\delta^\efb_{\mn}\delta^{kl}_{\ma\mb}\delta^{IJ}_{\mc\md}\\
&\mbox{for}&\efa\ge28,\;\efb<28,\;(\ma,\mb)=Z(\mm-28),\;(\mc,\md)=Z(\mn)
\end{array}
\end{equation}

Here, we are using the auxiliary function~$Z$ to translate integer
indices for the adjoint representation of~$\mathfrak{so}(8)$ to
ordered pairs of indices in the defining representation, with
index-counting starting at zero,
\begin{equation}
\label{eq:Z}
\begin{array}{l}
Z(i\cdot 8+j-(i+1)(i+2)/2)=(i, j),\\
\mbox{i.e.\ $Z(0)=(0,1),\,Z(1)=(0,2),\ldots,\,Z(27)=(6,7)$.}
\end{array}
\end{equation}

The ``input data'' are the 70~$\phi_\mn$ coefficients of
non-compact~$\mathfrak{e}_{7(7)}$ generators~$g^{(\mn)}$.
Even as in this work, we only use the non-compact
and~$\mathfrak{so}(8)$ generators of~$\mathfrak{e}_{7(7)}$,
we give a complete construction of the~$133$ $56\times 56$
generator matrices in appendix~\ref{app:e7}, mostly to
ensure that all subsequent investigations into alternative gaugings
can all use the same definitions.

\subsubsection{Equilibria of the equations of motion}

When looking for viable 11-dimensional field configurations of
supergravity that correspond to vacua of a four-dimensional theory,
one is asking for solutions to the dynamical
equations of motion in which, from the four dimensional perspective,
all directional quantities are zero (since a ``vacuum'' should not
have a preferred spatial direction) -- so, we can set all
four-dimensional gauge boson field strengths to zero, i.e.\ we are here
not interested in
``electrovacuum''~\xcite{Rainich124}{https://doi.org/10.1090/S0002-9947-1925-1501302-6}
type solutions. Also, in this analysis, we set all fermionic (spin-1/2
matter and spin-3/2 gravitino) fields to zero. We do not consider
fermion condensates here.
This leaves us with the need to pick a ground state on the
70-dimensional manifold parameterized by the Higgs-Boson-like scalars
of the theory. Conceptually, one would want to look for minima of the
scalar potential, but the actual story is slightly more involved
here~\xcite{warner1984some}{https://linkinghub.elsevier.com/retrieve/pii/0550321384902864}.

\subsubsection{Vacuum stability}

While the equations of motion for the scalar fields (and fields
coupling to them) require the gradient of the potential to vanish in a
vacuum configuration, it so turns out that viable vacuum states
correspond not just to minima, but also some saddle points (and even a
maximum at the origin!) in the potential. This is due to the value of
the scalar potential playing the role of a cosmological constant in
these models. So, for a negative cosmological constant, our vacuum
will have the geometry of a space of constant negative curvature -- an
Anti-de Sitter (AdS) space. When studying stability with respect to
small localized scalar field perturbations of finite total energy, one
has to take into account that the spatial variation of such a
perturbation can not be made arbitrarily small in an AdS background
geometry. So, if a localized perturbation of a spatially constant
background scalar field at a saddle point (or maximum) can decrease
potential energy, the spatial gradient will lead to an increase in
kinetic energy that cannot be made arbitrarily small. One finds that,
overall, one can have (perturbative) stability even at a non-minimum
critical point (i.e.\ $\nabla V(\phi_0) = 0$) as long as there is no
direction~$\delta\phi$ for which the 2nd derivative of the scalar
potential (in a parametrization that gives a ``conventionally
normalized'' kinetic
term~$\mathcal{L}_{kin}=\frac{1}{2}\left(\delta\phi\right)^2$)
is smaller than a threshold known as the Breitenlohner-Freedman (BF)
bound~\xcite{breitenlohner1982stability}{https://doi.org/10.1016/0003-4916(82)90116-6}:
\begin{equation}
\label{eq:BFBound}
m^2L^2=-\frac{1}{2}(d-1)(d-2)\frac{V''(\phi_0)}{V(\phi_0)}\ge -\frac{1}{4}(d-1)^2,
\end{equation}
which for~$d=4$ is~$-9/4=-2.25$.
Here,~$L$ is the AdS radius,~$L^2=m_0^{-2}=-3/V(\phi_0)$~\xcite{nicolai2012consistent}{https://doi.org/10.1007/JHEP03(2012)099}.
Loosely speaking, ``masslessness'' does not correspond to zero eigenvalues
of the mass matrix in the curved~AdS background. For a representation theoretic
perspective and explanation, cf.~\xcite{deWit:2002vz}{https://arxiv.org/abs/hep-th/0212245}.

In fact, it so turns out that for standard $SO(8)$~supergravity (and
many other Kaluza-Klein models), the potential does not seem to have any
minima at all, but there are saddle points that give rise to AdS
backgrounds in which this bound is satisfied. In particular, any
background geometry with some residual supersymmetry will be stable and
not violate this bound. To date, there is only a single known critical
point of the scalar potential of $SO(8)$ supergravity that corresponds
to a stable non-supersymmetric AdS background~\xcite{Fischbacher:2010ec}{https://arxiv.org/abs/1010.4910}~\xcite{godazgar2015so}{https://doi.org/10.1007/JHEP01(2015)056}~\xcite{warner1984some}{https://doi.org/10.1016/0550-3213(84)90286-4}.
While even this detailed investigation, which presents
many more critical points, did not manage to reveal any other stable
non-supersymmetric solutions, and there are good reasons to believe that
they are indeed rare~\xcite{borghese2012minimal}{https://doi.org/10.1007/JHEP07(2012)034},
there are indications that the method
used here to search for solutions tends to (unfortunately) somewhat
avoid parameter space regions that do correspond to stable critical
points. This is, after all, how the new~$\mathcal{N}=1\,SO(3)$ vacuum
escaped discovery in earlier investigations.
So, the authors consider it possible (but unlikely) that there
still are other such solutions that hide very well.

\subsubsection{Finding Solutions}

Historically, the most powerful approach to find critical points of
supergravity potentials before a more effective strategy was presented
in~\xcite{fischbacher2009many}{https://doi.org/10.1007/s10714-008-0736-z}
was to introduce ``Euler angle style'' coordinate parameterizations of
interesting submanifolds of the scalar manifold that have been
selected according to group-theoretical considerations in such a way
that critical points on the submanifold also will be critical points
on the full manifold. While a full coordinate parameterization
of~$E_{7(7)}/(SU(8)/{\mathbb Z}_2)$ is easily seen to be well outside
computational reach, it is indeed feasible to consider the
subgroup~$SU(3)$ of $SO(8)$ in an~$SU(3)\subset SU(4)\subset
U(4)\subset SO(8)$ embedding and parameterize the six-dimensional
manifold of~$SU(3)$-invariant scalars. When Taylor expanding the full
70-dimensional potential around a point that is a critical point on
such a subgroup-invariant submanifold, the linear term has to vanish,
as the gradient then also decomposes into irreducible representations
of the selected subgroup, but cannot carry any contributions that are
not invariant under the chosen subgroup (since each term in the Taylor
expansion is). This strategy was used
in~\xcite{warner1983some}{https://doi.org/10.1016/0370-2693(83)90383-0}
to find all\footnote{There is a second way to embed~$SU(3)$
into~$SO(8)$, but this does not come with an invariant submanifold of
scalars.}  critical points with residual symmetry at least~$SU(3)$,
and led to the general belief that going substantially beyond this
analysis by picking a smaller subgroup of~$SO(8)$ would be possible in
principle, but technically very much infeasible, with perhaps only a
few possible exceptions. This is due to the combinatorial explosion in
algebraic complexity of explicit forms of coordinate-parametrized
potentials as the number of coordinates increases.

Now that we know many critical points that have very little or
even no continuous unbroken gauge symmetry at all, hindsight tells us
that insisting on a fully analytic approach to solve a ``discovery''-type
problem limited our view. While an analytic approach easily
becomes extremely complicated, all that complexity is eliminated by
instead working with numerically evaluated quantities, and focusing on
the use of backpropagation rather than analytic expressions in order
to obtain gradients. Once one has good numerical data, one can start
looking for corresponding exact expressions.

Critical points of the scalar potential correspond to (true or false)
vacuum solutions, i.e.\ field configurations for which all directed
quantities vanish, and the scalar fields do not experience any
acceleration. While false vacua are unstable with respect to some
small localized fluctuations that violate the BF
bound, and the vast majority of critical points of~$SO(8)$
supergravity are indeed observed to be of this type, they are
nevertheless interesting to study. In the past, we have learned much
from such solutions. For example, the study of the~$SO(7)$ critical
point~S0698771 in~\xcite{de1984new}{https://doi.org/10.1016/0370-2693(84)91611-3}
revealed the need to generalize the Freund-Rubin
ansatz to include a warp factor, while some of the new solutions
from~\xcite{fischbacher2010fourteen}{https://doi.org/10.1007/JHEP09(2010)068}
have been useful to identify and resolve subtleties in the uplifting from
four to eleven dimensions
in~\xcite{nicolai2012consistent}{https://doi.org/10.1007/JHEP03(2012)099}.
For some of the new solutions presented here, a deeper investigation into
the nature of accidental (i.e.\ unrelated to any obvious symmetry)
degeneracies in the mass spectra would seem appropriate.

So, while using the AdS/CFT correspondence to study e.g.\ condensed
matter phenomena is doubtful if the AdS side, when embedded into
M-Theory, has unstable modes (which may even be invisible in the
truncation, as is the case for the~$SU(4)$ solution), one would
nevertheless want to at least come to a deeper understanding of the
11-dimensional origin(s) of instability(-ies), perhaps even
looking for ways of stabilization,
cf.~e.g.~\xcite{bobev2010supergravity}{https://doi.org/10.1088/0264-9381/27/23/235013}.

The scalars transform as a (reducible, nontrivial) representation of
the gauge group~$SO(8)$, and a critical point with nonzero vacuum
expectation values for the scalars will hence break the gauge symmetry
to some subgroup of~$SO(8)$ via the Higgs effect. As the scalar
potential has an overall~$SO(8)$ rotational symmetry, a shift in the
scalar fields obtained by applying a small~$SO(8)$ rotation that
actually moves the critical point on the scalar manifold, i.e.\ some
generator of the~$SO(8)$ symmetry that is broken by the particular
choice of the solution on its~$SO(8)$ orbit, corresponds to a flat
direction in the potential. In the particle spectrum, these shifts
would hence correspond to massless scalar (``Goldstone'') particles,
which however for a broken local (gauge) symmetry get absorbed
(``eaten'') by the gauge field to form the extra (``longitudinal'')
spin-1 polarization state that a massive vector boson has over a
massless helicity-1 vector boson. Likewise, massless fermions get
absorbed by the gravitinos to produce missing gravitino polarization
states through the super-Higgs effect.

\subsection{TensorFlow to the rescue}

While we cannot use the supergravity potential directly as a ML loss
function (since we are looking for saddle points, and not minima), it
is possible to derive an expression that conceptually can serve as the
length-squared of the gradient, $S\colonequals |\nabla V(\phi)|^2$, which can be
used as a loss function and is reasonably easy to compute, cf.~$(2.21)$
in~\xcite{deWit:146408}{https://www.sciencedirect.com/science/article/pii/0550321384905170}:
% KEK: https://lib-extopc.kek.jp/preprints/PDF/1983/8310/8310048.pdf
\begin{equation}
\label{eq:stationaritycondition}
\begin{array}{lcl}
 S&\colonequals &|Q_{(+)}^{ijkl}|^2,\quad \text{where}\\
 Q_{(+)}^{ijkl}&=&Q^{ijkl}+\frac{1}{24}\epsilon^{ijklmnpq}Q_{mnpq},\quad\text{and}\\
 Q^{ijkl}&=&\bigl(\frac{3}{4}A_2{}_{m}{}^{ni'j'}A_2{}_n{}^{k'l'm}-A_1{}^{mi'}A_2{}_{m}{}^{j'k'l'}\bigr)\delta_{i'j'k'l'}^{ijkl}\;.
\end{array}
\end{equation}
The~$Q_{+}^{ijkl}$ is the (self-dual) change of the value of
the potential under an infinitesimal variation of the vielbein
when multiplying with an infinitesimal~$E_7$ element from the left,
i.e.\ we are not considering~$\delta V=V(\mathcal{V}(\phi+\delta\phi))-V(\mathcal{V}(\phi))$,
which would be the gradient of the potential with respect to the Higgs
fields~$\phi$, but use the~$E_7$ structure of the potential and
rather consider
\begin{equation}
\delta
V=V((1+\delta\mathcal{V})\cdot\mathcal{V}(\phi))-V(\mathcal{V}(\phi))
\end{equation}
i.e.\ the change of the potential with respect to a small $E_7$
rotation applied to the vielbein matrix~$\mathcal{V}$ from the left.
As the 70 parameters of~$\delta\mathcal{V}$ transform as self-dual
4-forms under the~$SU(8)$ subgroup of~$E_{7(7)}$, the self-dual part
of the tensor~$Q^{ijkl}$ that multiplies this variation to give the
change to the potential has to vanish at a critical
point. (This is also the variation one has to perform to get second
derivatives at a critical point that correspond to actual particle
masses, i.e.\ where the normalization of the kinetic term is the
conventional one.) Since we want to compute the tensors~$A_1,\,A_2$
anyway as part of the search procedure, e.g.\ to add a
supersymmetry-encouraging term to the loss function as discussed
later, this is straightforward to implement. A slightly less efficient
strategy would be to ask TensorFlow for the length-squared of the
gradient, which would (in ``classic'' TensorFlow
graph-mode) perform a backpropagating transformation on the
computational graph, costing roughly twice the memory, and six times
the computation time.

From the ML perspective, minimizing the
``stationarity violation''~$S$ of the potential then is a
problem of just tuning 70 ``learnable'' parameters so that the
stationarity condition is satisfied. While it is indeed possible to
use the rotational~$SO(8)$ symmetry of the potential to further reduce
this 70-dimensional optimization problem to a $70-28=42$-dimensional
one, performing the search in the full 70~dimensions instead seems to
make sense, as it is not very clear what a ``good'' random
distribution to sample starting points from would be. Furthermore,
even if one chooses to use~$SO(8)$ symmetry to (say) diagonalize the
35 pseudo-scalars, it may well happen that a critical point discovered
in this way is more easily understood in a presentation that
diagonalizes the 35 scalars. So, one should anyway always be able to
diagonalize any solution for any of these two representations.

Performing numerical optimization in some 70-dimensional space looks
like an unusually easy ML problem. Yet, there are some
peculiarities:

\begin{itemize}
\item We are not interested in one minimum of the loss function,
  but (ultimately) want to know all inequivalent ones.
\item The idea of ``stochastic gradient descent'' does not make sense
  in this setting: There is a well-defined gradient, but there are no
  ``examples to perform well on'', and therefore also no human-provided
  labels to tune towards.
\item The loss function takes a highly uncommon form. In particular,
  its computation involves exponentiating a complex matrix (in
  a differentiable way).
\item We are actually interested in high numerical accuracy in our
  numerically tuned ``training parameters''.
\end{itemize}

Our problem, then, is to:
\begin{itemize}
  \item numerically find solutions to the~$S=0$ stationarity condition~(\ref{eq:stationaritycondition}),
  \item canonicalize them to a form with few parameters, and
    obtain highly accurate numerical data, and
  \item extract information about physical properties (such as particle
    charges and masses) as well as (if possible) analytic expressions
    for the location of the solution.
\end{itemize}

Ideally, one would like the last step to at the very least produce
sufficiently accurate numerical data to leave little doubt about the
actual existence of a critical point -- even if its location and
properties are only approximately known. In the authors' view, seeing
that the stationarity condition is satisfied numerically to better
than~$10^{-100}$ (as was achievable for most of the new solutions) is
rather convincing.

Of the above steps, the first ``discovery'' step, when attempted without
an efficient computational framework that can do automated
backpropagation, would ask for manually re-writing Ricci-calculus code.
While this is certainly doable by hand (as has been demonstrated
with~\xcite{fischbacher2010fourteen}{https://doi.org/10.1007/JHEP09(2010)068}
and especially~\xcite{fischbacher2009many}{https://doi.org/10.1007/s10714-008-0736-z},
which was published including hand-backpropagated code), it requires both effort and
practice, and it certainly would be useful if this mechanical
transformation were automated -- especially when computations involve
steps such as matrix exponentiation. Also, debugging hand-written
gradient backpropagation code is often tedious, but at least
straightforward, since one can always check the claimed sensitivitites
in the backward pass by ad-hoc injecting an~$\epsilon$ change into the
associated quantity in the forward pass and observing the actual
sensitivity.

Here, TensorFlow can help in these ways:

\begin{itemize}
\item We only need to write code for the computation of the loss function.
      All code that then computes the gradient efficiently is generated
      automatically.
\item It becomes almost trivial to do exploration that requires
      computing gradients for scalar(!) quantities that are themselves
      defined in terms of gradients.
\item Tensor arithmetic be executed on hardware that has been optimized to
      perform well on such tasks, such as in particular GPUs.
\item Google Colab sandbox notebooks~\xcite{googlecolab}{https://colab.sandbox.google.com}
      simplify TensorFlow based code sharing and collaboration.
\end{itemize}

While TensorFlow also allows executing code on specialized Machine
Learning hardware, such as Google's Tensor Processing Units
(TPUs)~\xcite{jouppi2017datacenter}{https://doi.org/10.1145/3079856.3080246},
this is at present not an interesting option for this research here,
since ML applications generally can work with much lower
numerical accuracy than what is needed in this work, and so there is
not a strong economic incentive towards high numerical precision for
TPUs. Similarly, while quantum field theoretic problems often involve
somewhat sparse tensors (in particular due to sparsity of Gamma
matrices), the general trend in ML seems to be away from designs
that rely on sparsely populated tensors, and so trying to exploit
sparseness to improve computational efficiency when solving field
theory problems like the ones studied here with TensorFlow may often
not be worthwhile.

The second point above is interesting. As is known from the general
theory of reverse-mode automatic differentiation of
algorithms~\xcite{Speelpenning:1980:CFP:909337}{https://archive.org/details/compilingfastpar1002spee},
it is always possible to compute the gradient of a scalar function
that is described by an algorithm in a way that needs no more than
some small constant~$k$ times the effort for evaluating the original
function, \emph{independent of the number of components of the
gradient!} In practice,~$k$ somewhat depends on e.g.\ cache
performance, and one typically finds~$k\sim5$, but
\emph{never}~$k\ge 10$.

\subsubsection{Simplifying basic analysis}

For this work, masses of the scalars had to be determined in order to
check whether any modes violate the BF
bound~(\ref{eq:BFBound}). Still, no code had to be written to implement the
mass matrix formula, eq.~(2.25)
from~\xcite{deWit:146408}{https://www.sciencedirect.com/science/article/pii/0550321384905170},
\begin{equation}
\label{eq:massmatrix}
\begin{array}{lcl}
\mathcal{L}(\Sigma^2) &=& -\frac{1}{96}\,  \,g^{\mu \nu} \partial_\mu  \Sigma_{ijkl} \, \partial_\nu\Sigma^{ijkl}  - \frac{g^2}{96}\, \Bigl( \bigl(\frac{2}{3}\, V + \frac{13}{72}\, \bigl| {{A_2}_\ell}^{ijk} \bigr|^2 \bigr) \, \Sigma_{ijkl} \,  \Sigma^{ijkl}   \\
&&+\bigl(6\,  {{A_2}_k}^{mni}   {{A_2}^j}_{mn\ell} - \frac{3}{2}\, {{A_2}_n}^{mij}   {{A_2}^{n}}_{mk\ell}  \bigr) \, \Sigma_{ijpq} \Sigma^{klpq}\\
&& -\frac{2}{3}\, {{A_2}^i}_{mnp} {{A_2}_{q}}^{jk\ell} \Sigma^{mnpq}\Sigma_{ijkl}\Bigr)\;.
\end{array}
\end{equation}
Rather, scalar masses were computed directly by just left-multiplying
the vielbein matrix with an exponentiated~$e_7$ generator
Taylor-expanded to 2nd order only, and then using TensorFlow's
{\tt tf.hessians()} function to obtain the mass matrix.
This performs 70~gradient computations each no more
than six times as expensive as one evaluation of the potential
starting from the unperturbed vielbein, rather than~$\sim 70^2$
evaluations of the potential. In this sense, this work provides an
independent confirmation for the correctness of~(\ref{eq:massmatrix}),
given that masses match values from the literature for critical points
known earlier.

As the potential is exactly known, our gradients are not noisy
estimates (as they usually are in ML), and it makes sense to employ an
optimization method that can utilize this, i.e.\ conjugate-gradient
optimization or BFGS
optimization~\xcite{nocedal2006numerical}{https://doi.org/10.1007/978-0-387-40065-5},
which both try to use subsequent
gradient evaluations to estimate the 2nd-order structure of the
objective function. One convenient way to use TensorFlow as a
``gradient machine'' for various such higher order optimization methods is provided
by the
{\tt tf.contrib.opt.ScipyOptimizerInterface()} helper function.
One must be aware, however, that for degenerate minima of the
objective function, these optimization methods are not expected to
always perform well very close to the minimum, and given the rather
special structure of the problem at hand, we may well encounter such
degenerate minima.

Starting at randomly chosen locations on the 70-dimensional scalar
manifold over and over again produces different critical points. For
this work, the authors solved about~$390\,000$ numerical minimization
problems, each producing a critical point, that afterwards were
de-duplicated. Two solutions were considered equivalent if both the
cosmological constant as well as the eigenvalue spectrum of
the~$A_{1}{}_{IJ}A_{1}{}^{JK}$ tensor were compatible to within the
estimated numerical accuracy of a solution candidate. There are some
cases of critical points with very similar cosmological constant, but
no degeneracies arise at the finesse provided by the~$Snnnnnnn$ naming
scheme that is used in this work for solutions.

Given location information for a solution-candidate that is good to
more than about five decimal digits, the discovery problem can be
considered solved, and one then has to deal with the subsequent
problem of finding a highly accurate -- ideally, analytic -- form. In
some cases, one finds that the geometry of a critical point is rather
special, making it hard for a higher order optimizer to produce an
accurate location. In such situations, it typically helps to run basic
gradient descent (still with hardware floating point accuracy) as a
post-processing step, which also can be done very efficiently
with TensorFlow.

Using this approach, different critical points of the scalar potential
get re-discovered many times over. One finds that the relative sizes
of ``basins of attraction'' for different solutions are very
different. While details do of course somewhat depend on the probability
distribution used to generate starting points, one observes (for
example) that the likelihood to end up at critical point~$S1400056$ is
about $100\times$ higher than the likelihood to end up at
the~$S1400000$ vacuum. Indeed, some of the solutions presented here
were seen only~\emph{once}\footnote{Specifically: $S2503105$, $S2547536$.}.
This makes it rather likely that just increasing the effort by another
factor~10 would produce further solutions. Figuratively speaking, we
suffer from some vacua strongly vacuuming in (pardon the pun) a large
region of search space.

\subsubsection{Loss function design}

Given this situation, one naturally would like to have alternative
approaches to investigate the structure of the scalar manifold. One
idea -- inspired by Morse Theory\footnote{One needs to keep in mind
that critical points of the (un-adulterated) potential may well be
degenerate, and not of the generic form required by Morse Theory. For
example, even when removing the~$SO(8)$ degeneracy~$S0800000$ has
extra flat-to-2nd-order
directions.}~\xcite{Morse1939}{https://doi.org/10.1007/BF01695535} --
is to look not for the minima of the scalar function that measures
stationarity violation, but its saddle points, and then determine how
following the gradient when starting from small perturbations along
special unstable directions (such as the principal axes of the
Hessian) carries one into different critical points of the
potential. As it is plausible that a critical point with a small basin
of attraction when minimizing stationarity violation may actually be
reachable by walking down from a saddle that has a large basin of
attraction in the search for such saddles, this change of perspective
may offer a way to improve the efficiency of the search for overlooked
critical points.

It turns out that implementing this idea in the most naive way is very
easy with TensorFlow, requiring only very little coding while (due to
backpropagation) still offering very good numerical performance. In
order to give an impression of how little effort this is indeed, we
show example code in appendix~\ref{app:watershed}. One notes that the
corresponding calculations involve~\emph{third} derivatives of the
potential (as the stationarity condition is a function of the
gradient, and in order to determine its saddle points, we look at its
gradient-squared, as well as the 2nd derivative (Hessian) of the
stationarity condition). Still, as long as these derivatives get
combined into intermediate scalar quantities (such as: length-squared
of a gradient), the basic insight of reverse mode automatic
differentiation holds, i.e.\ an extra derivative only multiplies the
computational effort by a factor of about six (but retaining high
numerical accuracy).

\begin{figure}
\centering
 \includegraphics[width=12cm,bb=0 0 1148 1103]{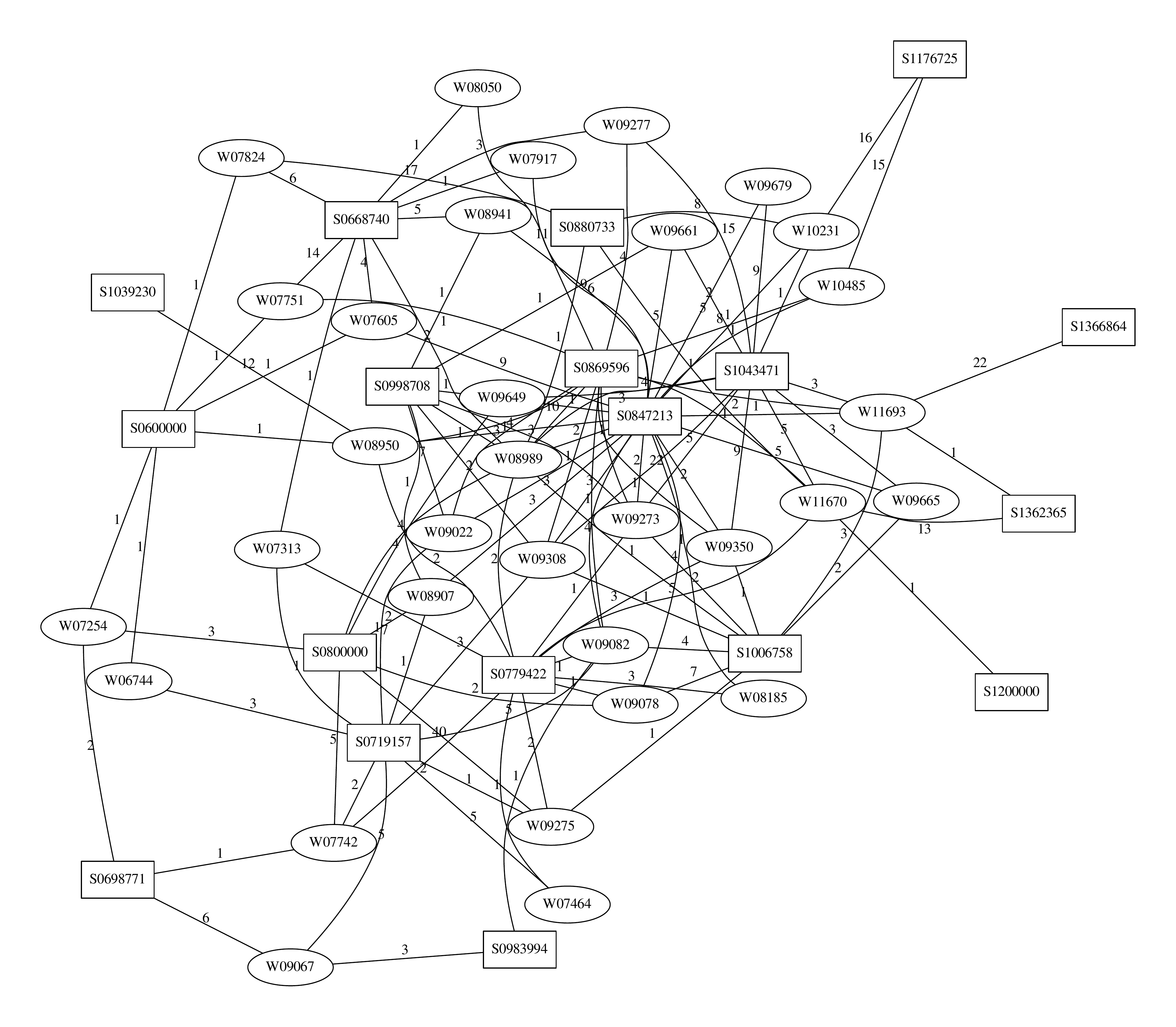}
\caption{
\label{fig:watershedmap}
A tiny part of the ``watershed map'' of
critical points of the stationarity condition.
Minima of the stationarity condition (at value zero) are (true and
false) vacua of the potential, and labeled using the naming scheme introduced
in~\protect{\cite{fischbacher2011encyclopedic}}.
Saddles are also labeled by cosmological constant, e.g.\ the saddle at
$-V/g^2\approx-7.605$ corresponds to~$W07605$.
Edge labels indicate how many gradient-parallel (in naive geometry)
paths along a principal axis run into a particular critical point.
}
\end{figure}

While naively following the gradient disrespects the underlying
symmetry of our 70-dimensional space\footnote{The gradient is an
element of the cotangent space, and, with our parametrization of
the~$e_7$ algebra, already at the origin, taking a step in the
corresponding coordinate-direction is conceptually wrong, as it needs
to be mapped back to an element of tangent space with the inverse
scalar product of the non-orthogonal basis used here.}, this may
actually help rather than harm the search, with an eye on the intended
purpose, by breaking up degeneracies in principal axes. With this
``naive'' saddle point approach, one observes that minimization is
much more likely to run into a saddle than a minimum of the
stationarity condition. Inspection makes it plausible that knowing the
height of the saddle as well as the value of the potential to three
digits after the point suffice to (mostly) deduplicate saddles, and
with this, one can produce a ``subway map'' of how one can cross from
one critical point to another via some saddle. Irrespective of whether
one uses the ``physically correct'' geometry on the scalar manifold or
not, a (mostly) complete map is too complex to be fully
visualized. Figure~\ref{fig:watershedmap} provides a glimpse on what a
tiny part of the graph looks like.

Generating~600 (non-unique) near-origin saddle points and then
analyzing their~$12\,000$ unstable principal axes did indeed confirm
that some critical points which are hard to find by minimizing the
stationarity condition are easier to obtain by this saddle point
method. In particular, the odds for hitting the non-supersymmetric
stable point raise from about~$1:20\,000$ to about~$1:600$. This
(limited) analysis did, however, not produce any new critical points
in the near-origin region where the search was performed. In the
authors' opinion, observing that a somewhat independent method only
reproduces the solutions found with a straightforward random search,
but fails to discover new ones, suggests that the list presented here
likely is the near-complete answer to the question what the critical
points of~$SO(8)$ supergravity are, at least in the near-origin
region. That is, the authors expect the long list to likely still miss
a few cases, perhaps even rather interesting ones\footnote{Soon
after the first release of a preprint of this article,
follow-up work~\cite{so3n1Upcoming} as a by-product indeed gave early
evidence for the existence of two further unstable critical points not
listed here,~$S2096313$ with~$SO(3)\times U(1)$ symmetry, and~$S2443607$
with~$SO(3)$ symmetry. These solutions will be discussed in the
upcoming article.}, but not to list only a small selection
of critical points that happen to be strongly attractive
in a random search. 

Likewise, TensorFlow makes it very simple to tweak loss functions in
order to search for points on the scalar manifold with specific
desired properties. Clearly, one would like to know whether the
current work now gives a complete list of the supersymmetric vacua
of~$SO(8)$ supergravity. While the methods employed here are
insufficient to stringently prove this, it is very easy to tune the
search to strongly favor supersymmetric critical points. A
straightforward way to do this is to replace the
length-squared-of-the-gradient loss function~$L_0=|\nabla V|^2$ with a
loss function that includes another term which is zero for
supersymmetric solutions only. The obvious idea here is that, for a
supersymmetric solution, there needs to be a massless gravitino,
i.e.\ some vector~$\eta_K$ such that
\begin{equation}
L_{S}\colonequals \Bigl|A_1{}^{IJ}A_1{}_{IK}\eta^K + \frac{V}{6g^2}\eta^K\Bigr|^2=0.
\end{equation}

Due to the~$SO(8)$ symmetry of the potential, we do not have to
compute an eigenvector in a differentiable way here, but can simply
fix~$\eta_K=\delta_{K0}$ without loss of generality. Using not~$L_0$
but~$L_0+\lambda L_s$ with a BFGS~optimizer is indeed observed to be
extremely effective for finding supersymmetric solutions.
Taking~$\lambda\sim10$, and starting from a randomly picked
70-dimensional vector (with coordinates drawn from a normal
distribution), with uniform distribution of a length multiplier, one
observes that numerical optimization occasionally does get caught in a
new local minimum with~$L_0>0$ (i.e.\ not a critical point), but
otherwise manages to find each of the known supersymmetric vacua
multiple times with in less than an hour of computing time on
moderately recent hardware. This approach also unearths one additional
supersymmetric vacuum (which also is found many times over) that
has~$\mathcal{N}=1$ supersymmetry and breaks~$SO(8)$ to~$SO(3)$. This
is the solution named~S1384096, see
section~\ref{sec:solutionsguide} and the appendices for
properties. Running this search for a day on a single computer
produced~$7150$ supersymmetric solutions, with each of the now five
solutions ($SO(8)$, $G_2$, $SU(3)\times U(1)$, $U(1)\times U(1)$,
$SO(3)$) being discovered many times over, the lowest count
being~318 for~S0600000.

Is it also possible to directly encode BF-stability as a
ML loss function, and hence directly search for stable vacua in a
similar way? In principle, this can be implemented e.g.\ by
adding to the stationarity-violation loss~$L_0$ another non-negative
contribution~$L_{BF}$ that can only be zero if the scalar mass
matrix~$S$ with all eigenvalues shifted up by the
BF bound is positive semidefinite, i.e.
\begin{equation}
L_{BF} \colonequals  \sum_{A, B}\Bigl|S_{AB} + \frac{9}{4}I_{AB} - \Lambda_{CA}\Lambda_{CB}\Bigr|^2,
\end{equation}
where one introduces a lower triangular matrix~$\Lambda$
of~$70\cdot 71/2=2485$ trainable parameters that will, when minimizing
the loss, try to (Cholesky-)factorize the shifted mass
matrix. Unfortunately, the associated cost that comes with this
large increase in the number of training parameters means
that loss minimization becomes (in comparison) painfully slow. One
notes that the mass matrix in general unfortunately is not
$35^+,35^-$-block-diagonal.  We anticipate that this technique might
become useful for problems with smaller scalar sectors, such as
perhaps maximal gauged~$D=5$ supergravity, but not for~$SO(8)$
supergravity in~$D=4$. Still, there are other minor (fixable)
annoyances with the basic form of this loss contribution,
such as bad behavior in the~$V>0$ region, and the new term
driving the search too fast towards the origin.

There are many more ways in which being able to effortlessly engineer
loss functions might help. For example, it might be feasible to multiply
the stationarity-violation with an extra factor that increases as
search approaches known strong attractors, effectively reducing the
size of their basin of attraction. One obvious way in which this could
be realized would be to add factors of the form
\begin{equation}
|Q_{ijkl}|^2\cdot\left(\sum_{n, p} f\left({\rm tr}\,\left(A_{1\,ij}A_{1}{}^{ik}\right)^p-c_{n, p}\right)\right)
\end{equation}
where the sum over~$p$ runs over (the first few) powers of the gravitino mass
matrix that we use to ``fingerprint'' solutions, the~$c_{n, p}$ is the
corresponding known fingerprint-value for the~$n$-th known
too-attractive solution that should be punished in the search,
and~$f$ is some function with $f(x)\approx 1$ away
from~$0$ and~$f(x)\gg 1$ near~$0$. Some experimenting
will be needed to find an approach that does not create
many new nonzero minima of the loss function.

\subsection{Canonicalization} \label{sec:canonicalization}

For any critical point obtained by a numerical search, the~$SO(8)$
symmetry of the scalar potential allows us to freely pick an arbitrary
point on its~$SO(8)$ orbit as an equivalent presentation. Naturally,
one would want to use a form that allows describing the solution with
a minimal number of parameters. This is not only desirable for
typographic compactness, but also establishes the connection with
simple exact analytic descriptions of these critical points. Setting
an additional coordinate on~$E_{7(7)}/(SU(8)/{\mathbb Z}_2$ to zero
corresponds to imposing an extra algebraic constraint on the solution,
and using sufficiently many such constraints to eliminate all freedom to rotate
a solution produces a 56-bein matrix with only algebraic entries,
since the defining properties of the 56-bein, i.e.\ belonging
to~$E_{7(7)}/(SU(8)/{\mathbb Z}_2)$, can also be expressed through algebraic
constraints. Specificaly, the 56-bein respects the symplectic invariant
of~$Sp(56)$ as well as  Cartan's quartic invariant
of~$E_{7(7)}$~(e.g.\ (B.4b)
in~\xcite{cremmer1979so}{https://doi.org/10.1016/0550-3213(79)90331-6}),
\begin{equation}
  \begin{array}{lcl}
    \mathbf{56}&\to&(\mathbf{28}, \mathbf{\overline{28}}): (x_{ij}, y^{kl})\\
    I_4&=&x_{ij}y^{jk}x_{kl}y^{li}-\frac{1}{4}\left(x_{ij}y^{ij}\right)^2+\\
    &&\frac{1}{96}\left(\epsilon_{ijklmnpq}y^{ij}y^{kl}y^{mn}y^{pq}+\epsilon^{ijklmnpq}x_{ij}x_{kl}x_{mn}x_{pq}\right)\;.
    \end{array}
\end{equation}

At the Lie algebra level (i.e.\ prior to exponentiation),
this then means that there are exact analytic expressions
for the coordinate-parameters describing a given solution,
typically of the form~$\{\mbox{algebraic number}\}\times\log\{\mbox{algebraic number}\}$.
Given that the actual 56-bein entries may well be determined by
rather complicated intersections of many algebraic varieties, actually
finding algebraic forms may well be computationally out of reach in
some cases (i.e.\ one may well imagine to encounter zeros of
irreducible polynomials of degrees well beyond 1000).
Still, for some of the new solutions described here, the authors were
able (with reasonable computational effort) to determine analytic
expressions from high-precision numerics alone. Each such
expression is correct with overwhelming likelihood.

This procedure starts with first obtaining high-precision
(hundreds to thousands of correct digits) numerical data for
quantities that are known to be algebraic (i.e.\ vielbein entries and
derived quantities, such as the cosmological constant).
Unfortunately for us, as extremely high numerical accuracy is
generally not very relevant for ML, TensorFlow does not
support tensor arithmetics with higher numerical precision than
what common hardware can provide, i.e.\ IEEE-754 double precision
floating point.  In that sense, from the perspective of M-Theory
research, TensorFlow perhaps is best thought of as a ``discovery
machine'' and not a ``precision machine'', carrying over terminology
from accelerator physics
(e.g.~\xcite{dittmaierprecision}{https://www.slac.stanford.edu/econf/C060619/present/DITTMAIER1.PDF},~\xcite{collins2008large}{https://www.scientificamerican.com/article/the-discovery-machine-hadron-collider/}).

In principle, it would be doable to run already the ``discovery''
computation with adjustable accuracy, computing e.g.\ the algebraic
entries of~$\mathcal{V}$ to hundreds of decimal digits. This technique
has partly been employed in
~\xcite{fischbacher2010fourteen}{https://doi.org/10.1007/JHEP09(2010)068},
bases on highly performant compiled Common Lisp code in conjunction
with an adapter library that allows a common generic limited-precision
numerical optimizer to work in an high precision setting, but all
this generally involves carefully performing the code transformations
needed for sensitivity backpropagation by hand\footnote{At the time of
this writing, trying to combine the ``mpmath''~\xcite{mpmath}{http://mpmath.org/} and
``autograd''~\xcite{maclaurin2016modeling}{https://github.com/HIPS/autograd}
Python libraries in order to achieve this does not work.}.  This
technique that produces algebraic expressions in a fully automatized
way should hence not (yet) be considered as being easy to apply and
widely accessible. It hence makes sense to aim for a clearer
separation of the ``discovery'' and ``precision'' steps.

\subsection{Parameter-reducing heuristics}

An approximate location of a solution obtained by the ``discovery''
step, once suitably rotated to be coordinate-aligned to the largest
possible extent (using the procedure described further on), gives us
an idea about what coordinates on the scalar manifold can be set to
zero, and what others can likely be set to identical values (or simple
rational multiples of one another). One finds that most solutions are
very non-generic and allow very many such simplifying linear
identities.  In terms of automated processing of many
solution-candidates, this then requires code that tries some basic
heuristics (and automatically abandons them when they turn out to not
actually hold at high precision). The basic process is to go through
all coordinates, check if an observed coordinate is close to another
one seen earlier (or a simple rational multiple thereof, or zero)
within some tolerance limit~$\tau$ such as~$10^{-3}, 10^{-4},
10^{-5}$, etc. If so, the observed coincidence is assumed to hold, and
codified in a linear model-parameters-to-solution-coordinates matrix
map. If the attempt to improve accuracy based on such a model map runs
into a dead end, the process is restarted with a less permissive
tolerance limit~$\tau$. This ``automated heuristic modeling'' step
typically reduces a 70-dimensional optimization problem to a much more
manageable problem in~2--20 or so parameters, for which obtaining
high-accuracy data is very often feasible even without having fast
gradient computation available. The most important techniques here are
using a multidimensional Newton solver (as provided by the
``mpmath''~\xcite{mpmath}{http://mpmath.org/} package), and
high-precision Nelder-Mead optimization, which is feasible for up to
about~$14$ parameters. If both these techniques fail, it is sometimes
useful to use basic fixed-scaling (TensorFlow-based) gradient descent
with hardware numerics to turn a solution that is good to eight digits
into one that is good to at least twelve. One finds that basic
gradient descent with some simple heuristic to make learning rates
adaptive indeed seems to work better for this problem than any of the
more advanced minimization methods that are currently popular in ML
applications, e.g.\ Adam, RMSProp, AdaGrad, FTRL. In some situations,
the parameter reducing heuristic produced a problematic canonical
form, and one has to start over with canonicalization after applying
a random~$SO(8)$ rotation to the solution.

This ``distillation'' step produces a high degree of evidence for the
existence of a particular critical point (the length of the
potential's gradient having been shown numerically to reach values
typically below~$10^{-20}$), as well as a first highly accurate
location described by only a few numbers, and also information on
whether the solution is sufficiently well-behaved (i.e.\ non-degenerate)
for the multidimensional Newton method to allow quick determination of
coordinates and physical properties such as the cosmological constant
to an accuracy of hundreds of digits.

\subsection{Coordinate-aligning rotations}

As a point on the scalar manifold can be described by giving two
symmetric traceless matrices, one carrying the~${\bf 35}_s$ and one
carrying the~${\bf 35}_c$ representation of~$\mathfrak{so}(8)$,
and we can always use a~$SO(8)$ rotation to diagonalize one of these,
the effective dimension of the scalar manifold relevant for finding critical points
is reduced to~$70-28=42$. Still, even if one exploited this symmetry
from the start and only looked for critical points for which one
of~${\bf 35}_s, {\bf 35}_c$ is diagonal, this does not eliminate the
need to numerically canonicalize a solution, as any degeneracy in the
entries of the diagonal matrix would leave some residual rotational
symmetry that can be used to reduce the number of non-zero entries in
the other diagonal matrix. For example, if the parameters in~${\bf 35}_s$
can be brought into the form~${\rm diag}(3A, 3A, -A, -A, -A, -A, -A, -A)$,
this still leaves a residual symmetry of~$SO(2)\times SO(6)$
which must be fixed by imposing algebraic constraints on the~${\bf 35}_c$ in
order to make the entries of the 56-bein matrix algebraic.

As it hence is difficult, in a numerics-based search, to avoid the
need for a ``canonicalization'' step that eliminates residual
rotational freedom, we may just as well fish for solutions in
full 70-dimensional parameter space. While our particular choice
of~$e_7$ generators leaves us with a non-diagonal scalar product
on the 70-dimensional manifold of scalars\footnote{It is orthonormal
up to an overall factor, and the non-orthogonal inner products of the
basis vectors that correspond to the
${\rm diag}(1, -1, 0, \ldots, 0)$,
${\rm diag}(0, 1, -1, 0, \ldots, 0)$ parts of the symmetric
traceless matrices.}, we nevertheless start
the search with a 70-vector picked at random from a distribution
that is isotropic with respect to the coordinate-basis, not the
restricted~$e_7$ Killing form. This choice is apparently
``good enough'' to find many new solutions.

For elements of a 35-dimensional irreducible representation
of~$SO(8)$, it is easy to numerically find a rotation~$G_d$ that
diagonalizes the corresponding symmetric traceless matrix. The
orthonormal eigenbasis serves this purpose if we multiply the last
eigenvector with~$\pm 1$ in order to ensure a positive determinant.
Knowing how to diagonalize, say,~${\bf 35}_s$, how do we then find the
corresponding action
on the~${\bf 35}_c$ (and vice-versa)? Logarithmizing the group element to
obtain the algebra element is out of the question, but we can employ a
higher-dimensional generalization of Davenport chained
rotations~\xcite{davenport1973rotations}{https://dx.doi.org/10.2514/3.6842}~\xcite{wittenburg2003decomposition}{https://doi.org/10.1023/A:1023389218547}
to write~$G_d$ as a product of a sequence
of up to 28 rotations in coordinate-aligned planes,
\begin{equation}
  G_d=R_{67}(\alpha_{67})\cdots R_{57}(\alpha_{57})\cdots R_{56}(\alpha_{56})\cdots\cdot R_{02}(\alpha_{02})\cdot R_{01}(\alpha_{01}),
  \end{equation}
each of which, when moved to the left size, cancels another
off-diagonal entry of~$G_d$ without destroying earlier such
reductions\footnote{For better numerical stability, one should
  re-order processing of row-entries by absolute magnitude.}.
Using this presentation, we can then proceed to
logarithmize each factor
\begin{equation}
R_{\gamma\delta}(\alpha_{\gamma\delta})=\exp\left(r_{\gamma\delta}\cdot\alpha_{\gamma\delta}\right)
\end{equation}
and find the corresponding Lie algebra action on the~$8_c$
representation by employing
the~$\gamma^{\alpha\beta}_{\dot\gamma\dot\delta}$ invariant. Lifting
this to an action on~${\bf 35}_c\subset ({\bf 8}_c\times {\bf 8}_c)$ and exponentiating,
we can readily determine the action of~$G_d$ on~${\bf 35}_c$.  An
alternative approach would be to run numerical minimization starting
from Lie algebra elements that get exponentiated to obtain group
actions on the~${\bf 35}_s$ and~${\bf 35}_c$ with an objective function that
punishes off-diagonal elements for the matrix carrying the~${\bf 35}_s$
representation.

Once the~${\bf 35}_s$ has been diagonalized by any such method that also
gives us the effect of the rotation which was employed on the~${\bf 35}_c$,
we can proceed to determine the residual subalgebra of~$\mathfrak{so}(8)$ that
keeps the diagonalized~${\bf 35}_s$ unchanged. We can employ this subalgebra
to reduce the number of off-diagonal entries for the matrix carrying
the~${\bf 35}_c$ representation, but in general not completely. Also, this
residual symmetry group will typically be rather small -- such
as~$SO(3)\times SO(2)$. It hence makes sense to consider this step as
a somewhat low-dimensional numerical optimization problem. In general,
problems that maximize the number of zero entries in a matrix often
are hard, but here, a common sparseness-encouraging ML
technique works reasonably well: we use the~$L_1$-norm of the
off-diagonal matrix entries as a loss function.

After this sparseness-encouraging rotation, which will in general have
put many more than 28~coefficients to zero, we proceed by making an
automated guess for the form of the symmetric matrices carrying
the~${\bf 35}_s$ and~${\bf 35}_c$ representations as described.

\subsection{``Algebraization''}

Once an highly-accurate numerical value for a known-to-be-algebraic
parameter has been found, one can use an integer relation algorithm such as
PSLQ~\xcite{bailey2009pslq}{https://doi.org/10.2172/963658}
to find a polynomial of which this is a zero and that seems plausible,
i.e.\ the total information content of its coefficients is much smaller
than the information content of the known digits of the
parameter. This works well for up to a few thousand decimal digits.
More specifically, one scans for a set of integer
coefficients~$c_j, j\in\{0,1,\ldots,N-1\}, |c_j|<10^d$
such that for a number~$\xi$ known to~$D$ (perhaps $D\sim300$) digits after the point,
with~$D>N\cdot d+30$, we have~$|\sum_j c_j\xi^j| < 10^{-(Nd+10)}$. If
such a polynomial is found, we can easily determine its actual zero
to~$D$ digits after the point, and check if this polynomial correctly
predicts many further digits of~$\xi$ that were known
but not used to find the~$c_j$. Naturally, if there is a single
candidate polynomial that was found by using 300~known good digits of
precision manages to predict the next 50~digits, we would expect this
to happen purely by chance to be~$\sim 1:10^{50}$. So, by using a
large enough reservoir of extra precision, we can make the likelihood
to accidentally predict an incorrect algebraic expression
fantastically small. While this still
does not constitute a strict mathematical proof, it would be rather
unreasonable to disbelieve the result. Of course, it will in many
cases then be possible to independently establish the validity of a
claimed exact expression, but this will generally require ingenuity
and effort beyond what can easily be automated to process scores of
solutions.

\subsection{Tweaks to the basic procedure}

In general, it makes sense to use moderately refined numerical data
(e.g.\ known to be good to 14 digits of accuracy) and then repeat
the entire procedure that starts with finding a
coordinate-aligning~$\mathfrak{so}(8)$ rotation from cleaned up numerical data,
as limitations on numerical precision of input data from the
ML library data may have caused the (imperfect)
heuristic that suggests a low-dimensional model to miss some
reduction opportunities. In particular, starting from an already
partially parameter-reduced model will lead to rather short products
of Davenport chained rotations, which are numerically much better
behaved than those seen in the initial reduction step.

Once highly-accurate and nicely coordinate-aligned numerical data are
available, inverse symbolic computation methods can automatically
search for exact expressions for physically relevant properties such
as the cosmological constant, coordinates, particle masses and charges
(and hence also residual supersymmetry), and others.

\subsection{Extracting the physics}

Unbroken continuous gauge symmetries can be determined numerically by
mostly straightforward methods, and Lie group theory then allows an automatic
classification. In a first step, one determines the space~$h$
of~$\mathfrak{so}(8)$ generators that leave a given solution invariant and splits
it into orthogonal pieces w.r.t. the Cartan-Killing metric~$h=[h, h]+h'$.
This separates off the generators of the~$U(1)^N$ part of the
residual symmetry. For all new solutions described here, the dimension
of the semisimple~$[h, h]=:\tilde h$ part is 0, 3, or 6, so the
corresponding non-abelian symmetry algebra can only
be~$k\cdot \mathfrak{so(3)}, k\in\{0,1,2\}$, but perhaps embedded
into~$\mathfrak{so}(8)$ in different
ways. The authors' automated analysis handles this case by looking for
a maximal set of~$p$ orthogonal and commuting $\tilde h$-elements~$\tilde h_{c,j}, j\in\{0,\ldots,p-1\}$ (which thanks to
$\mathfrak{so}(8)$ being the algebra of a compact Lie group are simultaneously
diagonalizable in the adjoint representation), then splitting~$\tilde h$
into subspaces that are simultaneous eigenspaces for these~$p$
generators, labeled by $p$-dimensional vectors of eigenvalues. After
defining a subspace of positive roots and identifying simple roots,
taking the commutators of raising and lowering operators associated
with simple roots produces a basis for the Cartan subalgebra that is
useful for numerically identifying angular momentum spectra, which
for~$k\cdot \mathfrak{so}(3)$ are straightforward to map to the irreducible
representation content.  For~$U(1)^N$ generators, the~$N=2$ case will
in general require finding a rotation that coordinate-aligns the
charge lattice.  In any case, we scale every~$U(1)$ generator~$u$ such
that~$\exp(2\pi u)$ is the identity on all particle states, while
no~$\lambda u$ with~$0<\lambda<1$ also has this property. For
solutions whose gauge group has a single~$U(1)$, we use superscripts
to indicate electric charge of particle states, while for solutions
with~$U(1)\times U(1)$ abelian gauge symmetry, we use super- and
sub-scripts for the two different types of charges.

Having split the unbroken symmetry in this way, we can proceed to present
branching rules for the~$\mathbf{8}_{v,s,c}$ as well as~$\mathbf{28}$,
and decompose particle mass-eigenstates for the spin-$1/2$ fermions,
spin-$3/2$ gravitinos, and spin-$0$ scalars in terms of irreducible
representations of the residual gauge algebra.  For the scalars, we first
try to split mass eigenspaces into orthogonal subspaces of the parity
operator~$P$ that is~$+\mathbf{I}$ on~${\bf 35}_{s}$ and is~$-\mathbf{I}$ on~${\bf 35}_{c}$.
However, the analysis presented here only splits off the pure~$P=1$
and~$P=-1$ subspaces and subsequently merges their joint orthogonal
complement into one subspace. Mass eigenstates are then determined on these
three subspaces, and marked with a superscript of~$(s)$,~$(c)$, or~$(m)$
for the ``mixed'' case, respectively. This step is numerically slightly
problematic, as it so turns out that, for some critical points, we find
$P=1-\epsilon$ eigenstates with~$\epsilon<10^{-3}$. Hence,
auto-identification of some mass eigenstate as a pure scalar or pure
pseudoscalar state may be somewhat unreliable. It is nevertheless
reassuring that decomposition of weights into irreducible representations
is observed to always succeed, which it would not if some~$P$-eigenvectors
were wrongly assigned to subspaces.  All particle masses are known with an
accuracy of more than eight digits, but it may happen that different
particle masses look the same when truncated to three digits after the
point for presentation. This explains why some solutions list the same mass
as belonging to more than one mass-subspace. This happens
e.g.\ for~$S1039624$, which lists scalars with $m^2/m_0^2$
of~$-2.417^m_{\mathbf{1}^{++++}+\mathbf{1}^{----}},-2.417^m_\mathbf{1}$,
rather than~$-2.417^m_{\mathbf{1}^{++++}+\mathbf{1}^{----}+\mathbf{1}}$
i.e.\ there is a two-dimensional subspace of mass eigenstates with electric
charges~$\pm 4$ and another one-dimensional mass eigenstate with a
too-close-to-be-discriminated-at-presentation-accuracy mass.
For a solution to be perturbatively stable, no scalar mass-eigenstate
must have~$m^2/m_0^2$ that violates the
BF bound of~$-9/4$. Unstable mass-eigenstates
that violate this bound are marked with an asterisk~$(*)$ in the tables.

As none of the new solutions has a residual gauge group that
contains a simple group of rank~$\ge 2$, and so apparently all such
solutions already have been identified and studied in detail
in the 80's~\xcite{warner1983some}{https://doi.org/10.1016/0370-2693(83)90383-0},
there is no compelling reason to automate assignment of quantum
numbers to such larger symmetry groups.

% Unfortunately, here, we only have the inspirehep.net target link.
% Details are discussed around (4.12):
% https://lib-extopc.kek.jp/preprints/PDF/1983/8308/8308355.pdf
Residual supersymmetry can be
identified~\xcite{deWit:1983dmg}{https://inspirehep.net/record/191218} numerically by using Singular
Value Decomposition~(SVD) to look for solutions of
\begin{equation}
  \eta^{i} A_{2}{}_{i}^{jkl} = 0\;.
\end{equation}
The corresponding gravitino states are marked with an asterisk in the tables.

\section{A guide to the new solutions}\label{sec:solutionsguide}

Detailed data on all the solutions obtained in a large TensorFlow based
cluster search are presented in appendix~\ref{app:summary}.

The structure of the location data presented in these tables is
explained in section~\ref{sec:canonicalization}. Given the sheer
amount of data (masses and charges for 26k particles!), it
makes sense to include a lookup table that lists the most important
properties. This is presented in appendix~\ref{app:summary}. Let us in
this section highlight some specific examples with interesting
properties.

{\bf S1384096} is a new stable vacuum with~$\mathcal{N}=1$
supersymmetry. This is clearly the most remarkable new
discovery. As the 10-parameter form presented here resisted all
attempts to increase numerical accuracy by employing a
multidimensional Newton solver, a computationally rather expensive
Nelder-Mead optimization had to be performed to extract 500+ digits of
numerical accuracy. With this, PSLQ based analysis
(using 400 digits of accuracy to predict 120 further digits) was able to
determine the cosmological constant algebraically as a root
of the following (rather remarkable!) polynomial:
\begin{equation}
  \begin{array}{lcl}
    v&\colonequals  &-V/g^2\approx-13.84096,\qquad w\colonequals  v^4\\
    &&\\
    2^{20} \cdot 3^{30}&=&5^{15}\, w^3 - 2^8\cdot 3^4 \cdot 7 \cdot 53 \cdot 107 \cdot
    887 \cdot 1567\, w^2 + 2^{15}\cdot 3^{17} \cdot 210719\, w\\
    &&\\
    \implies v&=&\frac{2}{3125}\cdot 13500^{1/4}\cdot
    5^{1/2}\cdot\\
    &&\{2^{1/3}\cdot(11731979383924735786651611328125\cdot{129}^{1/2}\\
    &&+ 4882181729086557805429315734818179)^{1/3} + \\
    &&22836248051085301205852\cdot2^{2/3}\cdot D^{-1/3} + 220704046052\}^{1/4}\\
    \text{where}&&\\
    D&\colonequals &11731979383924735786651611328125\cdot{129}^{1/2}+\\
    &&4882181729086557805429315734818179.
\end{array}
\end{equation}

Further, the gravitino masses are roots of these algebraic expressions:
\begin{equation}
\begin{array}{lcll}
\mu\colonequals m^2/m_0^2[\psi]&=&\{\mbox{A zero of\ldots}\}&\\
&&\mu - 1&(\mu=1.0),\\
&&9\,\mu^3-48\,\mu^2+80\,\mu-48&(\mu\approx 2.90620272),\\
&&4\,\mu^3-28\,\mu^2+64\,\mu-49&(\mu\approx 3.17965204),\\
&&36\,\mu^3-216\,\mu^2+412\,\mu-321&(\mu\approx 3.41145711).\\
\end{array}
\end{equation}

The gauge group~$SO(3)$ is embedded in a triality-invariant way as the
diagonal subgroup of a~$SO(3)\times SO(3)\subset SO(8)$.  There are
ten~$SO(3)$-invariant scalars which all have different masses apart
from a pair of two with~$m=0$. A detailed study of the analytic
properties of this new solution is in progress~\cite{so3n1Upcoming}.
Remarkably,~{\bf S1384135}, which has only a minimally lower
cosmological constant has a Gravitino state that~\emph{almost} would
satisfy the Killing spinor equation (no other Gravitino in the long
list comes this close) and also has its gauge group -- here,~$U(1)$ --
embedded in a triality-invariant way. The mass spectra of these two
solutions are very similar.

One notes that all the cosmological constant polynomials that have
been identified have rather special form which, given their
2-3-5-factorizations, somehow seems to be suggestive of an underlying
$E_8$ structure.

{\bf S1600000} and {\bf S1800000} extend the known list of solutions
with rational (even integer) cosmological constant by two new entries,
the others being {\bf S0600000} with $\mathcal{N}=8\; SO(8)$ symmetry,
{\bf S0800000} with $SU(4)$ symmetry, {\bf S1200000} with
$\mathcal{N}=1\; U(1)\times U(1)$ symmetry, and the only known stable
non-supersymmetric point {\bf S1400000}, which has~$SO(3)\times SO(3)$
gauge symmetry (listed as~$SO(4)$ in the summary table).
Despite both {\bf S1600000} and {\bf S1800000} having
no residual gauge symmetry, they have rather remarkable and currently
unexplained degeneracies in particular in the fermion mass spectra.
Such ``accidental degeneracies'' also occur for some other critical
points, such as {\bf S1046018} or~{\bf S1176725}.

The {\bf S0847213} solution is the only ``modern'' critical point with
residual gauge symmetry for which the~$\mathbf{8}_{v,s,c}$ branching
does \emph{not} have any residual triality symmetry. Also, it is
the only case with a 3-component gauge group.

The {\bf S1039230} solution has been first described
in~\xcite{Borghese:2013dja}{https://doi.org/10.1007/JHEP05(2013)107}.
In total, there are three known critical points with residual gauge
symmetry~$SO(4)$.  The gauge group embedding has $V\leftrightarrow S$
triality symmetry for this solution, while it has $V\leftrightarrow C$
triality symmetry for the~{\bf S0880733} solution and
$S\leftrightarrow C$ symmetry for the (BF-stable!)~{\bf S1400000}
solution.

{\bf S2099419} and {\bf S2099422} are a pair of critical points with
extremely similar but nevertheless different particle properties and
also cosmological constant. There are a few other examples for
critical points with very small difference in the cosmological
constant and similar properties, such as the pair~{\bf S2511744}-{\bf
S2512364}, or {\bf S3254262}-{\bf S3254576}.

For almost every solution, the number of scalar modes
with~$m^2/m_0^2=0$ matches the number of broken~$\mathfrak{so}(8)$
generators, as it has to according to Goldstone's theorem. Two
solutions have extra massless (w.r.t. ``naive mass'', not
AdS-massless) modes that survive being eaten by gauge bosons: {\bf
S0800000} (i.e.\ the~$SU(4)$ critical point) and {\bf S1200000}
(i.e.\ the~$\mathcal{N}=1\; U(1)\times U(1)$ vacuum).

Solutions~{\bf S1200000},~{\bf S2279257} and~{\bf S2279859} are listed
in the summary table with symmetry~$U(1)_1$, since in these cases, the
charge of all~$U(1)$-charged particles is the same in
magnitude. Still, these particle-states are listed as having
charges~${\bf 1}^{++}$ etc. in the tables for typographic reasons, due
to the~${\bf 8}_{s, c}$ representations branching to~$4\times{\bf
1}^{+}+4\times{\bf 1}^{-}$, but the spectrum not having any particles
transforming under these representations.

While this expanded list did not uncover any other stable
non-supersymmetric critical points beyond the known~$SO(3)\times
SO(3)$ solution~{\bf S1400000}, we observe a another instance of the
phenomenon that instabilities can become invisible when truncating a
solution to the scalar submanifold that is invariant under the
unbroken gauge group. This was first observed and discussed in detail
for~{\bf S0800000}
in~\xcite{bobev2010supergravity}{https://doi.org/10.1088/0264-9381/27/23/235013}.
There, the unstable scalars come as a single 20-dimensional
irreducible representation of~$SU(4)$, and the question was raised
whether one could project out these unstable modes with an orbifold
construction which would have to use a non-abelian discrete subgroup
of~$SU(4)$ other than the Weyl group (since in these cases, the {\bf
20'} decomposes with singlets). This phenomenon also happens for
the~$SO(7)$ critical points~{\bf S0668740} and~{\bf S0698771}, where
the unstable modes transform as a~{\bf 27}, as well as for~{\bf
S1424025}, where the unstable modes form a single~{\bf 5} of~$SO(3)$.
Clearly, the analysis of all viable discrete subgroups of the gauge
group under which this irreducible representation branches without
singlets is greatly simplified by going from~$SU(4)$ to~$SO(3)$, where
there is an obvious candidate symmetry (namely the icosahedral group)
under which ${\bf 5}\to{\bf 5}$.

Remarkably,~{\bf S2416856} has~\emph{no} residual gauge symmetry, but
six unstable scalar modes with 1+2+3 mass-degeneracy. While trying to
find a way to stabilize this solution seems hopeless here, it would be
interesting to understand what symmetry is responsible for the
accidental mass degeneracies in the unstable modes.

The solutions~{\bf S1195898} and {\bf S2503105} feature a peculiar
$SO(3)$~gauge symmetry which is embedded in a triality-invariant way,
with branching~${\bf 8}_{v,s,c}\to{\bf 3}+{\bf 5}$. In both cases,
there are only two $SO(3)$-invariant scalars, but unfortunately, one
of them is unstable. This is also the only instability of these
solutions. Remarkably, the minimal polynomial for the cosmological
constant is \emph{identical} for these two solutions, so this seems to
be an example of the algebraic equations for a vanishing gradient
allowing a pair of real solutions that are Galois conjugates. It
is noteworthy that even in the large numerical search performed here,
{\bf S2503105} was a chance hit that only showed up~\emph{once}. Had
it been missed, it is conceivable that it ultimately might have been
found in a detailed study of~{\bf S1195898} as an algebraically
equivalent solution. So, looking for Galois conjugates seems to be one
new method to fill possible gaps in the list of solutions.

The same Galois-doubling phenomenon occurs also for the pair~{\bf
S1039624}--{\bf S1402217}, which both also have identically embedded
gauge group. One might speculate that Galois conjugates may well
occur more often, in particular with solutions without residual
symmetry and a cosmological constant that is not known
algebraically. Perhaps then, it may be possible to extract algebraic
numbers that are easier to identify from pairs of numerically known
cosmological constants of critical points with identically embedded
gauge group.

A rather unique property of the~{\bf S4168086}-solution will be the
topic of a short follow-up article.

\section{Conclusions and outlook}

Some problems in the world are very amenable to ML based
approaches, others not so much. As we are, during the current ML
revolution, working on finding out which is which, we sometimes
encounter pleasant surprises. Being able to present an elegant way to
address a fundamental need in quantum gravity research -- the need to
be able to analyze potentials on complicated high-dimensional scalar
manifolds -- certainly is one of these.

More work remains to be done that analyzes other relevant cases using
the methods presented here, in particular the scalar sector of maximal~$D=5$ gauged
supergravity~\xcite{gunaydin1985gauged}{https://doi.org/10.1016/0370-2693(85)90361-2},
i.e.\ the AdS side of the best studied case of the AdS/CFT
correspondence, as well as $CSO(p,q,r)$-gaugings of four-dimensional
maximal supergravity, plus their dyonic
variants~\xcite{dall2012evidence}{https://doi.org/10.1103/PhysRevLett.109.201301},
and other gaugings discussed
in~\xcite{de2007maximal}{https://doi.org/10.1088/1126-6708/2007/06/049},
as well as gauged supergravities in three and two dimensions.

Also, while the present article is a major step forward towards a
proven-exhaustive classification of the critical points of~$SO(8)$
supergravity, this challenging problem is still out of reach. It is
quite likely that the rather peculiar form of the minimal
polynomials of the cosmological
constant that could be obtained for about three dozen critical points
(including all solutions that can be described with up to six
parameters) is a very major clue which the authors currently do not
know how to utilize. Nevertheless, having an algebraic expression for
the cosmological constant typically means that it is also feasible to
find algebraic expressions for the entries of the 56-bein matrix, and
hence quite a bit of the work required to uplift a critical point to a
solution of the equations of motion of 11-dimensional supergravity can
be automated. Not surprisingly, giving exact expressions for
coordinates (which are not algebraic, but
typically~$\{\text{algebraic}\}\cdot{\rm log}\,\{\text{algebraic}\}$)
is quite a bit harder (in fact, with the techniques employed here,
this could only be achieved for a few coordinates), but might actually
be unnecessary to answer most questions as long as the vielbein entries
are known exactly and one can show that the vielbein indeed is an element
of~$E_{7(7)}$. Clearly, it would be very interesting to know in
particular for all the supersymmetric vacua what 11-dimensional
geometries these solutions correspond to -- and also what the
CFT renormalization group flows on M2-branes between the corresponding
fixed points look line.

For all solutions presented here, numerical data on their location are
available in the source package of the arXiv.org preprint of this
article. In many cases, the authors have been able to numerically
reduce the violation of the stationarity condition to much less
than~$10^{-1000}$, and for each claimed critical point, it would be
unreasonable to doubt its existence. Unfortunately, going from
machine accuracy as obtained with TensorFlow to high accuracy
is a somewhat messy process as the heuristics to guess a
low-dimensional model as well as the attempt to use a multi-dimensional
Newton solver do occasionally fail (for quite a few independent
reasons) and require manual intervention. Due to limited time and a
rather large number of cases to analyze, the quality of numerical
data is rather uneven, typically providing 100+ good digits,
but sometimes only providing as few as 16. With a bit of
numerical experimenting, it typically is possible to obtain
1000+ good digits for any given solution within about two days
of work.

Quite a few of the solutions feature (occasionally large) accidental
degeneracies in the spectrum that should be understood and may point
to exploitable symmetry properties. It may well be that the
accidental~$U(1)\not\subset SO(8)$
``background round $S^7$ diffeomorphism'' symmetry of the~$S1400000$
solution discussed in~\xcite{godazgar2015so}{https://doi.org/10.1007/JHEP01(2015)056},
is a first example of such an extra symmetry. Also, it is interesting
to observe that, out of all the solutions newly discovered by
employing numerical methods, there is a single one for which the
residual symmetry group is \emph{not} embedded in a way that has
triality symmetry. Clearly, triality seems to play a rather important
role, and so re-phrasing the problem in octonionic language might
reveal additional structure.

In this work, we have only scratched the surface on cleverly
engineering loss functions in order to extract additional information
from the scalar potential. We expect that, with some ingenuity, much
more is possible here, perhaps even allowing an efficient direct scan
for BF-stable critical points.

Given that even this long list will still have some gaps, what are the
most promising approaches to fill them? One strategy still is to
parametrize and study interesting submanifolds, such as the manifold
of scalars that are invariant under the~$SO(3)$ gauge group of the new
supersymmetric vacuum. Also, assuming that there indeed are perfect
algebraic ``Galois twin'' solutions (i.e.\ not only the
cosmological constant, but also all particle masses are related by
root flipping), it may be possible to discover new solutions by
obtaining algebraic expressions for some at present only numerically
known solutions and then check whether root-flipping can produce new
real solutions. A very promising further approach may be to study the
extremal structure of the ``dyonic'' variants described
in~\xcite{dall2012evidence}{https://doi.org/10.1103/PhysRevLett.109.201301}
and study the fate of critical points when changing the
new~$\omega$-parameter.
In~\xcite{Borghese:2013dja}{https://doi.org/10.1007/JHEP05(2013)107},
this method found the third~$SO(4)$ critical point~$S1039230$ before
it was independently rediscovered here, and it may well reveal
further solutions.

% Parting words.

Historically, the potential of~$SO(8)$ supergravity was first written down
in~\xcite{de1982nwithlocal}{https://doi.org/10.1016/0370-2693(82)91194-7}. About fourty years later, we
now have what is likely the almost-complete list of critical points, and very likely the complete
list of supersymmetric vacua of this theory. Given that some of the ideas that enabled this analysis
were about as old as the original problem but largely unknown in the String Theory community, and
were identified as useful and had to be brought together as part of one of the authors' personal
journey, it seems likely that other technically hard problems in String Theory also would benefit
from more exchange with other fields of research. We hope that by aiming to keep the introduction in
this article accessible to readers with a technical background who are not experts in field theory,
we might attract the attention of experts who can contribute missing puzzle pieces that we are not
even aware of yet, perhaps even allowing a completeness proof of the
list of solutions (perhaps after filling the last gaps).

\noindent {\bf Acknowledgments}

\noindent
It is a pleasure to thank our managers, first and foremost Jyrki
Alakuijala, Rahul Sukthankar, and Jay Yagnik for support on this
rather exotic side project, which turned into a rather unusual Machine
Learning adventure. T.F.~would like to thank Hermann Nicolai and the
Albert Einstein Institute in Potsdam for hospitality and useful
discussions during the final phase of this project, and also Krzysztof
Pilch, Nikolay Bobev, Adolfo Guarino, and Michael Duff for feedback.

\appendix

\vskip3em

{\Large\noindent {\bf Appendix}}

\vskip3em

\renewcommand{\theequation}{\Alph{section}.\arabic{equation}}
\renewcommand{\thesection}{\Alph{section}}

\section{$E_{7}$ Conventions} \label{app:e7}

Spin(8) triality provides some intuitive insights into the question why the
exceptional Lie group~$E_7$ exists. Let us start from the observation
that one can, for example, understand the algebra rotations in nine spatial
dimensions, $\mathfrak{so}(9)$, as an expansion of the group of rotations in eight spatial
dimensions,~$\mathfrak{so}(8)$, by eight extra elements~$r_{k8}$ that, when
scaled and exponentiated, rotate each of the coordinate axes of
eight-dimensional space against the ninth axis,
\begin{equation}
  \begin{array}{lcl}
    R_{k8}(\alpha)_{mn}&=&\exp(\alpha r_{k8})_{mn}\\
    &=&(\delta_{km}\delta_{kn}+\delta_{8m}\delta_{8n})\cos\alpha+(\delta_{km}\delta_{8n}-\delta_{8m}\delta_{kn})\sin\alpha
    \end{array}\;.
\end{equation}
and themselves transform under~$\mathfrak{so}(8)$ as eight-dimensional vectors,
while giving rise to other~$\mathfrak{so}(8)$ rotations in their
commutator,~$[r_{08},r_{18}]=-2r_{01}$ so that we get an overall
structure of~$[g,g] \sim g$, $[g,h] \sim h$, $[h,h] \sim g$. Likewise, one can
obtain the 63-dimensional algebra~$\mathfrak{sl}(8)$ from the 28-dimensional
algebra of~$\mathfrak{so}(8)$ by adding 35 generators that transform as symmetric
traceless matrices under~$\mathfrak{so}(8)$. One easily sees how for the
$8$-dimensional vector representation of~$\mathfrak{sl}(8)$, the generators can
be expressed in a basis of symmetric and anti-symmetric traceless
matrices, where the latter form the subalgebra of~$\mathfrak{so}(8)$ and the
former are the generators that extend this to~$\mathfrak{sl}(8)$. One notes that
one also gets a 63-dimensional real algebra (over complex matrices)
that closes if one multiplied each of the extra 35 generators
with~$i$, so that commutator relations are of the
form~$[g,g] \sim g$, $[g,ih] \sim ih$, $[ih,ih] \sim -g$. The extra minus sign
here also shows up in the signature of the quadratic invariant that
can be formed from the generators, the Killing
form,~$G_{mn}\colonequals {\rm Tr}\, (g_m g_n)$.
Applying this ``Weyl unitarity trick'' to~$\mathfrak{sl}(8)$ produces the Lie
algebra of the \emph{compact} Lie group~$SU(8)$, i.e.~$\mathfrak{su}(8)$.

Now, the~$\mathfrak{so}(8)$ (or ``$\mathfrak{spin}(8)$'')
algebra is special in that there is a
non-abelian~$S_3$ symmetry that acts on its irreducible
representations. This symmetry permutes the roles of the
three inequivalent eight-dimensional irreducible representations,
the ``vectors'', ``spinors'', and ``co-spinors''. Let us consider
extending~$\mathfrak{so}(8)$ with an eight-dimensional vector representation~$v$
as before in order to get~$\mathfrak{so}(9)$, but simultaneously with eight-dimensional
spinors~$s$ and co-spinors~$c$ in such a way that we get commutator
relations of the form~$[g,g] \sim g$, $[g,v] \sim v$, $[v,v] \sim g$, $[v,s] \sim c$ plus
the corresponding cyclically triality-rotated forms such as~$[s,s] \sim g$,
$[s,c] \sim v$, etc, by employing the invariant
tensor~$\gamma^i_{\alpha\dot\beta}$ of~$so(8)$ that implements a
non-degenerate generalized product of eight-dimensional
representations. There is a unique way to work out commutator
relations such that the Jacobi identity~$[a, [b, c]] + \{{\rm cyclic}\}=0$ holds,
and this gives the Lie algebra of the $28+3\cdot 8=52$-dimensional
exceptional group~$F_4$. Performing a similar construction that
extends~$\mathfrak{so}(8)$ in a triality-symmetric way now with the symmetric
traceless matrices over the vectors, spinors and co-spinors produces
the~$28+35_{v}+35_{s}+35_{c}=133$-dimensional Lie algebra
of~$\mathfrak{e}_7$. It is possible to choose signs such that one of the~$35$-dimensional
representations extends~$\mathfrak{so}(8)$ to~$\mathfrak{su}(8)$, while the other two each
extend $\mathfrak{so}(8)$ to~$\mathfrak{sl}(8)$. This then gives the noncompact real
form~$\mathfrak{e}_{7(7)}$ that shows up in four-dimensional
$N=8$~supergravity. So, in a sense, the~$\mathfrak{e}_7$ algebra can be thought
of as a generalized algebra of rotations with ``siamese triplet''
structure, three organisms co-joined at the~$\mathfrak{so}(8)$ heart, but
functioning as one whole body. In order to make this work
self-contained, we spell out the conventions underlying our
construction of~$\mathfrak{e}_{7(7)}$ in detail.
These
match~\xcite{fischbacher2010fourteen}{https://doi.org/10.1007/JHEP09(2010)068},
apart from index counting always being 0-based the present article.

We start from the $\mathfrak{so}(8)$
``Pauli-matrices''~$\gamma^i_{\alpha\dot\beta}$
in the conventions of Green, Schwarz, and
Witten~\xcite{green150superstring}{https://doi.org/10.1017/CBO9781139248563},
but with all indices shifted down by~$1$, as it makes much more sense
in a computational setting to consistently start index counting at
zero. The nonzero entries of~$\gamma^i_{\alpha\dot\beta}$ are all~$\pm
1$, and we list them in compact form~$i\alpha\dot\alpha_{\pm}$, so
e.g.~$357_{-}$ translates as~$\gamma^{i=3}_{\alpha=5,\dot\beta=7}=-1$, etc.

\begin{equation}
{\tiny
\begin{array}{llllllll}
007_{+}&016_{-}&025_{-}&034_{+}&043_{-}&052_{+}&061_{+}&070_{-}\\
101_{+}&110_{-}&123_{-}&132_{+}&145_{+}&154_{-}&167_{-}&176_{+}\\
204_{+}&215_{-}&226_{+}&237_{-}&240_{-}&251_{+}&262_{-}&273_{+}\\
302_{+}&313_{+}&320_{-}&331_{-}&346_{-}&357_{-}&364_{+}&375_{+}\\
403_{+}&412_{-}&421_{+}&430_{-}&447_{+}&456_{-}&465_{+}&474_{-}\\
505_{+}&514_{+}&527_{+}&536_{+}&541_{-}&550_{-}&563_{-}&572_{-}\\
606_{+}&617_{+}&624_{-}&635_{-}&642_{+}&653_{+}&660_{-}&671_{-}\\
700_{+}&711_{+}&722_{+}&733_{+}&744_{+}&755_{+}&766_{+}&777_{+}
\end{array}}
\end{equation}

The~$Spin(8)$ traceless symmetric matrices over the vectors and
spinors are ``bosonic'' objects, i.e.\ cannot discriminate between
a~$360$-degrees rotation and the identity, so they must be expressible
in terms of~$SO(8)$ representations alone. It so turns out that
the~$\bf{35}_{s}$ and~$\bf{35}_{c}$ are equivalent to the self-dual,
respectively anti self-dual four-forms of~$SO(8)$, and the
corresponding~$Spin(8)$-invariant dictionaries are the tensors
\begin{eqnarray}
\gamma^{ijkl}_{\alpha\beta}&=&\gamma^m_{\alpha\dot\gamma}\gamma^n_{\gamma\dot\gamma}\gamma^p_{\gamma\dot\epsilon}\gamma^q_{\beta\dot\epsilon}\delta^{ijkl}_{mnpq}\\
\gamma^{ijkl}_{\dot\alpha\dot\beta}&=&\gamma^m_{\gamma\dot\alpha}\gamma^n_{\gamma\dot\gamma}\gamma^p_{\epsilon\dot\gamma}\gamma^q_{\epsilon\dot\beta}\delta^{ijkl}_{mnpq}.
\end{eqnarray}

We choose the basis for $e_{7(7)}$ such that the last~$28$ generators
(elements $\#105$ to $\#132$) form the $so(8)$ subalgebra,
elements~$\#0$ to $\#34$ transform as symmetric traceless matrices
over the spinors~$\mathbf{35}_s$, elements~$\#35$ to $\#69$ as
symmetric traceless matrices over the co-spinors ~$\mathbf{35}_c$,
and elements~$\#70$ to~$\#104$ as symmetric traceless matrices over
the vectors. Thus, the last~$63$ elements form the
sub-algebra~$\mathfrak{su}(8)$, and the first~$70$ elements the~$70$
non-compact directions of the coset manifold of supergravity scalars.
An~$E_7$ adjoint index~$\mathcal A$ splits as (with the underline
representing a single index that can be associated with a pair
of~$\mathfrak{so}(8)$-indices):
\begin{equation}
\mathcal{A}\rightarrow \underline{(\alpha\beta)} + \underline{(\dot\gamma\dot\delta)} + \underline{(ij)} + \underline{[kl]}
\end{equation}

We furthermore choose the basis for~$\mathfrak{so}(8)$ in such a way that
element~$105+n$, when acting on the vector representation of~$\mathfrak{so}(8)$,
would be represented as the rotation
matrix~$(r_{\underline{[jk]}})^m{}_p=\delta^j_m \delta^k_p-\delta^j_p
\delta^k_m$, i.e.\ the rotation that takes the~$k$-direction into
the~$j$-direction, with~$0\le j\le k\le 7$, where
$n=j\cdot8+k-(j+1)(j+2)/2$.
Hence,~$\mathfrak{e}_7$ basis element~$\#105$ corresponds to the rotation~$r_{\underline{[01]}}$,
basis element~$\#106=r_{\underline{[02]}}$, etc. (lexicographically ordered).

For the three $35$-dimensional symmetric traceless
irreducible~$\mathfrak{so}(8)$ representations, we use the convention that
the first~$7$ basis elements correspond to the diagonal
matrices~$\rm diag(1,-1,0,\ldots,0)$, $\rm diag(0,1,-1,0,\ldots,0)$,
$\rm diag(0,\ldots,0,1,-1)$ (in that order), while element~$7+n$
corresponds to the matrix~$(S_{\underline{(jk)}})^m{}_p=\delta^j_m \delta^k_p+\delta^j_p\delta^k_m$, again with~$0\le j\le k\le 7$,
and also $n=j\cdot8+k-(j+1)(j+2)/2$ -- and with a likewise
lexicographical order for the corresponding non-diagonal
parts of~$S_{\underline{(\alpha\beta)}}$
and~$S_{\underline{(\dot\alpha\dot\beta)}}$.
With these conventions, the~$e_7$
symmetric bilinear form obtained from the fundamental
representation,~$g_{\eaa\eab}=T_{\eaa\efc}{}^\efd T_{\eab\efd}{}^\efc$
is almost diagonal, with entries~$+96$ for~$\efb=\efc<70$,
entries~$-96$ for~$\efb=\efc\ge70$, entries~$-48$ for
the non-orthogonal generators corresponding to the diagonal
parts of the symmetric traceless matrices over the spinors
and co-spinors
(i.e.\ $g_{\eaa=0\,\eab=1},\;g_{\eaa=3\,\eab=2}\;g_{\eaa=35\,\eab=36}$,
etc.), and entries~$+48$ for $G_{\eaa=70\,\eab=71}$, etc. for
the diagonal part of the~${\bf 35}_v$ representation.

In order to define the 56-bein, we need explicit generators for the
pseudoreal~$56$-dimensional fundamental representation of~$\mathfrak{e}_7$.  In
the expressions below, the Einstein summation convention does not
apply for ``technical'' auxiliary indices that are set in typewriter
font and do not belong to irreducible representations).

Given~$T^{(E7)}{}_{\eaa\efb}{}^{\efc}$,
the~$\mathfrak{e}_7$ generator matrices~$g$ used to define the scalar potential
are~$\left(g^{(\mn)}\right)^{\efc}{}_{\efb}=\left(T^{(E7)}{}_{\eaa=\mn}\right)_{\efb}{}^{\efc}$
for~$n=0,\ldots,69$. These $56\times56$ matrices $\left(g^{(\mn)}\right)$ look as
follows (with~$Z(n)$ given by~(\ref{eq:Z})):

\begin{equation}
\begin{array}{lcl}
%%%%%%%%%%%%%%%%%%%%%%%%%%%%%%%%%
T^{(SU(8))}{}_{\Ua\ub}{}^\uc&=&\left\{
\begin{array}{lcl}
+i&\mbox{for}&\Ua=\ub=\uc< 7\\
-i&\mbox{for}&\Ua+1=\ub=\uc< 7\\
+i&\mbox{for}&7\le\Ua<35,\;(\mm,\mn)=Z(\Ua-7), \ub=\mm, \uc=\mn\\
+i&\mbox{for}&7\le\Ua<35,\;(\mm,\mn)=Z(\Ua-7), \ub=\mn, \uc=\mm\\
+1&\mbox{for}&35\le\Ua,\; (\mm,\mn)=Z(\Ua-35), \ub=\mm, \uc=\mn\\
-1&\mbox{for}&35\le\Ua,\; (\mm,\mn)=Z(\Ua-35), \ub=\mn, \uc=\mm\\
\end{array}\right.\\
&&\\
S^{(SO(8))}{}_{\sstab}^{\va\vb\vc\vd}&=&\left\{
\begin{array}{lcl}
\gamma^{\va\vb\vc\vd}_{\sa\sb}(
\delta^\sa_{\mm}\delta^\sb_{\mm}
-\delta^\sa_{\mn}\delta^\sb_{\mn})
&\mbox{for}&\sstab=\mm=\mn-1<7\\
\gamma^{\va\vb\vc\vd}_{\sa\sb}(\delta^\sa_{\mm}\delta^\sb_{\mn}+\delta^\sa_{\mn}\delta^\sb_{\mm})&\mbox{for}&\sstab\ge7,\;{(\mm,\mn)}=Z(\sstab-7)
\end{array}\right.\\
&&\\
C^{(SO(8))}{}_\sctab^{\va\vb\vc\vd}&=&\left\{
\begin{array}{lcl}
\gamma^{\va\vb\vc\vd}_{\ca\cb}(
\delta^{\ca}_{\mm}\delta^{\cb}_{\mm}
-\delta^{\ca}_{\mn}\delta^{\cb}_{\mn})
&\mbox{for}&\sctab=\mm=\mn-1<\\
\gamma^{\va\vb\vc\vd}_{\ca\cb}(\delta^{\ca}_{\mm}\delta^{\cb}_{\mn}+\delta^{\ca}_{\mn}\delta^{\cb}_{\mm})&\mbox{for}&\sctab\ge7,\;{(\mm,\mn)}=Z(\sctab-7)
\end{array}\right.\\
&&\\
T^{(E7)}{}_{\eaa\efb}{}^{\efc}&=&\left\{
\begin{array}{lcl}
\frac{1}{8}S^{(SO(8))}{}_{\sstab}^{\va\vb\vc\vd}(\delta^\va_\mm\delta^\vb_\mn-\delta^\va_\mn\delta^\vb_\mm)(\delta^\vc_\mp\delta^\vd_\mq-\delta^\vc_\mq\delta^\vd_\mp)\qquad\mbox{for}&&\\
\qquad\eaa<35,\;\sstab=\eaa,&&\\
\qquad\efb\ge28,\;(\mp,\mq)=Z(\efb-28),&&\\
\qquad\efc<28,\;(\mm,\mn)=Z(\efc)&&\\
\hline\\
\frac{1}{8}S^{(SO(8))}{}_{\sstab}^{\vc\vd\va\vb}(\delta^\vc_\mm\delta^\vd_\mn-\delta^\vc_\mn\delta^\vd_\mm)(\delta^\va_\mp\delta^\vb_\mq-\delta^\va_\mq\delta^\vb_\mp)\qquad\mbox{for}&&\\
\qquad\eaa<35,\;\sstab=\eaa,&&\\
\qquad\efb<28,\;(\mp,\mq)=Z(\efb),&&\\
\qquad\efc\ge28,\;(\mm,\mn)=Z(\efc-28)&&\\
\hline\\
\frac{i}{8}C^{(SO(8))}{}_{\sctab}^{\va\vb\vc\vd}(\delta^\va_\mm\delta^\vb_\mn-\delta^\va_\mn\delta^\vb_\mm)(\delta^\vc_\mp\delta^\vd_\mq-\delta^\vc_\mq\delta^\vd_\mp)\qquad\mbox{for}&&\\
\qquad35\le\eaa<70,\;\sctab=\eaa-35,&&\\
\qquad\efb\ge28,\;(\mp,\mq)=Z(\efb-28),&&\\
\qquad\efc<28,\;(\mm,\mn)=Z(\efc)&&\\
\hline\\
-\frac{i}{8}C^{(SO(8))}{}_{\sctab}^{\vc\vd\va\vb}(\delta^\vc_\mm\delta^\vd_\mn-\delta^\vc_\mn\delta^\vd_\mm)(\delta^\va_\mp\delta^\vb_\mq-\delta^\va_\mq\delta^\vb_\mp)\qquad\mbox{for}&&\\
\qquad35\le\eaa<70,\;\sctab=\eaa-35,&&\\
\qquad\efb<28,\;(\mp,\mq)=Z(\efb),&&\\
\qquad\efc\ge28,\;(\mm,\mn)=Z(\efc-28)&&\\
\hline\\
%%%%%%%%%%%%%%%%%%%%%%%%%%%%%%%%%%%%%%%%%%%%%%%
2\,T^{(SU(8))}{}_{\Ua\ub}{}^\uc\left(%
  \delta^\mq_\mn\delta^\ub_\mm\delta_\uc^\mp
 +\delta^\mp_\mm\delta^\ub_\mn\delta_\uc^\mq
 -\delta^\mp_\mn\delta^\ub_\mm\delta_\uc^\mq
 -\delta^\mq_\mm\delta^\ub_\mn\delta_\uc^\mp
\right) \qquad\mbox{for}&&\\
\qquad\eaa\ge70,\;\Ua=\eaa-70&&\\
\qquad\efb<28,\;(\mp,\mq)=Z(\efb),&&\\
\qquad\efc<28,\;(\mm,\mn)=Z(\efc)&&\\
\hline\\
2\,\left(T^{(SU(8))}{}_{\Ua\ub}{}^\uc\right)^*\left(%
  \delta^\mq_\mn\delta^\ub_\mm\delta_\uc^\mp
 +\delta^\mp_\mm\delta^\ub_\mn\delta_\uc^\mq
 -\delta^\mp_\mn\delta^\ub_\mm\delta_\uc^\mq
 -\delta^\mq_\mm\delta^\ub_\mn\delta_\uc^\mp
\right) \qquad\mbox{for}&&\\
\qquad\eaa\ge70,\;\Ua=\eaa-70&&\\
\qquad\efb\ge28,\;(\mp,\mq)=Z(\efb-28),&&\\
\qquad\efc\ge28,\;(\mm,\mn)=Z(\efc-28)&&\\
\end{array}\right.\end{array}
\end{equation}

\section{The Octonions and the $\mathfrak{spin}(8)$ invariant $\gamma^i_{\alpha\dot\beta}$} \label{app:octonions}

This section provides a simple and fully self-contained instructive
example for using TensorFlow to numerically solve tensorial algebraic
constraints with very little mental effort. The basic techniques are
essentially the same as the ones used for the main part of this
work. Also, this section provides a detailed answer to the question
how two common conventions for Octonions and $\mathfrak{spin}(8)$
gamma matrices are related.

The Lie group~$Spin(8)$ has three inequivalent irreducible
eight-dimensional representations, the vectors~$\mathbf{8}_v$, for
which we use indices~$i,j,k,\ldots$, the spinors~$\mathbf{8}_s$,
indexed with~$\alpha,\beta,\ldots$, and the co-spinors~$\mathbf{8}_c$,
indexed with~$\dot\alpha,\dot\beta,\ldots$. The $\mathfrak{spin}(8)$
invariant~$\gamma^i_{\alpha\dot\beta}$ provides a unique way to map
spinors and co-spinors to vectors,
$\phi(\cdot,\cdot): \mathbf{8}_s\times \mathbf{8}_c\to\mathbf{8}_v:
(s, c)\to v$ such that (e.g.) for any non-zero element~$S$ of
$\mathbf{8}_s$, the map $\mathbf{8}_c\to\mathbf{8}_v:c\mapsto\phi(S, c)$ is nondegenerate.

This means that we can use~$\gamma^i_{\alpha\dot\beta}$ to define an
invertible 8-dimensional product, that is, an 8-dimensional real division
algebra. Now, using e.g.\ an explicit form of the~$(8, 8, 8)$
tensor~$\gamma^i_{\alpha\dot\beta}$ such as the one given
in~\xcite{green150superstring}{https://doi.org/10.1017/CBO9781139248563},
eq.~(5.B.3), which has no reason to
know about the division algebra interpretation, some arbitrary choice has
been made for the vector space bases of
the~$\mathbf{8}_v$,~$\mathbf{8}_s$,~$\mathbf{8}_c$ representations.
Without loss of generality, we can identify the chosen basis
of~$\mathbf{8}_v$ with the basis of the ``output'' vector space of
the octonionic product as it is defined
in~\xcite{Baez:2001dm}{https://doi.org/10.1090/S0273-0979-01-00934-X},
with the octonionic imaginary units~$e_1\ldots e_7$ satisfying:
\begin{equation}
\begin{array}{l}
e_je_j = -1,\quad e_je_k=-e_ke_j\,(j\neq k),\quad e_1\cdot e_2=e_4\\
e_i\cdot e_j = e_k \implies e_{i+1}\cdot e_{j+1} = e_{k+1}\;\text{and}\;e_{2i}\cdot e_{2j} = e_{2k}\quad\text{(mod~$7$)}.
\end{array}
\end{equation}

Then, as we want to retain orthonormality (but not necessarily
handedness), we can try to find one~$O(8)$ element that changes the
basis of~$\mathbf{8}_s$, and another~$O(8)$ element that changes the
basis of~$\mathbf{8}_c$ in order to precisely align the entries of
the $(8, 8, 8)$-tensor~$\gamma^i_{\alpha\dot\beta}$ with the $(8, 8,
8)$-tensor of the octonionic multiplication table. This objective
provides~$8^3=512$ constraints, while we only have~$2\cdot 28$
continuous parameters (two eight-dimensional rotations) to do the
alignment. Hence, it is quite a nontrivial statement that the stated
objective can indeed be achieved. (It so turns out that there are
discrete choices for this problem that differ in the way how signs are
distributed.) For this instructional example, we are extra lazy and do
not even try to employ a proper parametrization of~$O(8)$ elements
(such as e.g.\ via a Cayley transform~$M=(I-A)(I+A)^{-1}$).  Rather, we
design our optimization problem such that we in principle allow
arbitrary elements of~$GL(8)$, but introduce a term in our objective
function that punishes deviations from orthonormality.

The important point about the piece of code shown below is that, while
this solves a numerical optimization problem in~$2\cdot 64=128$
parameters both with very good performance and accuracy, nowhere did
the need arise for us to provide any code that computes
gradients. This is all handled by the TensorFlow framework.

\lstset{style=pystyle}
\begin{lstlisting}[language=Python,basicstyle=\scriptsize]
import numpy
import tensorflow as tf


def get_gamma_vsc():
  """Computes SO(8) gamma-matrices."""
  # Conventions match Green, Schwarz, Witten's.
  entries = (
      "007+ 016- 025- 034+ 043- 052+ 061+ 070- "
      "101+ 110- 123- 132+ 145+ 154- 167- 176+ "
      "204+ 215- 226+ 237- 240- 251+ 262- 273+ "
      "302+ 313+ 320- 331- 346- 357- 364+ 375+ "
      "403+ 412- 421+ 430- 447+ 456- 465+ 474- "
      "505+ 514+ 527+ 536+ 541- 550- 563- 572- "
      "606+ 617+ 624- 635- 642+ 653+ 660- 671- "
      "700+ 711+ 722+ 733+ 744+ 755+ 766+ 777+")
  ret = numpy.zeros([8, 8, 8])
  for ijkc in entries.split():
    ijk = tuple(map(int, ijkc[:-1]))
    ret[ijk] = +1 if ijkc[-1] == '+' else -1
  return ret


def get_octonion_mult_table():
  """Computes the octonionic multiplication table"""
  # Cf. diagram at: http://math.ucr.edu/home/baez/octonions/
  ret = numpy.zeros([8, 8, 8])
  fano_lines = "124 156 137 235 267 346 457"
  for n in range(1, 8):
      ret[0, n, n] = -1
      ret[n, n, 0] = ret[n, 0, n] = 1
  ret[0, 0, 0] = 1
  for cijk in fano_lines.split():
    ijk = map(int, cijk)
    for p, q, r in ((0, 1, 2), (1, 2, 0), (2, 0, 1)):
      # Note that we have to `go against the direction of the arrows'
      # to make the correspondence work.
      ret[ijk[r], ijk[p], ijk[q]] = -1
      ret[ijk[r], ijk[q], ijk[p]] = +1
  return ret


def find_transforms():
  with tf.Graph().as_default():
    # Ensure reproducibility by seeding random number generators.
    tf.set_random_seed(0)
    transforms = tf.get_variable('transforms', shape=(2, 8, 8),
                                 dtype=tf.float64,
                                 trainable=True,
                                 initializer=tf.random_normal_initializer())
    id8 = tf.constant(numpy.eye(8), dtype=tf.float64)
    gamma = tf.constant(get_gamma_vsc(), dtype=tf.float64)
    otable = tf.constant(get_octonion_mult_table(),
                         dtype=tf.float64)
    # Transform gamma matrices step-by-step, since tf.einsum() does not
    # do SQL-like query planning optimization.
    rotated_gamma = tf.einsum(
        'vAb,bB->vAB', tf.einsum('vab,aA->vAb', gamma, transforms[0]),
        transforms[1])
    delta_mult = rotated_gamma - otable
    delta_ortho_s = tf.einsum('ab,cb->ac',
                              transforms[0], transforms[0]) - id8
    delta_ortho_c = tf.einsum('ab,cb->ac',
                              transforms[1], transforms[1]) - id8
    # This 'loss' function punishes deviations of the rotated gamma matrices
    # from the octonionic multiplication table, and also deviations of the
    # spinor and cospinor transformation matrices from orthogonality.
    loss = (tf.nn.l2_loss(delta_mult) +
            tf.nn.l2_loss(delta_ortho_s) + tf.nn.l2_loss(delta_ortho_c))
    opt = tf.contrib.opt.ScipyOptimizerInterface(
        loss, options=dict(maxiter=1000))
    with tf.Session() as sess:
      sess.run(tf.global_variables_initializer())
      opt.minimize(session=sess)
      return sess.run([loss, transforms])


loss, transforms = find_transforms()
print('Loss: %.6g, Transforms:\n%r\n' % (
    loss, numpy.round(transforms, decimals=5)))

## Prints:
# Loss: 2.19694e-11, Transforms:
# array([[[ 0.5, -0. , -0. ,  0.5,  0.5,  0. ,  0.5,  0. ],
#         [-0.5,  0. ,  0. ,  0.5, -0.5,  0. ,  0.5,  0. ],
#         [-0. , -0.5,  0.5, -0. , -0. ,  0.5,  0. ,  0.5],
#         [ 0. ,  0.5, -0.5, -0. , -0. ,  0.5, -0. ,  0.5],
#         [-0.5,  0. , -0. , -0.5,  0.5,  0. ,  0.5,  0. ],
#         [ 0.5, -0. ,  0. , -0.5, -0.5, -0. ,  0.5, -0. ],
#         [ 0. , -0.5, -0.5,  0. ,  0. , -0.5, -0. ,  0.5],
#         [-0. ,  0.5,  0.5,  0. , -0. , -0.5, -0. ,  0.5]],
#
#        [[-0. ,  0.5,  0.5,  0. ,  0. , -0.5,  0. ,  0.5],
#         [ 0. ,  0.5,  0.5, -0. , -0. ,  0.5,  0. , -0.5],
#         [ 0.5, -0. ,  0. ,  0.5,  0.5,  0. , -0.5,  0. ],
#         [ 0.5,  0. , -0. , -0.5,  0.5,  0. ,  0.5, -0. ],
#         [-0. , -0.5,  0.5, -0. ,  0. , -0.5, -0. , -0.5],
#         [-0. , -0.5,  0.5, -0. ,  0. ,  0.5,  0. ,  0.5],
#         [ 0.5,  0. ,  0. ,  0.5, -0.5, -0. ,  0.5, -0. ],
#         [ 0.5, -0. , -0. , -0.5, -0.5, -0. , -0.5, -0. ]]])
\end{lstlisting}

\section{TensorFlow code for Watershed Analysis} \label{app:watershed}

Finding saddles in the stationarity
condition~(\ref{eq:stationaritycondition}) as well as their principal
axes can in principle be done symbolically, at the level of tensor
equations (requiring substantial effort), or by manually performing
the backpropagation code transformation, and then once again on the
backpropagated code (requiring substantial effort). The code below
illustrates how TensorFlow allows one to achieve this objective with
minimal additional coding effort (27 lines of code) if one already has
code to compute the stationarity condition violation.  The code shown
here sets up a TensorFlow session context (roughly: an association
between a graph describing tensor arithmetic operations and hardware
resources) and then executes a sequence of independent
minimization-searches as specified, serializing results to the
filesystem.

\lstset{style=pystyle}
\begin{lstlisting}[language=Python,basicstyle=\scriptsize]
def do_watershed_descent(seeds_scales=[(0, 0.1)],
                         out='watershed_{}.pickle'):
  graph = tf.Graph()
  with graph.as_default():
    tf_scalar_evaluator = scalar_sector.get_tf_scalar_evaluator()
    t_input = tf.placeholder(tf.float64, shape=[70])
    t_v70 = tf.Variable(
        initial_value=numpy.zeros([70]), trainable=True, dtype=tf.float64)
    op_assign_input = tf.assign(t_v70, t_input)
    sinfo = tf_scalar_evaluator(tf.cast(t_v70, tf.complex128))
    t_potential = sinfo.potential
    t_stationarity = sinfo.stationarity
    t_grad_stationarity = tf.gradients(t_stationarity, t_v70)[0]
    t_loss = tf.tensordot(t_grad_stationarity, t_grad_stationarity, 1)
    t_stationarity_hessian = tf.hessians([t_stationarity], [t_v70])[0]
    optimizer = tf.contrib.opt.ScipyOptimizerInterface(t_loss)
    with tf.Session() as sess:
      sess.run([tf.global_variables_initializer()])
      for n, (seed, scale) in enumerate(seeds_scales):
        rng = numpy.random.RandomState(seed=seed)
        v70 = rng.normal(size=70, scale=scale)
        sess.run([op_assign_input], feed_dict={t_input: v70})
        optimizer.minimize(sess)
        n_info = sess.run([t_potential, t_stationarity, t_loss,
                           t_v70, t_stationarity_hessian])
        with open(out.format(n), 'w') as h:
          pickle.dump(n_info, h)
\end{lstlisting}

\section{Overview over the solutions} \label{app:summary}

In this table, we list all known critical points of~$SO(8)$
supergravity ordered by negative cosmological constant~$-V/g^2$. The
table's columns are:

\begin{tabular}{ll}
N&Number of the solution\\
S&Tag of the solution, based on the truncated integer part of $-V/g^2\cdot10^5$.\\
$\mathcal{N}$&Number of unbroken supersymmetries. Empty for unstable vacua.\\
$H$&The residual gauge symmetry Lie group~$H$.\\
$T$&Triality-invariance of the $H\subseteq SO(8)$ embedding.\\
$M$&The dimension of the~$H$-invariant submanifold of the scalar manifold.\\
$P$&Number of parameters of the solution.\\
$D$&Degree of the cosmological constant's minimal polynomial.\\
$A$&Numerical accuracy.\\
C&Citations, major articles that covered this solution.
\end{tabular}

The tag in the~$S$-column uses the truncated, rather than rounded,
value of the potential, in alignment with earlier articles that use
this naming scheme -- which has become necessary due to the large
number of solutions without any unbroken gauge
symmetry. The $\mathcal{N}$-column is empty for
(BF-)unstable critical points, and shows the
number of unbroken supersymmetries for stable critical
points. Supersymmetric solutions are automatically stable. The
$H$-columns lists residual gauge symmetry (ignoring extra discrete factors),
where we make an effort here to be specific about the actual group. So, if a~$\mathfrak{so}(3)$
subalgebra is embedded into~$\mathfrak{so}(8)$ in such a way that some particle
state transforms as a spinor, we call the gauge group~``$Spin(3)$'',
otherwise~``$SO(3)$''. For $SO(3)\times SO(3)\equiv SO(4)$, we use the
name~``$SO(4)$'' if all particle states can be arranged into
representations of~$SO(4)$ (which would not hold e.g.\ for an
isolated~$({\bf 3}, {\bf 1})$ representation), and $SO(3)\times SO(3)$
otherwise. For~$U(1)$ factors, we indicate with a subscript the largest observed
particle charge once the generator has been re-scaled to the minimal
length that makes all charges integral.
So, for example, the gauge group of {\bf S0880733} that is associated with
the Lie algebra~$\mathfrak{so}(3)+\mathfrak{so}(3)$
is~``$SO(4)$'', while the gauge group of {\bf S1075828} is~``$Spin(3)\times U(1)_{4}$''.
The $T$-column indicates what subgroup of the ``triality'' outer
automorphism group $S_3$ of~$\mathfrak{so}(8)$ the embedding of~$H$ is
invariant under. Here, ``$VSC$'' means full triality invariance, while
``$SC$'' means invariance under a~$S\leftrightarrow C$ exchange, etc.
The ~$M, P, D$-columns provide different rough measures of the
complexity of the solution. The~$M$-column shows the dimension of
the~$H$-invariant submanifold of the scalar manifold, if there is some
residual gauge symmetry~$H$. If one wanted to
find exact expressions for the location and properties of a solution
without resorting to inverse symbolic computation, one would want to
coordinate-parametrize this manifold~$M$. A critical point of the
restricted potential on~$M$ is then guaranteed to also be a critical
point on the whole scalar
manifold~\xcite{warner1983some}{https://doi.org/10.1016/0370-2693(83)90383-0}.
The~$P$-column lists the number of different numerical parameters used
in this work to describe the solution. It is possible that, in some
cases, one can use a~$SO(8)$ rotation to find an alternative form with
even fewer parameters, so this value only gives an upper bound on how
many coordinate-parameters are necessary. The~$D$-column shows the
degree of the smallest polynomial with integer coefficients that could
be found which has a zero at~$-V/g^2$ -- if algebraic identification
of the cosmological constant was successful given the number of
available digits. An entry~$a$ indicates an a-th order polynomial,
while an entry~$a^b$ indicates an $a$-th order polynomial in~$x^b$.
The~$A$-column shows the decimal logarithm of the residual value
of the stationarity condition~$|Q_{ijkl}|^2$. So, an entry of~$100$ means
that~$|Q_{ijkl}|^2<10^{-100}$ for the numerical location data that
have been made available alongside the preprint of this work. The~$p$
column shows the page number on which the solution can be found, and
the~$C$-column lists major articles in which the corresponding
critical point is discussed. A ``*'' indicates a new discovery.

Given the sheer number of almost 200~critical points, and observing
that detailed data on each of these typically fills more than half a
page, the total amount of data is too large to be included in full in
a journal publication. Hence, the authors decided to make both raw
data and typeset summaries available as ancillary files to this
article's preprint on arXiv.org. The URL for a typeset PDF document
that describes the properties of critical point \texttt{Snnnnnnn} is:
\[
\mbox{\texttt{https://arxiv.org/src/1906.00207v4/anc/extrema/Snnnnnnn/physics.pdf}}
\]
while the URL for raw position data is:
\[
\mbox{\texttt{https://arxiv.org/src/1906.00207v4/anc/extrema/Snnnnnnn/location.py.txt}}
\]

In online versions of this article, the overview table should have
working clickable links to the corresponding detailed physics
summaries and high-accuracy scalar parameters.\footnote{
An expanded version of this work with a consolidated table of detailed
properties of critical points is available as a pre-publication
preprint of this work on arXiv.org
at \href{https://arxiv.org/abs/1906.00207v3}{\texttt{https://arxiv.org/abs/1906.00207v4}}}.

For each solution, the detailed table shows the location in the form of two
symmetric traceless matrices,~$M_{\alpha\beta}$
and~$M_{\dot\alpha\dot\beta}$, which have been rotated to maximize
(but in some cases perhaps not globally) the number of zero
entries. The normalization of these matrices is such that
the~$\mathfrak{e}_7$ generator~$G$ in the fundamental~${\bf
56}$-representation that is parameterized by the
matrices~$M_{\alpha\beta}, M_{\dot\alpha\dot \beta}$ satisfies
\begin{equation}
{\rm tr}\, G\cdot G = 48\cdot\left(M_{\alpha\beta}M_{\alpha\beta} + M_{\dot\alpha\dot\beta}M_{\dot\alpha\dot\beta}\right).
\end{equation}

In the literature, locations of critical points are typically given in~$\phi_{ijkl}$ four-form language,
e.g.~(4.7) in~\xcite{deWit:1983dmg}{https://inspirehep.net/record/191218} for the solution~S1400000,
\begin{equation}
\phi_{ijkl}={\rm artanh}\,(2/\sqrt5)\Bigl(\left(\delta_{ijkl}^{1234}+\delta_{ijkl}^{5678}\right)+i\left(\delta_{ijkl}^{1235}-\delta_{ijkl}^{4678}\right)\Bigr).
\end{equation}
The dictionary to translate between~$\phi_{ijkl}$ and~$M_{\alpha\beta},\,M_{\dot\alpha\dot\beta}$ reads\footnote{For~$\mathfrak{spin}(8)$
with the obvious basis choices for representation spaces, there is no difference between upper and lower indices.}:
\begin{equation}
\sqrt{2}\cdot\phi_{ijkl}=M_{\alpha\beta}\gamma_{\alpha\beta}^{ijkl}+i\cdot M_{\dot\alpha\dot\beta}\gamma_{\dot\alpha\dot\beta}^{ijkl}
\end{equation}

The detailed location data on arXiv.org allow checking all claims made here about
the existence of particular critical points to numerical machine
precision with the code that has been made available
alongside~\xcite{Fischbacher:2010ki}{https://arxiv.org/e-print/1007.0600}. For
some critical points, the numerical data provided are accurate to
beyond 1000~digits, which should in some cases be sufficient to obtain
algebraic expressions for the 56-bein matrix entries via inverse
symbolic computation techniques, hence allowing automation(!) of the
uplifting to 11~dimensions along the lines of the
construction presented in~\xcite{nicolai2012consistent}{https://doi.org/10.1007/JHEP03(2012)099},~\xcite{godazgar2015so}{https://doi.org/10.1007/JHEP01(2015)056}.
Unfortunately, for about~40 critical points, the authors were not able
to obtain data that in accuracy go substantially beyond numerical
hardware (i.e.\ IEEE-754 double float) precision.

For the scalars~$\phi$, the tables list masses $m^2/m_0^2[\phi]=m^2L^2$ relative to the
AdS radius~$L^2=-3/V$ according to eq.~(\ref{eq:BFBound}) and~(\ref{eq:massmatrix}).
The ``naive'' gravitino masses~$m^2/m_0^2[\psi]$ listed are the eigenvalues
of~$2\,g^2L^2 A_{1\,ij}A_{1}{}^{ik}$, so unbroken supersymmetry shows
as a mass-squared eigenvalue of~$+1$. Likewise, fermion masses~$m^2/m_0^2[\chi]$
are eigenvalues of~$2\,g^2L^2 \cdot\frac{1}{6}A_{3\,ijk\,lmn}A_{3}{}^{ijk\,pqr}$.

The code that was written to automate gauge group representation assignment for the
particle spectra only processes gauge groups whose Lie algebra
is of the form~$\mathfrak{so}(3)^a+\mathfrak{u}(1)^b$, which must hold for
all new solutions, as all the solutions with symmetry at least~$\mathfrak{su}(3)$
have been classified a long time ago~\xcite{warner1983some}{https://doi.org/10.1016/0370-2693(83)90383-0},
and there are no critical points with residual
symmetry~$\mathfrak{su}(3)\not\supseteq\mathfrak{s}\subseteq\mathfrak{so}(5)$.
So, data on the spectrum of these points were added by hand,
based on~\xcite{bobev2010supergravity}{https://doi.org/10.1088/0264-9381/27/23/235013}
and~\xcite{meissner2015standard}{https://doi.org/10.1103/PhysRevD.91.065029}.

\newcommand{\citeff}{\xcite{fischbacher2010fourteen}{https://doi.org/10.1007/JHEP09(2010)068}}
\newcommand{\citefe}{\xcite{fischbacher2011encyclopedic}{http://arxiv.org/abs/arXiv:1109.1424}}
\newcommand{\citebgr}{\xcite{Borghese:2013dja}{https://doi.org/10.1007/JHEP05(2013)107}}
\newcommand{\arxivpageref}[1]{\arxivonly{\pageref{#1}}}

{\scriptsize\begin{longtable}{RRRGGRRLRRG}
\mbox{N}&\mbox{S}&\mathcal{N}&\mbox{H}&\mbox{T}&\mbox{M}&\mbox{P}&\mbox{D}&\mbox{A}&\arxivonly{\mbox{p}}&\mbox{C}\\
\hline
1&\href{https://arxiv.org/src/1906.00207v4/anc/extrema/S0600000/physics.pdf}{S0600000}&8&SO(8)&VSC&0&\href{https://arxiv.org/src/1906.0207v4/anc/extrema/S0600000/location.py.txt}{0}&1&\infty&\arxivpageref{S:S0600000}&\mbox{\xcite{de1982n}{https://doi.org/10.1016/0550-3213(82)90120-1}\xcite{warner1984some}{https://doi.org/10.1016/0550-3213(84)90286-4}}\\
2&\href{https://arxiv.org/src/1906.00207v4/anc/extrema/S0668740/physics.pdf}{S0668740}&&SO(7)&VC&1&\href{https://arxiv.org/src/1906.0207v4/anc/extrema/S0668740/location.py.txt}{1}&1^4&2399&\arxivpageref{S:S0668740}&\mbox{\xcite{warner1983some}{https://doi.org/10.1016/0370-2693(83)90383-0}}\\
3&\href{https://arxiv.org/src/1906.00207v4/anc/extrema/S0698771/physics.pdf}{S0698771}&&SO(7)&VS&1&\href{https://arxiv.org/src/1906.0207v4/anc/extrema/S0698771/location.py.txt}{1}&1^2&2400&\arxivpageref{S:S0698771}&\mbox{\xcite{Englert:1982vs}{https://doi.org/10.1016/0370-2693(82)90684-0}}\\
4&\href{https://arxiv.org/src/1906.00207v4/anc/extrema/S0719157/physics.pdf}{S0719157}&1&G_2&VSC&2&\href{https://arxiv.org/src/1906.0207v4/anc/extrema/S0719157/location.py.txt}{2}&1^4&2399&\arxivpageref{S:S0719157}&\mbox{\xcite{warner1983some}{https://doi.org/10.1016/0370-2693(83)90383-0}}\\
5&\href{https://arxiv.org/src/1906.00207v4/anc/extrema/S0779422/physics.pdf}{S0779422}&2&SU(3)\times U(1)_6&-&4&\href{https://arxiv.org/src/1906.0207v4/anc/extrema/S0779422/location.py.txt}{2}&1^2&2400&\arxivpageref{S:S0779422}&\mbox{\xcite{warner1983some}{https://doi.org/10.1016/0370-2693(83)90383-0}}\\
\hline
6&\href{https://arxiv.org/src/1906.00207v4/anc/extrema/S0800000/physics.pdf}{S0800000}&&SU(4)&VSC&2&\href{https://arxiv.org/src/1906.0207v4/anc/extrema/S0800000/location.py.txt}{1}&1&2400&\arxivpageref{S:S0800000}&\mbox{\xcite{warner1983some}{https://doi.org/10.1016/0370-2693(83)90383-0}}\\
7&\href{https://arxiv.org/src/1906.00207v4/anc/extrema/S0847213/physics.pdf}{S0847213}&&Spin(3)\times U(1)_2\times U(1)_2&-&8&\href{https://arxiv.org/src/1906.0207v4/anc/extrema/S0847213/location.py.txt}{3}&2&1978&\arxivpageref{S:S0847213}&\mbox{\citefe}\\
8&\href{https://arxiv.org/src/1906.00207v4/anc/extrema/S0869596/physics.pdf}{S0869596}&&SO(3)\times U(1)_4&VS&4&\href{https://arxiv.org/src/1906.0207v4/anc/extrema/S0869596/location.py.txt}{2}&2^2&2398&\arxivpageref{S:S0869596}&\mbox{\citefe}\\
9&\href{https://arxiv.org/src/1906.00207v4/anc/extrema/S0880733/physics.pdf}{S0880733}&&SO(4)&VC&6&\href{https://arxiv.org/src/1906.0207v4/anc/extrema/S0880733/location.py.txt}{2}&2^2&2398&\arxivpageref{S:S0880733}&\mbox{\citefe}\\
10&\href{https://arxiv.org/src/1906.00207v4/anc/extrema/S0983994/physics.pdf}{S0983994}&&SO(3)\times U(1)_4&VC&4&\href{https://arxiv.org/src/1906.0207v4/anc/extrema/S0983994/location.py.txt}{3}&1^4&2399&\arxivpageref{S:S0983994}&\mbox{\citefe}\\
\hline
11&\href{https://arxiv.org/src/1906.00207v4/anc/extrema/S0998708/physics.pdf}{S0998708}&&U(1)_4&VSC&14&\href{https://arxiv.org/src/1906.0207v4/anc/extrema/S0998708/location.py.txt}{9}&&2399&\arxivpageref{S:S0998708}&\mbox{\citeff}\\
12&\href{https://arxiv.org/src/1906.00207v4/anc/extrema/S1006758/physics.pdf}{S1006758}&&U(1)_2&VSC&26&\href{https://arxiv.org/src/1906.0207v4/anc/extrema/S1006758/location.py.txt}{10}&&257&\arxivpageref{S:S1006758}&\mbox{\citefe}\\
13&\href{https://arxiv.org/src/1906.00207v4/anc/extrema/S1039230/physics.pdf}{S1039230}&&SO(4)&VS&6&\href{https://arxiv.org/src/1906.0207v4/anc/extrema/S1039230/location.py.txt}{2}&1^2&2398&\arxivpageref{S:S1039230}&\mbox{\citebgr}\\
14&\href{https://arxiv.org/src/1906.00207v4/anc/extrema/S1039624/physics.pdf}{S1039624}&&U(1)_6&VC&20&\href{https://arxiv.org/src/1906.0207v4/anc/extrema/S1039624/location.py.txt}{6}&5^4&227&\arxivpageref{S:S1039624}&\mbox{\citefe}\\
15&\href{https://arxiv.org/src/1906.00207v4/anc/extrema/S1043471/physics.pdf}{S1043471}&&&&&\href{https://arxiv.org/src/1906.0207v4/anc/extrema/S1043471/location.py.txt}{10}&&2399&\arxivpageref{S:S1043471}&\mbox{\citeff}\\
\hline
16&\href{https://arxiv.org/src/1906.00207v4/anc/extrema/S1046017/physics.pdf}{S1046017}&&&&&\href{https://arxiv.org/src/1906.0207v4/anc/extrema/S1046017/location.py.txt}{6}&&219&\arxivpageref{S:S1046017}&\mbox{\citefe}\\
17&\href{https://arxiv.org/src/1906.00207v4/anc/extrema/S1067475/physics.pdf}{S1067475}&&U(1)_2\times U(1)_4&VC&12&\href{https://arxiv.org/src/1906.0207v4/anc/extrema/S1067475/location.py.txt}{5}&3^4&2398&\arxivpageref{S:S1067475}&\mbox{\citeff}\\
18&\href{https://arxiv.org/src/1906.00207v4/anc/extrema/S1068971/physics.pdf}{S1068971}&&U(1)_4\times U(1)_4&VS&18&\href{https://arxiv.org/src/1906.0207v4/anc/extrema/S1068971/location.py.txt}{7}&5^2&2399&\arxivpageref{S:S1068971}&\mbox{\citefe}\\
19&\href{https://arxiv.org/src/1906.00207v4/anc/extrema/S1075828/physics.pdf}{S1075828}&&Spin(3)\times U(1)_4&VS&10&\href{https://arxiv.org/src/1906.0207v4/anc/extrema/S1075828/location.py.txt}{5}&1^2&256&\arxivpageref{S:S1075828}&\mbox{\citefe}\\
20&\href{https://arxiv.org/src/1906.00207v4/anc/extrema/S1165685/physics.pdf}{S1165685}&&U(1)_4\times U(1)_4&VS&18&\href{https://arxiv.org/src/1906.0207v4/anc/extrema/S1165685/location.py.txt}{1}&2&2398&\arxivpageref{S:S1165685}&\mbox{\citeff}\\
\hline
21&\href{https://arxiv.org/src/1906.00207v4/anc/extrema/S1176725/physics.pdf}{S1176725}&&&&&\href{https://arxiv.org/src/1906.0207v4/anc/extrema/S1176725/location.py.txt}{4}&4^4&2399&\arxivpageref{S:S1176725}&\mbox{\citefe}\\
22&\href{https://arxiv.org/src/1906.00207v4/anc/extrema/S1195898/physics.pdf}{S1195898}&&SO(3)&VSC&2&\href{https://arxiv.org/src/1906.0207v4/anc/extrema/S1195898/location.py.txt}{3}&3^4&2398&\arxivpageref{S:S1195898}&\mbox{\citefe}\\
23&\href{https://arxiv.org/src/1906.00207v4/anc/extrema/S1200000/physics.pdf}{S1200000}&1&U(1)_1\times U(1)_1&SC&14&\href{https://arxiv.org/src/1906.0207v4/anc/extrema/S1200000/location.py.txt}{2}&1&2399&\arxivpageref{S:S1200000}&\mbox{\citeff}\\
24&\href{https://arxiv.org/src/1906.00207v4/anc/extrema/S1212986/physics.pdf}{S1212986}&&U(1)_4&VSC&14&\href{https://arxiv.org/src/1906.0207v4/anc/extrema/S1212986/location.py.txt}{6}&&257&\arxivpageref{S:S1212986}&\mbox{\citefe}\\
25&\href{https://arxiv.org/src/1906.00207v4/anc/extrema/S1271622/physics.pdf}{S1271622}&&Spin(3)&VS&20&\href{https://arxiv.org/src/1906.0207v4/anc/extrema/S1271622/location.py.txt}{3}&3&2398&\arxivpageref{S:S1271622}&\mbox{\citefe}\\
\hline
26&\href{https://arxiv.org/src/1906.00207v4/anc/extrema/S1301601/physics.pdf}{S1301601}&&U(1)_4\times U(1)_4&VC&18&\href{https://arxiv.org/src/1906.0207v4/anc/extrema/S1301601/location.py.txt}{4}&3^4&2397&\arxivpageref{S:S1301601}&\mbox{\citefe}\\
27&\href{https://arxiv.org/src/1906.00207v4/anc/extrema/S1360892/physics.pdf}{S1360892}&&U(1)_6\times U(1)_2&SC&18&\href{https://arxiv.org/src/1906.0207v4/anc/extrema/S1360892/location.py.txt}{5}&3^2&238&\arxivpageref{S:S1360892}&\mbox{\citeff}\\
28&\href{https://arxiv.org/src/1906.00207v4/anc/extrema/S1362365/physics.pdf}{S1362365}&&U(1)_2&VSC&26&\href{https://arxiv.org/src/1906.0207v4/anc/extrema/S1362365/location.py.txt}{10}&&241&\arxivpageref{S:S1362365}&\mbox{\citefe}\\
29&\href{https://arxiv.org/src/1906.00207v4/anc/extrema/S1363782/physics.pdf}{S1363782}&&&&&\href{https://arxiv.org/src/1906.0207v4/anc/extrema/S1363782/location.py.txt}{18}&&230&\arxivpageref{S:S1363782}&\mbox{\citefe}\\
30&\href{https://arxiv.org/src/1906.00207v4/anc/extrema/S1366864/physics.pdf}{S1366864}&&&&&\href{https://arxiv.org/src/1906.0207v4/anc/extrema/S1366864/location.py.txt}{9}&&24&\arxivpageref{S:S1366864}&\mbox{\citefe}\\
\hline
31&\href{https://arxiv.org/src/1906.00207v4/anc/extrema/S1367611/physics.pdf}{S1367611}&&&&&\href{https://arxiv.org/src/1906.0207v4/anc/extrema/S1367611/location.py.txt}{13}&&38&\arxivpageref{S:S1367611}&\mbox{\citeff}\\
32&\href{https://arxiv.org/src/1906.00207v4/anc/extrema/S1379439/physics.pdf}{S1379439}&&&&&\href{https://arxiv.org/src/1906.0207v4/anc/extrema/S1379439/location.py.txt}{10}&&2398&\arxivpageref{S:S1379439}&\mbox{\citefe}\\
33&\href{https://arxiv.org/src/1906.00207v4/anc/extrema/S1384096/physics.pdf}{S1384096}&1&SO(3)&VSC&10&\href{https://arxiv.org/src/1906.0207v4/anc/extrema/S1384096/location.py.txt}{6}&3^4&522&\arxivpageref{S:S1384096}&*\\
34&\href{https://arxiv.org/src/1906.00207v4/anc/extrema/S1384135/physics.pdf}{S1384135}&&U(1)_2&VSC&26&\href{https://arxiv.org/src/1906.0207v4/anc/extrema/S1384135/location.py.txt}{11}&&227&\arxivpageref{S:S1384135}&\mbox{\citefe}\\
35&\href{https://arxiv.org/src/1906.00207v4/anc/extrema/S1400000/physics.pdf}{S1400000}&0&SO(4)&SC&4&\href{https://arxiv.org/src/1906.0207v4/anc/extrema/S1400000/location.py.txt}{1}&1&2397&\arxivpageref{S:S1400000}&\mbox{\xcite{warner1984some}{https://doi.org/10.1016/0550-3213(84)90286-4}}\\
\hline
36&\href{https://arxiv.org/src/1906.00207v4/anc/extrema/S1400056/physics.pdf}{S1400056}&&&&&\href{https://arxiv.org/src/1906.0207v4/anc/extrema/S1400056/location.py.txt}{18}&&149&\arxivpageref{S:S1400056}&\mbox{\citefe}\\
37&\href{https://arxiv.org/src/1906.00207v4/anc/extrema/S1402217/physics.pdf}{S1402217}&&U(1)_6&VC&20&\href{https://arxiv.org/src/1906.0207v4/anc/extrema/S1402217/location.py.txt}{6}&5^4&245&\arxivpageref{S:S1402217}&\mbox{\citefe}\\
38&\href{https://arxiv.org/src/1906.00207v4/anc/extrema/S1424025/physics.pdf}{S1424025}&&SO(3)&VSC&10&\href{https://arxiv.org/src/1906.0207v4/anc/extrema/S1424025/location.py.txt}{6}&&232&\arxivpageref{S:S1424025}&\mbox{\citefe}\\
39&\href{https://arxiv.org/src/1906.00207v4/anc/extrema/S1441574/physics.pdf}{S1441574}&&U(1)_4&VC&36&\href{https://arxiv.org/src/1906.0207v4/anc/extrema/S1441574/location.py.txt}{11}&&148&\arxivpageref{S:S1441574}&\mbox{\citefe}\\
40&\href{https://arxiv.org/src/1906.00207v4/anc/extrema/S1442018/physics.pdf}{S1442018}&&U(1)_4&VC&36&\href{https://arxiv.org/src/1906.0207v4/anc/extrema/S1442018/location.py.txt}{12}&&228&\arxivpageref{S:S1442018}&\mbox{\citefe}\\
\hline
41&\href{https://arxiv.org/src/1906.00207v4/anc/extrema/S1443834/physics.pdf}{S1443834}&&U(1)_6&VS&20&\href{https://arxiv.org/src/1906.0207v4/anc/extrema/S1443834/location.py.txt}{10}&&23&\arxivpageref{S:S1443834}&\mbox{\citefe}\\
42&\href{https://arxiv.org/src/1906.00207v4/anc/extrema/S1464498/physics.pdf}{S1464498}&&&&&\href{https://arxiv.org/src/1906.0207v4/anc/extrema/S1464498/location.py.txt}{7}&4^2&2398&\arxivpageref{S:S1464498}&\mbox{\citefe}\\
43&\href{https://arxiv.org/src/1906.00207v4/anc/extrema/S1465354/physics.pdf}{S1465354}&&&&&\href{https://arxiv.org/src/1906.0207v4/anc/extrema/S1465354/location.py.txt}{19}&&24&\arxivpageref{S:S1465354}&\mbox{\citefe}\\
44&\href{https://arxiv.org/src/1906.00207v4/anc/extrema/S1469693/physics.pdf}{S1469693}&&U(1)_2&VSC&26&\href{https://arxiv.org/src/1906.0207v4/anc/extrema/S1469693/location.py.txt}{6}&1^2&2398&\arxivpageref{S:S1469693}&\mbox{\citefe}\\
45&\href{https://arxiv.org/src/1906.00207v4/anc/extrema/S1470986/physics.pdf}{S1470986}&&&&&\href{https://arxiv.org/src/1906.0207v4/anc/extrema/S1470986/location.py.txt}{8}&&2397&\arxivpageref{S:S1470986}&*\\
\hline
46&\href{https://arxiv.org/src/1906.00207v4/anc/extrema/S1473607/physics.pdf}{S1473607}&&&&&\href{https://arxiv.org/src/1906.0207v4/anc/extrema/S1473607/location.py.txt}{8}&7^4&2398&\arxivpageref{S:S1473607}&*\\
47&\href{https://arxiv.org/src/1906.00207v4/anc/extrema/S1474271/physics.pdf}{S1474271}&&&&&\href{https://arxiv.org/src/1906.0207v4/anc/extrema/S1474271/location.py.txt}{11}&&2398&\arxivpageref{S:S1474271}&*\\
48&\href{https://arxiv.org/src/1906.00207v4/anc/extrema/S1477609/physics.pdf}{S1477609}&&&&&\href{https://arxiv.org/src/1906.0207v4/anc/extrema/S1477609/location.py.txt}{6}&&228&\arxivpageref{S:S1477609}&*\\
49&\href{https://arxiv.org/src/1906.00207v4/anc/extrema/S1497038/physics.pdf}{S1497038}&&U(1)_2&VSC&26&\href{https://arxiv.org/src/1906.0207v4/anc/extrema/S1497038/location.py.txt}{12}&&17&\arxivpageref{S:S1497038}&\mbox{\citeff}\\
50&\href{https://arxiv.org/src/1906.00207v4/anc/extrema/S1514242/physics.pdf}{S1514242}&&&&&\href{https://arxiv.org/src/1906.0207v4/anc/extrema/S1514242/location.py.txt}{8}&&226&\arxivpageref{S:S1514242}&*\\
\hline
51&\href{https://arxiv.org/src/1906.00207v4/anc/extrema/S1571627/physics.pdf}{S1571627}&&U(1)_4&VS&36&\href{https://arxiv.org/src/1906.0207v4/anc/extrema/S1571627/location.py.txt}{5}&11^2&2397&\arxivpageref{S:S1571627}&*\\
52&\href{https://arxiv.org/src/1906.00207v4/anc/extrema/S1586253/physics.pdf}{S1586253}&&&&&\href{https://arxiv.org/src/1906.0207v4/anc/extrema/S1586253/location.py.txt}{15}&&237&\arxivpageref{S:S1586253}&*\\
53&\href{https://arxiv.org/src/1906.00207v4/anc/extrema/S1588533/physics.pdf}{S1588533}&&&&&\href{https://arxiv.org/src/1906.0207v4/anc/extrema/S1588533/location.py.txt}{12}&&38&\arxivpageref{S:S1588533}&*\\
54&\href{https://arxiv.org/src/1906.00207v4/anc/extrema/S1596185/physics.pdf}{S1596185}&&&&&\href{https://arxiv.org/src/1906.0207v4/anc/extrema/S1596185/location.py.txt}{11}&&216&\arxivpageref{S:S1596185}&*\\
55&\href{https://arxiv.org/src/1906.00207v4/anc/extrema/S1600000/physics.pdf}{S1600000}&&&&&\href{https://arxiv.org/src/1906.0207v4/anc/extrema/S1600000/location.py.txt}{3}&1&2398&\arxivpageref{S:S1600000}&*\\
\hline
56&\href{https://arxiv.org/src/1906.00207v4/anc/extrema/S1603495/physics.pdf}{S1603495}&&&&&\href{https://arxiv.org/src/1906.0207v4/anc/extrema/S1603495/location.py.txt}{10}&&2398&\arxivpageref{S:S1603495}&*\\
57&\href{https://arxiv.org/src/1906.00207v4/anc/extrema/S1609267/physics.pdf}{S1609267}&&&&&\href{https://arxiv.org/src/1906.0207v4/anc/extrema/S1609267/location.py.txt}{16}&&2397&\arxivpageref{S:S1609267}&*\\
58&\href{https://arxiv.org/src/1906.00207v4/anc/extrema/S1611549/physics.pdf}{S1611549}&&U(1)_6&VS&20&\href{https://arxiv.org/src/1906.0207v4/anc/extrema/S1611549/location.py.txt}{6}&&2398&\arxivpageref{S:S1611549}&*\\
59&\href{https://arxiv.org/src/1906.00207v4/anc/extrema/S1624005/physics.pdf}{S1624005}&&&&&\href{https://arxiv.org/src/1906.0207v4/anc/extrema/S1624005/location.py.txt}{13}&&596&\arxivpageref{S:S1624005}&*\\
60&\href{https://arxiv.org/src/1906.00207v4/anc/extrema/S1624009/physics.pdf}{S1624009}&&&&&\href{https://arxiv.org/src/1906.0207v4/anc/extrema/S1624009/location.py.txt}{10}&&204&\arxivpageref{S:S1624009}&*\\
\hline
61&\href{https://arxiv.org/src/1906.00207v4/anc/extrema/S1627388/physics.pdf}{S1627388}&&&&&\href{https://arxiv.org/src/1906.0207v4/anc/extrema/S1627388/location.py.txt}{13}&&2398&\arxivpageref{S:S1627388}&*\\
62&\href{https://arxiv.org/src/1906.00207v4/anc/extrema/S1637792/physics.pdf}{S1637792}&&&&&\href{https://arxiv.org/src/1906.0207v4/anc/extrema/S1637792/location.py.txt}{11}&&17&\arxivpageref{S:S1637792}&*\\
63&\href{https://arxiv.org/src/1906.00207v4/anc/extrema/S1637802/physics.pdf}{S1637802}&&&&&\href{https://arxiv.org/src/1906.0207v4/anc/extrema/S1637802/location.py.txt}{7}&&257&\arxivpageref{S:S1637802}&*\\
64&\href{https://arxiv.org/src/1906.00207v4/anc/extrema/S1641445/physics.pdf}{S1641445}&&&&&\href{https://arxiv.org/src/1906.0207v4/anc/extrema/S1641445/location.py.txt}{18}&&24&\arxivpageref{S:S1641445}&\mbox{\citeff}\\
65&\href{https://arxiv.org/src/1906.00207v4/anc/extrema/S1650530/physics.pdf}{S1650530}&&&&&\href{https://arxiv.org/src/1906.0207v4/anc/extrema/S1650530/location.py.txt}{16}&&146&\arxivpageref{S:S1650530}&*\\
\hline
66&\href{https://arxiv.org/src/1906.00207v4/anc/extrema/S1650772/physics.pdf}{S1650772}&&&&&\href{https://arxiv.org/src/1906.0207v4/anc/extrema/S1650772/location.py.txt}{14}&&256&\arxivpageref{S:S1650772}&*\\
67&\href{https://arxiv.org/src/1906.00207v4/anc/extrema/S1652031/physics.pdf}{S1652031}&&&&&\href{https://arxiv.org/src/1906.0207v4/anc/extrema/S1652031/location.py.txt}{18}&&200&\arxivpageref{S:S1652031}&*\\
68&\href{https://arxiv.org/src/1906.00207v4/anc/extrema/S1652212/physics.pdf}{S1652212}&&&&&\href{https://arxiv.org/src/1906.0207v4/anc/extrema/S1652212/location.py.txt}{8}&&2397&\arxivpageref{S:S1652212}&*\\
69&\href{https://arxiv.org/src/1906.00207v4/anc/extrema/S1671973/physics.pdf}{S1671973}&&U(1)_4&VS&36&\href{https://arxiv.org/src/1906.0207v4/anc/extrema/S1671973/location.py.txt}{7}&&2398&\arxivpageref{S:S1671973}&*\\
70&\href{https://arxiv.org/src/1906.00207v4/anc/extrema/S1691171/physics.pdf}{S1691171}&&&&&\href{https://arxiv.org/src/1906.0207v4/anc/extrema/S1691171/location.py.txt}{12}&&216&\arxivpageref{S:S1691171}&*\\
\hline
71&\href{https://arxiv.org/src/1906.00207v4/anc/extrema/S1775878/physics.pdf}{S1775878}&&&&&\href{https://arxiv.org/src/1906.0207v4/anc/extrema/S1775878/location.py.txt}{10}&&26&\arxivpageref{S:S1775878}&*\\
72&\href{https://arxiv.org/src/1906.00207v4/anc/extrema/S1775885/physics.pdf}{S1775885}&&&&&\href{https://arxiv.org/src/1906.0207v4/anc/extrema/S1775885/location.py.txt}{16}&&204&\arxivpageref{S:S1775885}&*\\
73&\href{https://arxiv.org/src/1906.00207v4/anc/extrema/S1787646/physics.pdf}{S1787646}&&&&&\href{https://arxiv.org/src/1906.0207v4/anc/extrema/S1787646/location.py.txt}{13}&&26&\arxivpageref{S:S1787646}&\mbox{\citeff}\\
74&\href{https://arxiv.org/src/1906.00207v4/anc/extrema/S1800000/physics.pdf}{S1800000}&&&&&\href{https://arxiv.org/src/1906.0207v4/anc/extrema/S1800000/location.py.txt}{4}&1&258&\arxivpageref{S:S1800000}&*\\
75&\href{https://arxiv.org/src/1906.00207v4/anc/extrema/S1805269/physics.pdf}{S1805269}&&&&&\href{https://arxiv.org/src/1906.0207v4/anc/extrema/S1805269/location.py.txt}{9}&&26&\arxivpageref{S:S1805269}&\mbox{\citeff}\\
\hline
76&\href{https://arxiv.org/src/1906.00207v4/anc/extrema/S1810641/physics.pdf}{S1810641}&&U(1)_2&VSC&26&\href{https://arxiv.org/src/1906.0207v4/anc/extrema/S1810641/location.py.txt}{8}&&2396&\arxivpageref{S:S1810641}&*\\
77&\href{https://arxiv.org/src/1906.00207v4/anc/extrema/S1880177/physics.pdf}{S1880177}&&U(1)_4&VC&36&\href{https://arxiv.org/src/1906.0207v4/anc/extrema/S1880177/location.py.txt}{12}&&146&\arxivpageref{S:S1880177}&*\\
78&\href{https://arxiv.org/src/1906.00207v4/anc/extrema/S1889269/physics.pdf}{S1889269}&&U(1)_4&VC&36&\href{https://arxiv.org/src/1906.0207v4/anc/extrema/S1889269/location.py.txt}{10}&4^4&220&\arxivpageref{S:S1889269}&*\\
79&\href{https://arxiv.org/src/1906.00207v4/anc/extrema/S2006988/physics.pdf}{S2006988}&&U(1)_4&VSC&14&\href{https://arxiv.org/src/1906.0207v4/anc/extrema/S2006988/location.py.txt}{9}&&2397&\arxivpageref{S:S2006988}&*\\
80&\href{https://arxiv.org/src/1906.00207v4/anc/extrema/S2040656/physics.pdf}{S2040656}&&&&&\href{https://arxiv.org/src/1906.0207v4/anc/extrema/S2040656/location.py.txt}{26}&&21&\arxivpageref{S:S2040656}&*\\
\hline
81&\href{https://arxiv.org/src/1906.00207v4/anc/extrema/S2043647/physics.pdf}{S2043647}&&&&&\href{https://arxiv.org/src/1906.0207v4/anc/extrema/S2043647/location.py.txt}{23}&&22&\arxivpageref{S:S2043647}&*\\
82&\href{https://arxiv.org/src/1906.00207v4/anc/extrema/S2043965/physics.pdf}{S2043965}&&&&&\href{https://arxiv.org/src/1906.0207v4/anc/extrema/S2043965/location.py.txt}{19}&&21&\arxivpageref{S:S2043965}&*\\
83&\href{https://arxiv.org/src/1906.00207v4/anc/extrema/S2045402/physics.pdf}{S2045402}&&U(1)_4&VSC&14&\href{https://arxiv.org/src/1906.0207v4/anc/extrema/S2045402/location.py.txt}{9}&&2397&\arxivpageref{S:S2045402}&*\\
84&\href{https://arxiv.org/src/1906.00207v4/anc/extrema/S2054312/physics.pdf}{S2054312}&&&&&\href{https://arxiv.org/src/1906.0207v4/anc/extrema/S2054312/location.py.txt}{17}&&257&\arxivpageref{S:S2054312}&*\\
85&\href{https://arxiv.org/src/1906.00207v4/anc/extrema/S2054714/physics.pdf}{S2054714}&&U(1)_4&VSC&14&\href{https://arxiv.org/src/1906.0207v4/anc/extrema/S2054714/location.py.txt}{6}&&2397&\arxivpageref{S:S2054714}&*\\
\hline
86&\href{https://arxiv.org/src/1906.00207v4/anc/extrema/S2074862/physics.pdf}{S2074862}&&&&&\href{https://arxiv.org/src/1906.0207v4/anc/extrema/S2074862/location.py.txt}{13}&&837&\arxivpageref{S:S2074862}&*\\
87&\href{https://arxiv.org/src/1906.00207v4/anc/extrema/S2095412/physics.pdf}{S2095412}&&&&&\href{https://arxiv.org/src/1906.0207v4/anc/extrema/S2095412/location.py.txt}{15}&&2397&\arxivpageref{S:S2095412}&*\\
88&\href{https://arxiv.org/src/1906.00207v4/anc/extrema/S2099077/physics.pdf}{S2099077}&&U(1)_2\times U(1)_4&VS&12&\href{https://arxiv.org/src/1906.0207v4/anc/extrema/S2099077/location.py.txt}{6}&&256&\arxivpageref{S:S2099077}&*\\
89&\href{https://arxiv.org/src/1906.00207v4/anc/extrema/S2099419/physics.pdf}{S2099419}&&&&&\href{https://arxiv.org/src/1906.0207v4/anc/extrema/S2099419/location.py.txt}{16}&&195&\arxivpageref{S:S2099419}&*\\
90&\href{https://arxiv.org/src/1906.00207v4/anc/extrema/S2099421/physics.pdf}{S2099421}&&&&&\href{https://arxiv.org/src/1906.0207v4/anc/extrema/S2099421/location.py.txt}{12}&&22&\arxivpageref{S:S2099421}&*\\
\hline
91&\href{https://arxiv.org/src/1906.00207v4/anc/extrema/S2101265/physics.pdf}{S2101265}&&&&&\href{https://arxiv.org/src/1906.0207v4/anc/extrema/S2101265/location.py.txt}{13}&&21&\arxivpageref{S:S2101265}&*\\
92&\href{https://arxiv.org/src/1906.00207v4/anc/extrema/S2118474/physics.pdf}{S2118474}&&&&&\href{https://arxiv.org/src/1906.0207v4/anc/extrema/S2118474/location.py.txt}{16}&&243&\arxivpageref{S:S2118474}&*\\
93&\href{https://arxiv.org/src/1906.00207v4/anc/extrema/S2121742/physics.pdf}{S2121742}&&&&&\href{https://arxiv.org/src/1906.0207v4/anc/extrema/S2121742/location.py.txt}{13}&&235&\arxivpageref{S:S2121742}&*\\
94&\href{https://arxiv.org/src/1906.00207v4/anc/extrema/S2126597/physics.pdf}{S2126597}&&&&&\href{https://arxiv.org/src/1906.0207v4/anc/extrema/S2126597/location.py.txt}{9}&&74&\arxivpageref{S:S2126597}&\mbox{\citeff}\\
95&\href{https://arxiv.org/src/1906.00207v4/anc/extrema/S2135328/physics.pdf}{S2135328}&&&&&\href{https://arxiv.org/src/1906.0207v4/anc/extrema/S2135328/location.py.txt}{26}&&224&\arxivpageref{S:S2135328}&*\\
\hline
96&\href{https://arxiv.org/src/1906.00207v4/anc/extrema/S2135978/physics.pdf}{S2135978}&&&&&\href{https://arxiv.org/src/1906.0207v4/anc/extrema/S2135978/location.py.txt}{26}&&19&\arxivpageref{S:S2135978}&*\\
97&\href{https://arxiv.org/src/1906.00207v4/anc/extrema/S2140848/physics.pdf}{S2140848}&&&&&\href{https://arxiv.org/src/1906.0207v4/anc/extrema/S2140848/location.py.txt}{9}&&24&\arxivpageref{S:S2140848}&\mbox{\citeff}\\
98&\href{https://arxiv.org/src/1906.00207v4/anc/extrema/S2144361/physics.pdf}{S2144361}&&&&&\href{https://arxiv.org/src/1906.0207v4/anc/extrema/S2144361/location.py.txt}{27}&&18&\arxivpageref{S:S2144361}&*\\
99&\href{https://arxiv.org/src/1906.00207v4/anc/extrema/S2144749/physics.pdf}{S2144749}&&&&&\href{https://arxiv.org/src/1906.0207v4/anc/extrema/S2144749/location.py.txt}{27}&&18&\arxivpageref{S:S2144749}&*\\
100&\href{https://arxiv.org/src/1906.00207v4/anc/extrema/S2145935/physics.pdf}{S2145935}&&&&&\href{https://arxiv.org/src/1906.0207v4/anc/extrema/S2145935/location.py.txt}{18}&&149&\arxivpageref{S:S2145935}&*\\
\hline
101&\href{https://arxiv.org/src/1906.00207v4/anc/extrema/S2147655/physics.pdf}{S2147655}&&&&&\href{https://arxiv.org/src/1906.0207v4/anc/extrema/S2147655/location.py.txt}{18}&&222&\arxivpageref{S:S2147655}&*\\
102&\href{https://arxiv.org/src/1906.00207v4/anc/extrema/S2153574/physics.pdf}{S2153574}&&&&&\href{https://arxiv.org/src/1906.0207v4/anc/extrema/S2153574/location.py.txt}{17}&&2397&\arxivpageref{S:S2153574}&*\\
103&\href{https://arxiv.org/src/1906.00207v4/anc/extrema/S2154972/physics.pdf}{S2154972}&&&&&\href{https://arxiv.org/src/1906.0207v4/anc/extrema/S2154972/location.py.txt}{12}&&2397&\arxivpageref{S:S2154972}&*\\
104&\href{https://arxiv.org/src/1906.00207v4/anc/extrema/S2160231/physics.pdf}{S2160231}&&&&&\href{https://arxiv.org/src/1906.0207v4/anc/extrema/S2160231/location.py.txt}{18}&&226&\arxivpageref{S:S2160231}&*\\
105&\href{https://arxiv.org/src/1906.00207v4/anc/extrema/S2171187/physics.pdf}{S2171187}&&&&&\href{https://arxiv.org/src/1906.0207v4/anc/extrema/S2171187/location.py.txt}{13}&&236&\arxivpageref{S:S2171187}&*\\
\hline
106&\href{https://arxiv.org/src/1906.00207v4/anc/extrema/S2178669/physics.pdf}{S2178669}&&&&&\href{https://arxiv.org/src/1906.0207v4/anc/extrema/S2178669/location.py.txt}{11}&&257&\arxivpageref{S:S2178669}&*\\
107&\href{https://arxiv.org/src/1906.00207v4/anc/extrema/S2183495/physics.pdf}{S2183495}&&&&&\href{https://arxiv.org/src/1906.0207v4/anc/extrema/S2183495/location.py.txt}{18}&&234&\arxivpageref{S:S2183495}&*\\
108&\href{https://arxiv.org/src/1906.00207v4/anc/extrema/S2206714/physics.pdf}{S2206714}&&&&&\href{https://arxiv.org/src/1906.0207v4/anc/extrema/S2206714/location.py.txt}{10}&&2396&\arxivpageref{S:S2206714}&*\\
109&\href{https://arxiv.org/src/1906.00207v4/anc/extrema/S2215486/physics.pdf}{S2215486}&&U(1)_4&VSC&14&\href{https://arxiv.org/src/1906.0207v4/anc/extrema/S2215486/location.py.txt}{9}&&2397&\arxivpageref{S:S2215486}&*\\
110&\href{https://arxiv.org/src/1906.00207v4/anc/extrema/S2240836/physics.pdf}{S2240836}&&&&&\href{https://arxiv.org/src/1906.0207v4/anc/extrema/S2240836/location.py.txt}{18}&&21&\arxivpageref{S:S2240836}&*\\
\hline
111&\href{https://arxiv.org/src/1906.00207v4/anc/extrema/S2241557/physics.pdf}{S2241557}&&&&&\href{https://arxiv.org/src/1906.0207v4/anc/extrema/S2241557/location.py.txt}{11}&&2397&\arxivpageref{S:S2241557}&*\\
112&\href{https://arxiv.org/src/1906.00207v4/anc/extrema/S2262631/physics.pdf}{S2262631}&&&&&\href{https://arxiv.org/src/1906.0207v4/anc/extrema/S2262631/location.py.txt}{13}&&2397&\arxivpageref{S:S2262631}&*\\
113&\href{https://arxiv.org/src/1906.00207v4/anc/extrema/S2279257/physics.pdf}{S2279257}&&U(1)_1&SC&30&\href{https://arxiv.org/src/1906.0207v4/anc/extrema/S2279257/location.py.txt}{12}&&17&\arxivpageref{S:S2279257}&*\\
114&\href{https://arxiv.org/src/1906.00207v4/anc/extrema/S2279859/physics.pdf}{S2279859}&&U(1)_1&SC&30&\href{https://arxiv.org/src/1906.0207v4/anc/extrema/S2279859/location.py.txt}{7}&&216&\arxivpageref{S:S2279859}&*\\
115&\href{https://arxiv.org/src/1906.00207v4/anc/extrema/S2291945/physics.pdf}{S2291945}&&U(1)_6&VS&20&\href{https://arxiv.org/src/1906.0207v4/anc/extrema/S2291945/location.py.txt}{6}&&241&\arxivpageref{S:S2291945}&*\\
\hline
116&\href{https://arxiv.org/src/1906.00207v4/anc/extrema/S2308861/physics.pdf}{S2308861}&&&&&\href{https://arxiv.org/src/1906.0207v4/anc/extrema/S2308861/location.py.txt}{18}&&153&\arxivpageref{S:S2308861}&*\\
117&\href{https://arxiv.org/src/1906.00207v4/anc/extrema/S2309630/physics.pdf}{S2309630}&&&&&\href{https://arxiv.org/src/1906.0207v4/anc/extrema/S2309630/location.py.txt}{12}&&2397&\arxivpageref{S:S2309630}&*\\
118&\href{https://arxiv.org/src/1906.00207v4/anc/extrema/S2335620/physics.pdf}{S2335620}&&&&&\href{https://arxiv.org/src/1906.0207v4/anc/extrema/S2335620/location.py.txt}{18}&&244&\arxivpageref{S:S2335620}&*\\
119&\href{https://arxiv.org/src/1906.00207v4/anc/extrema/S2348985/physics.pdf}{S2348985}&&&&&\href{https://arxiv.org/src/1906.0207v4/anc/extrema/S2348985/location.py.txt}{12}&&28&\arxivpageref{S:S2348985}&*\\
120&\href{https://arxiv.org/src/1906.00207v4/anc/extrema/S2389014/physics.pdf}{S2389014}&&U(1)_4&VS&36&\href{https://arxiv.org/src/1906.0207v4/anc/extrema/S2389014/location.py.txt}{15}&&24&\arxivpageref{S:S2389014}&*\\
\hline
121&\href{https://arxiv.org/src/1906.00207v4/anc/extrema/S2389433/physics.pdf}{S2389433}&&U(1)_4&VS&36&\href{https://arxiv.org/src/1906.0207v4/anc/extrema/S2389433/location.py.txt}{9}&&2396&\arxivpageref{S:S2389433}&*\\
122&\href{https://arxiv.org/src/1906.00207v4/anc/extrema/S2395245/physics.pdf}{S2395245}&&&&&\href{https://arxiv.org/src/1906.0207v4/anc/extrema/S2395245/location.py.txt}{13}&&233&\arxivpageref{S:S2395245}&*\\
123&\href{https://arxiv.org/src/1906.00207v4/anc/extrema/S2411233/physics.pdf}{S2411233}&&&&&\href{https://arxiv.org/src/1906.0207v4/anc/extrema/S2411233/location.py.txt}{13}&&2396&\arxivpageref{S:S2411233}&*\\
124&\href{https://arxiv.org/src/1906.00207v4/anc/extrema/S2416856/physics.pdf}{S2416856}&&&&&\href{https://arxiv.org/src/1906.0207v4/anc/extrema/S2416856/location.py.txt}{6}&&193&\arxivpageref{S:S2416856}&*\\
125&\href{https://arxiv.org/src/1906.00207v4/anc/extrema/S2416940/physics.pdf}{S2416940}&&&&&\href{https://arxiv.org/src/1906.0207v4/anc/extrema/S2416940/location.py.txt}{9}&&226&\arxivpageref{S:S2416940}&*\\
\hline
126&\href{https://arxiv.org/src/1906.00207v4/anc/extrema/S2420275/physics.pdf}{S2420275}&&&&&\href{https://arxiv.org/src/1906.0207v4/anc/extrema/S2420275/location.py.txt}{17}&&836&\arxivpageref{S:S2420275}&*\\
127&\href{https://arxiv.org/src/1906.00207v4/anc/extrema/S2423138/physics.pdf}{S2423138}&&&&&\href{https://arxiv.org/src/1906.0207v4/anc/extrema/S2423138/location.py.txt}{13}&&2396&\arxivpageref{S:S2423138}&*\\
128&\href{https://arxiv.org/src/1906.00207v4/anc/extrema/S2435197/physics.pdf}{S2435197}&&U(1)_4&VSC&14&\href{https://arxiv.org/src/1906.0207v4/anc/extrema/S2435197/location.py.txt}{10}&&2397&\arxivpageref{S:S2435197}&*\\
129&\href{https://arxiv.org/src/1906.00207v4/anc/extrema/S2435907/physics.pdf}{S2435907}&&&&&\href{https://arxiv.org/src/1906.0207v4/anc/extrema/S2435907/location.py.txt}{11}&&207&\arxivpageref{S:S2435907}&*\\
130&\href{https://arxiv.org/src/1906.00207v4/anc/extrema/S2440304/physics.pdf}{S2440304}&&&&&\href{https://arxiv.org/src/1906.0207v4/anc/extrema/S2440304/location.py.txt}{16}&&230&\arxivpageref{S:S2440304}&*\\
\hline
131&\href{https://arxiv.org/src/1906.00207v4/anc/extrema/S2447241/physics.pdf}{S2447241}&&&&&\href{https://arxiv.org/src/1906.0207v4/anc/extrema/S2447241/location.py.txt}{18}&&226&\arxivpageref{S:S2447241}&*\\
132&\href{https://arxiv.org/src/1906.00207v4/anc/extrema/S2457396/physics.pdf}{S2457396}&&U(1)_2&VSC&26&\href{https://arxiv.org/src/1906.0207v4/anc/extrema/S2457396/location.py.txt}{10}&&225&\arxivpageref{S:S2457396}&*\\
133&\href{https://arxiv.org/src/1906.00207v4/anc/extrema/S2476056/physics.pdf}{S2476056}&&&&&\href{https://arxiv.org/src/1906.0207v4/anc/extrema/S2476056/location.py.txt}{26}&&17&\arxivpageref{S:S2476056}&*\\
134&\href{https://arxiv.org/src/1906.00207v4/anc/extrema/S2482062/physics.pdf}{S2482062}&&U(1)_2&VSC&26&\href{https://arxiv.org/src/1906.0207v4/anc/extrema/S2482062/location.py.txt}{13}&&219&\arxivpageref{S:S2482062}&*\\
135&\href{https://arxiv.org/src/1906.00207v4/anc/extrema/S2484339/physics.pdf}{S2484339}&&&&&\href{https://arxiv.org/src/1906.0207v4/anc/extrema/S2484339/location.py.txt}{26}&&18&\arxivpageref{S:S2484339}&*\\
\hline
136&\href{https://arxiv.org/src/1906.00207v4/anc/extrema/S2488182/physics.pdf}{S2488182}&&&&&\href{https://arxiv.org/src/1906.0207v4/anc/extrema/S2488182/location.py.txt}{17}&&214&\arxivpageref{S:S2488182}&*\\
137&\href{https://arxiv.org/src/1906.00207v4/anc/extrema/S2488241/physics.pdf}{S2488241}&&&&&\href{https://arxiv.org/src/1906.0207v4/anc/extrema/S2488241/location.py.txt}{26}&&203&\arxivpageref{S:S2488241}&*\\
138&\href{https://arxiv.org/src/1906.00207v4/anc/extrema/S2497037/physics.pdf}{S2497037}&&&&&\href{https://arxiv.org/src/1906.0207v4/anc/extrema/S2497037/location.py.txt}{18}&&17&\arxivpageref{S:S2497037}&*\\
139&\href{https://arxiv.org/src/1906.00207v4/anc/extrema/S2502436/physics.pdf}{S2502436}&&&&&\href{https://arxiv.org/src/1906.0207v4/anc/extrema/S2502436/location.py.txt}{18}&&19&\arxivpageref{S:S2502436}&*\\
140&\href{https://arxiv.org/src/1906.00207v4/anc/extrema/S2503105/physics.pdf}{S2503105}&&SO(3)&VSC&2&\href{https://arxiv.org/src/1906.0207v4/anc/extrema/S2503105/location.py.txt}{3}&3^4&2397&\arxivpageref{S:S2503105}&*\\
\hline
141&\href{https://arxiv.org/src/1906.00207v4/anc/extrema/S2503173/physics.pdf}{S2503173}&&&&&\href{https://arxiv.org/src/1906.0207v4/anc/extrema/S2503173/location.py.txt}{18}&&18&\arxivpageref{S:S2503173}&*\\
142&\href{https://arxiv.org/src/1906.00207v4/anc/extrema/S2511744/physics.pdf}{S2511744}&&U(1)_4&VS&36&\href{https://arxiv.org/src/1906.0207v4/anc/extrema/S2511744/location.py.txt}{9}&&2397&\arxivpageref{S:S2511744}&*\\
143&\href{https://arxiv.org/src/1906.00207v4/anc/extrema/S2512364/physics.pdf}{S2512364}&&&&&\href{https://arxiv.org/src/1906.0207v4/anc/extrema/S2512364/location.py.txt}{10}&&17&\arxivpageref{S:S2512364}&*\\
144&\href{https://arxiv.org/src/1906.00207v4/anc/extrema/S2514936/physics.pdf}{S2514936}&&&&&\href{https://arxiv.org/src/1906.0207v4/anc/extrema/S2514936/location.py.txt}{18}&&223&\arxivpageref{S:S2514936}&\mbox{\citeff}\\
145&\href{https://arxiv.org/src/1906.00207v4/anc/extrema/S2519962/physics.pdf}{S2519962}&&U(1)_2\times U(1)_4&VS&12&\href{https://arxiv.org/src/1906.0207v4/anc/extrema/S2519962/location.py.txt}{6}&&238&\arxivpageref{S:S2519962}&*\\
\hline
146&\href{https://arxiv.org/src/1906.00207v4/anc/extrema/S2522870/physics.pdf}{S2522870}&&U(1)_4&VSC&14&\href{https://arxiv.org/src/1906.0207v4/anc/extrema/S2522870/location.py.txt}{9}&&836&\arxivpageref{S:S2522870}&*\\
147&\href{https://arxiv.org/src/1906.00207v4/anc/extrema/S2535465/physics.pdf}{S2535465}&&U(1)_4&VC&36&\href{https://arxiv.org/src/1906.0207v4/anc/extrema/S2535465/location.py.txt}{15}&&17&\arxivpageref{S:S2535465}&*\\
148&\href{https://arxiv.org/src/1906.00207v4/anc/extrema/S2535919/physics.pdf}{S2535919}&&&&&\href{https://arxiv.org/src/1906.0207v4/anc/extrema/S2535919/location.py.txt}{17}&&218&\arxivpageref{S:S2535919}&*\\
149&\href{https://arxiv.org/src/1906.00207v4/anc/extrema/S2547536/physics.pdf}{S2547536}&&U(1)_2\times U(1)_4&VS&12&\href{https://arxiv.org/src/1906.0207v4/anc/extrema/S2547536/location.py.txt}{6}&&256&\arxivpageref{S:S2547536}&*\\
150&\href{https://arxiv.org/src/1906.00207v4/anc/extrema/S2550397/physics.pdf}{S2550397}&&&&&\href{https://arxiv.org/src/1906.0207v4/anc/extrema/S2550397/location.py.txt}{16}&&223&\arxivpageref{S:S2550397}&*\\
\hline
151&\href{https://arxiv.org/src/1906.00207v4/anc/extrema/S2551136/physics.pdf}{S2551136}&&&&&\href{https://arxiv.org/src/1906.0207v4/anc/extrema/S2551136/location.py.txt}{14}&&23&\arxivpageref{S:S2551136}&*\\
152&\href{https://arxiv.org/src/1906.00207v4/anc/extrema/S2566527/physics.pdf}{S2566527}&&&&&\href{https://arxiv.org/src/1906.0207v4/anc/extrema/S2566527/location.py.txt}{17}&&20&\arxivpageref{S:S2566527}&*\\
153&\href{https://arxiv.org/src/1906.00207v4/anc/extrema/S2584003/physics.pdf}{S2584003}&&&&&\href{https://arxiv.org/src/1906.0207v4/anc/extrema/S2584003/location.py.txt}{13}&&245&\arxivpageref{S:S2584003}&*\\
154&\href{https://arxiv.org/src/1906.00207v4/anc/extrema/S2595657/physics.pdf}{S2595657}&&U(1)_4&VC&36&\href{https://arxiv.org/src/1906.0207v4/anc/extrema/S2595657/location.py.txt}{12}&&2396&\arxivpageref{S:S2595657}&*\\
155&\href{https://arxiv.org/src/1906.00207v4/anc/extrema/S2625940/physics.pdf}{S2625940}&&&&&\href{https://arxiv.org/src/1906.0207v4/anc/extrema/S2625940/location.py.txt}{13}&&245&\arxivpageref{S:S2625940}&*\\
\hline
156&\href{https://arxiv.org/src/1906.00207v4/anc/extrema/S2644794/physics.pdf}{S2644794}&&&&&\href{https://arxiv.org/src/1906.0207v4/anc/extrema/S2644794/location.py.txt}{17}&&19&\arxivpageref{S:S2644794}&*\\
157&\href{https://arxiv.org/src/1906.00207v4/anc/extrema/S2648510/physics.pdf}{S2648510}&&&&&\href{https://arxiv.org/src/1906.0207v4/anc/extrema/S2648510/location.py.txt}{27}&&20&\arxivpageref{S:S2648510}&*\\
158&\href{https://arxiv.org/src/1906.00207v4/anc/extrema/S2650165/physics.pdf}{S2650165}&&U(1)_4&VSC&14&\href{https://arxiv.org/src/1906.0207v4/anc/extrema/S2650165/location.py.txt}{6}&&2396&\arxivpageref{S:S2650165}&*\\
159&\href{https://arxiv.org/src/1906.00207v4/anc/extrema/S2652206/physics.pdf}{S2652206}&&&&&\href{https://arxiv.org/src/1906.0207v4/anc/extrema/S2652206/location.py.txt}{18}&&225&\arxivpageref{S:S2652206}&*\\
160&\href{https://arxiv.org/src/1906.00207v4/anc/extrema/S2652377/physics.pdf}{S2652377}&&&&&\href{https://arxiv.org/src/1906.0207v4/anc/extrema/S2652377/location.py.txt}{26}&&16&\arxivpageref{S:S2652377}&*\\
\hline
161&\href{https://arxiv.org/src/1906.00207v4/anc/extrema/S2653753/physics.pdf}{S2653753}&&&&&\href{https://arxiv.org/src/1906.0207v4/anc/extrema/S2653753/location.py.txt}{17}&&235&\arxivpageref{S:S2653753}&*\\
162&\href{https://arxiv.org/src/1906.00207v4/anc/extrema/S2655334/physics.pdf}{S2655334}&&U(1)_2&VSC&26&\href{https://arxiv.org/src/1906.0207v4/anc/extrema/S2655334/location.py.txt}{14}&&76&\arxivpageref{S:S2655334}&*\\
163&\href{https://arxiv.org/src/1906.00207v4/anc/extrema/S2655918/physics.pdf}{S2655918}&&U(1)_4&VSC&14&\href{https://arxiv.org/src/1906.0207v4/anc/extrema/S2655918/location.py.txt}{9}&&227&\arxivpageref{S:S2655918}&*\\
164&\href{https://arxiv.org/src/1906.00207v4/anc/extrema/S2666807/physics.pdf}{S2666807}&&&&&\href{https://arxiv.org/src/1906.0207v4/anc/extrema/S2666807/location.py.txt}{13}&&225&\arxivpageref{S:S2666807}&*\\
165&\href{https://arxiv.org/src/1906.00207v4/anc/extrema/S2697505/physics.pdf}{S2697505}&&U(1)_4&VS&36&\href{https://arxiv.org/src/1906.0207v4/anc/extrema/S2697505/location.py.txt}{11}&&218&\arxivpageref{S:S2697505}&*\\
\hline
166&\href{https://arxiv.org/src/1906.00207v4/anc/extrema/S2702580/physics.pdf}{S2702580}&&&&&\href{https://arxiv.org/src/1906.0207v4/anc/extrema/S2702580/location.py.txt}{7}&3^2&2396&\arxivpageref{S:S2702580}&*\\
167&\href{https://arxiv.org/src/1906.00207v4/anc/extrema/S2705605/physics.pdf}{S2705605}&&&&&\href{https://arxiv.org/src/1906.0207v4/anc/extrema/S2705605/location.py.txt}{16}&&116&\arxivpageref{S:S2705605}&*\\
168&\href{https://arxiv.org/src/1906.00207v4/anc/extrema/S2707528/physics.pdf}{S2707528}&&U(1)_4&VS&36&\href{https://arxiv.org/src/1906.0207v4/anc/extrema/S2707528/location.py.txt}{9}&&255&\arxivpageref{S:S2707528}&*\\
169&\href{https://arxiv.org/src/1906.00207v4/anc/extrema/S2751936/physics.pdf}{S2751936}&&U(1)_6&VS&20&\href{https://arxiv.org/src/1906.0207v4/anc/extrema/S2751936/location.py.txt}{10}&&2396&\arxivpageref{S:S2751936}&*\\
170&\href{https://arxiv.org/src/1906.00207v4/anc/extrema/S2803140/physics.pdf}{S2803140}&&&&&\href{https://arxiv.org/src/1906.0207v4/anc/extrema/S2803140/location.py.txt}{8}&&2396&\arxivpageref{S:S2803140}&*\\
\hline
171&\href{https://arxiv.org/src/1906.00207v4/anc/extrema/S2849861/physics.pdf}{S2849861}&&&&&\href{https://arxiv.org/src/1906.0207v4/anc/extrema/S2849861/location.py.txt}{18}&&18&\arxivpageref{S:S2849861}&*\\
172&\href{https://arxiv.org/src/1906.00207v4/anc/extrema/S2855585/physics.pdf}{S2855585}&&U(1)_4&VS&36&\href{https://arxiv.org/src/1906.0207v4/anc/extrema/S2855585/location.py.txt}{11}&&2396&\arxivpageref{S:S2855585}&*\\
173&\href{https://arxiv.org/src/1906.00207v4/anc/extrema/S2874576/physics.pdf}{S2874576}&&&&&\href{https://arxiv.org/src/1906.0207v4/anc/extrema/S2874576/location.py.txt}{26}&&226&\arxivpageref{S:S2874576}&*\\
174&\href{https://arxiv.org/src/1906.00207v4/anc/extrema/S2952191/physics.pdf}{S2952191}&&U(1)_6&VS&20&\href{https://arxiv.org/src/1906.0207v4/anc/extrema/S2952191/location.py.txt}{6}&&2396&\arxivpageref{S:S2952191}&*\\
175&\href{https://arxiv.org/src/1906.00207v4/anc/extrema/S2980210/physics.pdf}{S2980210}&&U(1)_4&VS&36&\href{https://arxiv.org/src/1906.0207v4/anc/extrema/S2980210/location.py.txt}{15}&&75&\arxivpageref{S:S2980210}&*\\
\hline
176&\href{https://arxiv.org/src/1906.00207v4/anc/extrema/S3108383/physics.pdf}{S3108383}&&U(1)_4&VS&36&\href{https://arxiv.org/src/1906.0207v4/anc/extrema/S3108383/location.py.txt}{11}&&2396&\arxivpageref{S:S3108383}&*\\
177&\href{https://arxiv.org/src/1906.00207v4/anc/extrema/S3155402/physics.pdf}{S3155402}&&&&&\href{https://arxiv.org/src/1906.0207v4/anc/extrema/S3155402/location.py.txt}{18}&&230&\arxivpageref{S:S3155402}&*\\
178&\href{https://arxiv.org/src/1906.00207v4/anc/extrema/S3166153/physics.pdf}{S3166153}&&&&&\href{https://arxiv.org/src/1906.0207v4/anc/extrema/S3166153/location.py.txt}{12}&&2396&\arxivpageref{S:S3166153}&*\\
179&\href{https://arxiv.org/src/1906.00207v4/anc/extrema/S3254262/physics.pdf}{S3254262}&&&&&\href{https://arxiv.org/src/1906.0207v4/anc/extrema/S3254262/location.py.txt}{13}&&17&\arxivpageref{S:S3254262}&*\\
180&\href{https://arxiv.org/src/1906.00207v4/anc/extrema/S3254576/physics.pdf}{S3254576}&&&&&\href{https://arxiv.org/src/1906.0207v4/anc/extrema/S3254576/location.py.txt}{15}&&220&\arxivpageref{S:S3254576}&*\\
\hline
181&\href{https://arxiv.org/src/1906.00207v4/anc/extrema/S3254929/physics.pdf}{S3254929}&&&&&\href{https://arxiv.org/src/1906.0207v4/anc/extrema/S3254929/location.py.txt}{10}&&2396&\arxivpageref{S:S3254929}&*\\
182&\href{https://arxiv.org/src/1906.00207v4/anc/extrema/S3305153/physics.pdf}{S3305153}&&&&&\href{https://arxiv.org/src/1906.0207v4/anc/extrema/S3305153/location.py.txt}{9}&&2395&\arxivpageref{S:S3305153}&*\\
183&\href{https://arxiv.org/src/1906.00207v4/anc/extrema/S3498681/physics.pdf}{S3498681}&&&&&\href{https://arxiv.org/src/1906.0207v4/anc/extrema/S3498681/location.py.txt}{18}&&234&\arxivpageref{S:S3498681}&*\\
184&\href{https://arxiv.org/src/1906.00207v4/anc/extrema/S3560884/physics.pdf}{S3560884}&&&&&\href{https://arxiv.org/src/1906.0207v4/anc/extrema/S3560884/location.py.txt}{13}&&241&\arxivpageref{S:S3560884}&*\\
185&\href{https://arxiv.org/src/1906.00207v4/anc/extrema/S3640089/physics.pdf}{S3640089}&&U(1)_4&VS&36&\href{https://arxiv.org/src/1906.0207v4/anc/extrema/S3640089/location.py.txt}{11}&&233&\arxivpageref{S:S3640089}&*\\
\hline
186&\href{https://arxiv.org/src/1906.00207v4/anc/extrema/S3777270/physics.pdf}{S3777270}&&&&&\href{https://arxiv.org/src/1906.0207v4/anc/extrema/S3777270/location.py.txt}{16}&&226&\arxivpageref{S:S3777270}&*\\
187&\href{https://arxiv.org/src/1906.00207v4/anc/extrema/S3908381/physics.pdf}{S3908381}&&&&&\href{https://arxiv.org/src/1906.0207v4/anc/extrema/S3908381/location.py.txt}{18}&&234&\arxivpageref{S:S3908381}&*\\
188&\href{https://arxiv.org/src/1906.00207v4/anc/extrema/S4009882/physics.pdf}{S4009882}&&&&&\href{https://arxiv.org/src/1906.0207v4/anc/extrema/S4009882/location.py.txt}{19}&&236&\arxivpageref{S:S4009882}&*\\
189&\href{https://arxiv.org/src/1906.00207v4/anc/extrema/S4045562/physics.pdf}{S4045562}&&&&&\href{https://arxiv.org/src/1906.0207v4/anc/extrema/S4045562/location.py.txt}{18}&&232&\arxivpageref{S:S4045562}&*\\
190&\href{https://arxiv.org/src/1906.00207v4/anc/extrema/S4145121/physics.pdf}{S4145121}&&&&&\href{https://arxiv.org/src/1906.0207v4/anc/extrema/S4145121/location.py.txt}{18}&&21&\arxivpageref{S:S4145121}&*\\
\hline
191&\href{https://arxiv.org/src/1906.00207v4/anc/extrema/S4168086/physics.pdf}{S4168086}&&&&&\href{https://arxiv.org/src/1906.0207v4/anc/extrema/S4168086/location.py.txt}{9}&4^2&227&\arxivpageref{S:S4168086}&*\\
192&\href{https://arxiv.org/src/1906.00207v4/anc/extrema/S4599899/physics.pdf}{S4599899}&&&&&\href{https://arxiv.org/src/1906.0207v4/anc/extrema/S4599899/location.py.txt}{12}&&235&\arxivpageref{S:S4599899}&*\\
\end{longtable}}

\arxivonly{
\section{The list of solutions}

For convenience, the arXiv.org preprint version of this work also
contains the consolidated list of solutions, which would be too
voluminous for journal publication.

\input{solution_data.tex}
}

\bibliographystyle{JHEP}
\bibliography{sugra}

\end{document}